\documentclass[preprint,10pt]{article}
\usepackage[left=1in,top=1in,right=1in,bottom= 1in,includefoot,includehead]{geometry}

\usepackage{changepage}

\usepackage[utf8x]{inputenc}

\usepackage{textcomp,marvosym}

\usepackage{fixltx2e}

\usepackage{amsmath,amssymb}

\newcommand{\be}{\begin{equation}}
\newcommand{\ee}{\end{equation}}

\newcommand\ERtext{Erd\H{o}s-R\'{e}yni~}

\usepackage{cite}

\usepackage{nameref,hyperref}

\usepackage[right]{lineno}

\usepackage{microtype}
\DisableLigatures[f]{encoding = *, family = * }

\usepackage{color}

\usepackage[aboveskip=1pt,labelfont=bf,labelsep=period,justification=raggedright,singlelinecheck=off]{caption}

\usepackage[pdftex]{graphicx} 
\usepackage{hyperref}

\makeatletter
\renewcommand{\@biblabel}[1]{\quad#1.}
\makeatother

\date{}

\begin{document}
\vspace*{0.2in}

{\Large
\textbf\newline{Predicting how and when hidden neurons skew measured synaptic interactions} % Please use "title case" (capitalize all terms in the title except conjunctions, prepositions, and articles).
}
\newline
\\
Braden A. W. Brinkman\textsuperscript{1,2*},
Fred Rieke\textsuperscript{2,3},
Eric Shea-Brown\textsuperscript{1,2,3,4},
Michael A. Buice\textsuperscript{4,1}
\\

\bigskip
\noindent\textbf{1} Department of Applied Mathematics, University of Washington, Seattle, WA, 98195, USA
\\
\textbf{2} Department of Physiology and Biophysics, University of Washington, Seattle, WA, 98195, USA
\\
\textbf{3} Graduate Program in Neuroscience, University of Washington, Seattle, WA, 98195, USA
\\
\textbf{4} Allen Institute for Brain Science, Seattle, WA, 98109, USA
\\
\bigskip

\noindent * bradenb@uw.edu

\section*{Abstract}
\label{sec:abstract}
A major obstacle to understanding neural coding and computation is the fact that experimental recordings typically sample only a small fraction of the neurons in a circuit. Measured neural properties are skewed by interactions between recorded neurons and the ``hidden'' portion of the network. To properly interpret neural data and determine how biological structure gives rise to neural circuit function, we thus need a better understanding of the relationships between measured effective neural properties and the true underlying physiological properties. Here, we focus on how the effective spatiotemporal dynamics of the synaptic interactions between neurons are reshaped by coupling to unobserved neurons. We find that the effective interactions from a pre-synaptic neuron $r'$ to a post-synaptic neuron $r$ can be decomposed into a sum of the true interaction from $r'$ to $r$ plus corrections from every directed path from $r'$ to $r$ through unobserved neurons. Importantly, the resulting formula reveals when the hidden units have---or do not have---major effects on reshaping the interactions among observed neurons.  As a particular example of interest, we derive a formula for the impact of hidden units in random networks with ``strong'' coupling---connection weights that scale with $1/\sqrt{N}$, where $N$ is the network size, precisely the scaling observed in recent experiments. With this quantitative relationship between measured and true interactions, we can study how network properties shape effective interactions, which properties are relevant for neural computations, and how to manipulate effective interactions.

%\linenumbers

% Use "Eq" instead of "Equation" for equation citations.
\section*{Introduction}
Establishing relationships between a network's architecture and its function is a fundamental problem in neuroscience and network science in general. Not only is the architecture of a neural circuit intimately related to its function, but pathologies in wiring between neurons are believed to contribute significantly to circuit dysfunction 
\cite{BassettJNeurosci2008,KramerEpilepsyRes2008,SupekarPLOSCB2008,LoJNeurosci2010,ChavezPRL2010,DouwPLOSOne2010,vanDiessenNeuroImage2013,ReijmerNeurology2013,SeoPLOSOne2013,StamNatRevNeuro2014,WarrenPNAS2014,OldeDubbelinkBrain2014,BernhardEpBeh2015,MedagliaArxiv2017,HoneyNeuroImage2010}.

A major obstacle to uncovering structure-function relationships is the fact that most experiments can only directly observe  small fractions of an active network. 
State-of-the-art methods for determining connections between neurons in living networks infer them by fitting statistical models to neural spiking data \cite{Simoncelli2004,Paninski2004,Pillow2005,KulkarniNetworkCompNeuralSys2007,Pillow2008,Field2010,Vidne2012,Paninski2015,HuangJPhysA2015}. However, the fact that we cannot observe all neurons in a network means that the statistically inferred connections are ``effective'' connections, representing some dynamical relationship between the activity of nodes but not necessarily a true physical connection \cite{DahlhausMetrika2000,EichlerBiolCyber2003,PillowNIPS2007,StevensonCurrOpBio2008,StevensonPLOSCB2012,LiegeoisIEEE2015,HuangJPhysA2015,PetersRSSB2016,FotiMILETS2016}. Intuitively, reverberations through the network must contribute to these effective interactions; our goal in this work is to formalize this intuition and establish a quantitative relationship between measured effective interactions and the true synaptic interactions between neurons. With such a relationship in hand we can study the effective interactions generated by different choices of synaptic properties and circuit architectures, allowing us to not only improve interpretation of experimental measurements but also probe how circuit structure is tied to function. 

\begin{figure*}
 \centering
  \includegraphics[width=0.75\textwidth]{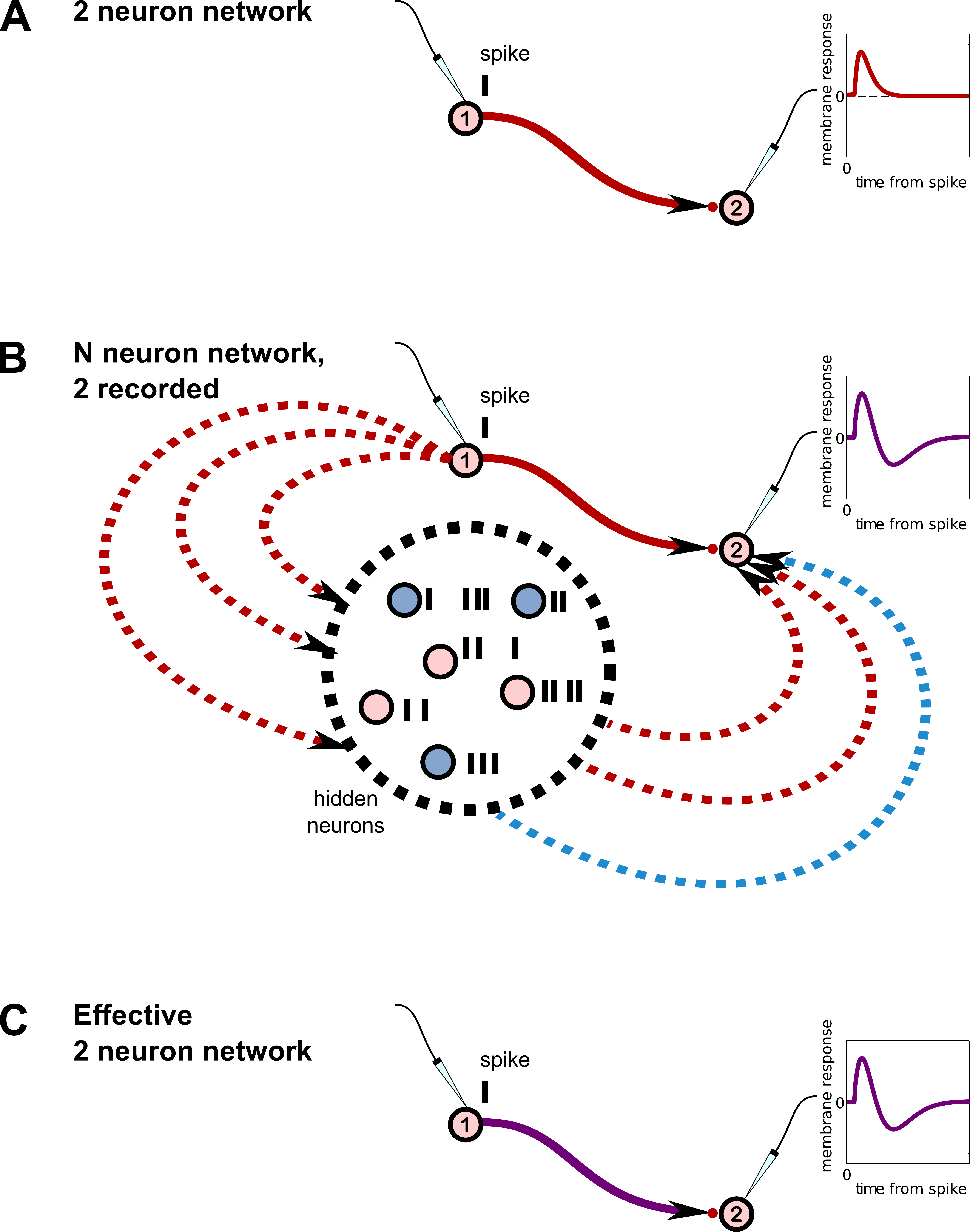}
  \caption{\textbf{The hidden unit problem.} \textbf{A.} In a hypothetical circuit consisting of just two recorded neurons (no hidden neurons), we can measure the strength and time course of the directed interactions between neurons by measuring the response of the post-synaptic neuron's membrane potential to a spike from the pre-synaptic neuron. \textbf{B.} Realistically, there are many more neurons in the network that are unrecorded and hence ``hidden.'' In this schematic, only two neurons are observed. The hidden neurons are driven by input from the presynaptic neuron labeled 1, and provide input to the recorded post-synaptic neuron labeled 2. Because the activity of the hidden neurons is not controlled, the membrane response reflects a combination of neuron 1's direct influence on neuron 2 and its indirect influence through the hidden network. \textbf{C.} The ``effective'' 2 neuron network observed experimentally.}
  \label{fig:introfigure}
\end{figure*}

The intuitive relationship between measured and effective interactions is demonstrated schematically in Fig.~\ref{fig:introfigure}. Fig.~\ref{fig:introfigure}A demonstrates that in a fully-sampled network the directed interactions between neurons---here, the change in membrane potential of the post-synaptic neuron after it receives a spike from the pre-synaptic neuron---can be measured directly, as observation of the complete population means different inputs to a neuron are not conflated. However, as shown in Fig.~\ref{fig:introfigure}B, the vastly more realistic scenario is that the recorded neurons are part of a larger network in which many neurons are unobserved or ``hidden.'' The response of the post-synaptic neuron 2 to a spike from pre-synaptic neuron 1 is a combination of both the direct response to neuron 1's input as well as input from the hidden network driven by neuron 1's spiking. Thus, the measured membrane response of neuron 2 due to a spike fired by neuron 1---which we term the ``effective interaction'' from neuron 1 to 2---may be quite different from the true interaction. It is well-known that circuit connections between recorded neurons, as drawn in Fig.~\ref{fig:introfigure}C, are at best effective circuits that encapsulate the effects of unobserved neurons, but are not necessarily indicative of the true circuit architecture. The formalized relationship we will establish in the Results is given in Fig.~\ref{fig:Jeffeqn}.

Even once we establish a relationship between the effective and true connections, we will in general not be able to use measurements of effective interactions to extrapolate back to a unique set of true connections; at best, we may be able to characterize some of the statistical properties of the full network. The obstacle is that several different networks---different both in terms of architecture and intrinsic neural properties---may give rise to the same network behaviors, a theme of much focus in the neuroscience literature\cite{Prinz2004,GutenkunstPLOSCB2007,ApgarMolBioSyst2010,GutierrezNeuron2013,FisherNeuron2013,MarderDevNeurobio2017}. That is, inferring the connections and intrinsic neural properties in a full network from activity recordings from a subset of neurons is in general an ill-posed problem, possessing several degenerate solutions. Several statistical inference methods have been constructed to attempt to infer the presence of, and connections to, hidden neurons \cite{PillowNIPS2007,DunnPRE2013,TyrchaMathBioEng2014,DunnArxiv2016}; the subset of the degenerate solutions that each of these methods finds will depend on the particular assumptions of the inference method (e.g., the regularization penalties applied). As an example, we demonstrate two small circuit motifs that give rise to nearly identical effective interactions, despite crucial differences between the circuits.

Understanding the effect of hidden neurons on small circuit motifs is only a small part of the hidden neuron puzzle, and a full understanding necessitates scaling up to large circuits containing many different motifs. Having an analytic relationship between true and effective interactions greatly facilitates such analyses by directly studying the structure of the relationship itself, rather than trying to extract insight indirectly through simulations. In particular, in going to large networks we focus on the degree to which hidden neurons skew measured interactions (Fig.~\ref{fig:Janalysis}), and how we can predict the features of effective interactions we expect to measure when recording from only a subset of neurons in a network with hypothesized true interactions (Fig.~\ref{fig:Jeffs}).

Establishing a theoretical relationship between measured and ``true'' interactions will thus enable us to study how one can alter the network properties to reshape the effective interactions, and will be of immediate importance not only for interpreting experimental measurements of synaptic interactions, but for elucidating their role in neural coding. Moreover, understanding how to shape effective interactions between neurons may yield new avenues for altering, in a principled way, the computations performed by a network, which could have applications for treating neurological diseases caused in part by pathological synaptic interactions. 

%%%%%%%%%%%%%%%%%%%%%%%%%%%%%%%%
%% RESULTS 
%%%%%%%%%%%%%%%%%%%%%%%%%%%%%%%%
\section*{Results}
\subsection*{Overview}

Our goal is to derive a relationship between the effective synaptic interactions between recorded neurons and the true synaptic interactions that would be obtained if the network were fully observed. This makes explicit how the synaptic interactions between neurons are modified by unobserved neurons in the network, and under what conditions these modifications are---or are not---significant. We derive this result first, using a probabilistic model of network activity in which all properties are known. We then build intuition by applying our result to two simple networks: a $3$-neuron feedforward-inhibition circuit in which we are able to qualitatively reproduce measurements by Pouille and Scanziani \cite{PouilleScience2001}, and a $4$-neuron circuit that demonstrates how degeneracies in hidden networks are handled within our framework. 

To extend our intuition to larger networks, we then study the effective interactions that would be observed in sparse random networks with $N$ cells and strong synaptic weights that scale as $1/\sqrt{N}$ \cite{vanVreeswijkScience1996,LitwinKumarNatNeuro2012,RosenbaumPRX2014,DeneveNatNeuro2016}, as has been recently observed experimentally \cite{BarralNatNeuro2016}. We show how unobserved neurons significantly reshape the effective synaptic interactions away from the ground-truth interactions. This is not the case with ``classical'' synaptic scaling, in which synaptic strengths are inversely proportional to the number of inputs they receive (assumed $\mathcal O(N)$), as we will also show. (The case of classical scaling has also been studied previously using a different approach in \cite{NykampSIAMJAM2005,NykampMathBiosci2007,NykampSIAMJAM2007,NykampPRE2008}).  

\subsection*{Model}

We model the full network of $N$ neurons as a nonlinear Hawkes process \cite{OckerPLOSCB2017}, commonly known as a ``Generalized linear (point process) model'' in neuroscience, and broadly used to fit neural activity data \cite{ChornoboyBiolCyber1988,Simoncelli2004,Paninski2004,Pillow2005,KulkarniNetworkCompNeuralSys2007,Pillow2008,Field2010,Vidne2012,Paninski2015}. Here we use it as a generative model for network activity, as it approximates common spiking models such as leaky integrate and fire systems driven by noisy inputs \cite{GerstnerBook,OstojicPLOSCB2011}, and is equivalent to current-based leaky integrate-and-fire models with soft-threshold (stochastic) spiking dynamics (see Methods). 

To derive an approximate model for an observed subset of the network, we partition the network into recorded neurons (labeled by indices $r$) and hidden neurons (labeled by indices $h$). Each recorded neuron has an instantaneous firing rate $\lambda_r(t)$ such that the probability that the neuron fires within a small time window $[t,t+dt]$ is $\lambda_r(t) dt$, when conditioned on the inputs to the neuron. The instantaneous firing rate in our model is
\begin{equation}
\lambda_r(t) = \lambda_0 \phi\left(\mu_r + \sum_{r'} J_{r,r'} \ast\dot{n}_{r'}(t) + \sum_{h} J_{r,h} \ast\dot{n}_{h}(t)\right),
\label{eqn:GLMrate}
\end{equation}
where $\lambda_0$ is a characteristic firing rate, $\phi(x)$ is a non-negative, continuous function, $\mu_r$ is a tonic drive that sets the baseline firing rate of the neuron, and $J_{i,j} \ast \dot{n}_j(t) \equiv \int_{-\infty}^\infty dt'~J_{i,j}(t-t') \dot{n}_j(t')$ is the convolution of the synaptic interaction (or ``spike filter'') $J_{i,j}(t)$ with spike train $\dot{n}_j(t)$ \emph{from} pre-synaptic neuron $j$ \emph{to} post-synaptic neuron $i$. In this work we take the tonic drive to be constant in time, and focus on the steady-state network activity in response to this drive. We consider interactions of the form $J_{i,j}(t) \equiv \mathcal J_{i,j}g_{j}(t)$, where the temporal waveforms $g_{j}(t)$ are normalized such that $\int_0^\infty dt~g_{j}(t) = 1$ for all neurons $j$. Because of this normalization, the weight $\mathcal J_{i,j}$ carries units of time. We include self-couplings $J_{i,i}(t)$ not to represent autapses, but to account for intrinsic neural properties such as refractory periods ($\mathcal J_{i,i} < 0$) or burstiness  ($\mathcal J_{i,i} > 0$). The firing rates for the hidden neurons follow the same expression with indices $h$ and $r$ interchanged.

We seek to describe the dynamics of the recorded neurons entirely in terms of their own set of spiking histories, eliminating the dependence on the activity of the hidden neurons. This demands calculating the effective membrane response of the recorded neurons by averaging out the activity of the hidden neurons, \emph{conditioned on the activity of the recorded neurons}. In practice this is intractable to perform exactly \cite{BraviArxiv2016,BraviArxiv2016b, BraviPRE2017}. Here, we use a mean field approximation to calculate the mean input from the hidden neurons (again, conditioned on the activity of the recorded neurons). The value of deriving such a relationship analytically, as opposed to simply numerically determining the effective interactions, is that the resulting expression will give us insight into how the effective interactions decompose into contributions of different network features, how tuning particular features shapes the effective interactions, and conditions under which we expect hidden units to skew our measurements of connectivity in large partially observed networks. 

As shown in detail in the Methods, the instantaneous firing rates of the recorded neurons can then be approximated as $$\lambda_r(t) \approx \lambda_0 \phi\left(\mu^{\rm eff}_r + \sum_{r'} J^{\rm eff}_{r,r'} \ast \dot{n}_{r'}(t) \right).$$
%\begin{linenomath*}
%$$\lambda_r(t) \approx \lambda_0 \phi\left(\mu^{\rm eff}_r + \sum_{r'} J^{\rm eff}_{r,r'} \ast \dot{n}_{r'}(t) \right).$$
%\end{linenomath*}
The effective baselines $\mu^{\rm eff}_r = \mu_r + \sum_{h}\mathcal J_{r,h} \nu_h$, are simply modulated by the net input to the neuron, so we do not focus on them here. The effective coupling filters are given in the frequency domain by
\begin{equation}
\hat{J}^{\rm eff}_{r,r'}(\omega) = \hat{J}_{r,r'}(\omega) + \sum_{h,h'} \hat{J}_{r,h}(\omega)\hat{\Gamma}_{h,h'}(\omega)\hat{J}_{h',r'}(\omega). 
\label{eqn:Jeffgeneral}
\end{equation}
These results hold for any pair of recorded neurons $r'$ and $r$, and any choice of network parameters for which the mean field steady state of the hidden network exists. Here, the $\nu_h$ are the steady-state mean firing rates of the hidden neurons and $\hat{\Gamma}_{h,h'}(\omega)$ is the linear response function of the hidden network to perturbations in the \emph{input}. That is, $\Gamma_{h,h'}(t-t')$ is the linear response of hidden neuron $h$ at time $t$ due to a perturbation to the input of neuron $h'$ at time $t'$, and incorporates the effects of $h'$ propagating its signal to $h$ through other hidden neurons, as demonstrated graphically in Fig.~\ref{fig:Jeffeqn}. Both $\nu_h$ and $\hat{\Gamma}_{h,h'}(\omega)$ are calculated \emph{in the absence of the recorded neurons}. In deriving these results, we have neglected both fluctuations around the mean input from the hidden neurons, as well as higher order filtering of the recorded neuron spikes. For details on the derivations and justification of approximations, see the Methods and Supporting Information (SI).

The effective coupling filters are what we would---in principle---measure experimentally if we observe only a subset of a network, for example by pairwise recordings shown schematically in Fig.~\ref{fig:introfigure}. For larger sets of recorded neurons, interactions between neurons are typically inferred using statistical methods, an extremely nontrivial task \cite{Simoncelli2004,Paninski2004,Pillow2005,KulkarniNetworkCompNeuralSys2007,PillowNIPS2007,Pillow2008,Field2010,Vidne2012,DunnPRE2013,TyrchaMathBioEng2014,Paninski2015}, and details of the fitting procedure could potentially further skew the inferred interactions away from what would be measured by controlled pairwise recordings. We will put aside these complications here, and assume we have access to an inference procedure that allows us to measure $J^{\rm eff}_{r,r'}(t)$ without error, so that we may focus on their properties and relationship to the ground-truth coupling filters.

\subsection*{Structure of effective coupling filters}

The ground-truth coupling filters $\hat{J}_{r,r'}(\omega)$ are modified by a correction term $\sum_{h,h'}\hat{J}_{r,h}(\omega) \hat{\Gamma}_{h,h'}(\omega)\hat{J}_{h',r'}(\omega)$. The linear response function $\hat{\Gamma}_{h,h'}(\omega)$ admits a series representation in terms of paths through the network through which neuron $r'$ is able to send a signal to neuron $r$ \emph{via hidden neurons only}.

\begin{figure*}
 \centering
  \includegraphics[width=\textwidth]{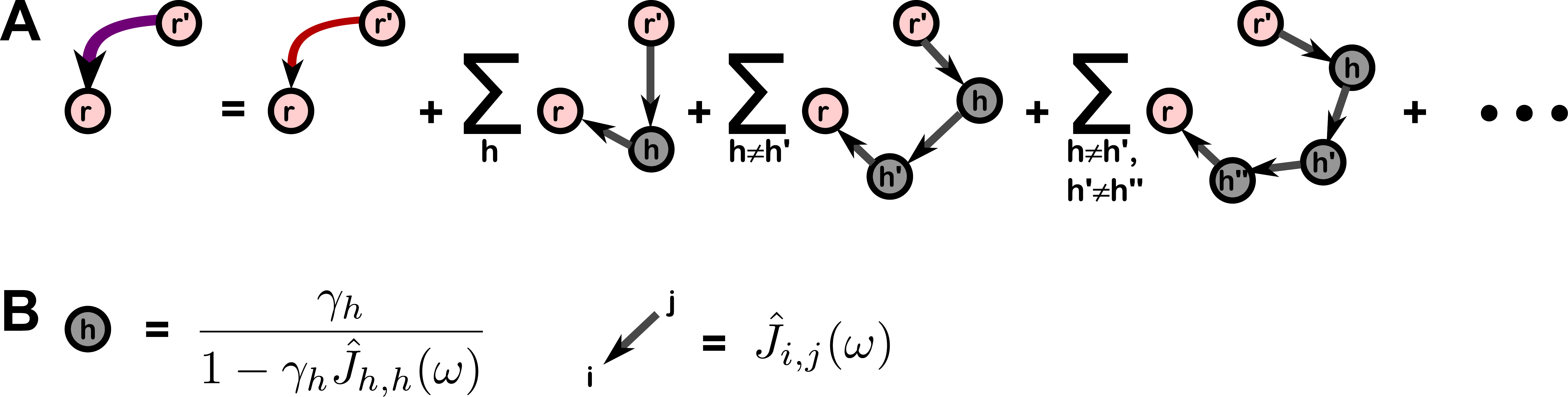}
  \caption{\textbf{Expansion of effective interactions into contributions from hidden paths.} \textbf{A.} Graphical representation of Eq.~(\ref{eqn:Jeffgeneral}). The linear response of the hidden network, $\hat{\Gamma}_{h,h'}(\omega)$, has been expanded as a series (corresponding to the grey hidden nodes and links between them), such that each term in the overall series can be interpreted as a contribution from a path through which the pre-synaptic neuron $r'$ is able to send a signal to post-synaptic neuron $r$ via 1, 2, etc. hidden neurons. This expression holds for any pair of neurons in the recorded subset. \textbf{B.} Quantitative expressions for each diagram in the series can be read off by assigning the shown factors for each hidden neuron node and each link between neurons, recorded or hidden, and multiplying them together. (No factor is assigned to the recorded neuron nodes). $\gamma_h$ is the gain of neuron $h$ and $\hat{J}_{i,j}(\omega)$ is the true interaction from $j$ to $i$ in the frequency domain.}
  \label{fig:Jeffeqn}
\end{figure*}

We may write down a set of ``Feynmanesque'' graphical rules for explicitly calculating terms in this series \cite{OckerPLOSCB2017}. First, we define the gain, $\gamma_h \equiv \lambda_0 \phi'\left(\mu_h + \sum_{h'}\mathcal J_{h,h'}\nu_{h'} \right)$. The contribution of each term can then be written down using the following rules, shown graphically in Fig.~\ref{fig:Jeffeqn}: \emph{i}) for the edge connecting recorded neuron $r'$ to a hidden neuron $h_i$, assign a factor $\hat{J}_{h_i,r'}(\omega)$; \emph{ii}) for each node corresponding to a hidden neuron $h_i$, assign a factor $\gamma_{h_i}/(1-\gamma_{h_i} \hat{J}_{h_i,h_i}(\omega))$; \emph{iii}) for each edge connecting hidden neurons $h_i\neq h_j$, assign a factor $\hat{J}_{h_j,h_i}(\omega)$; and \emph{iv}) for the edge connecting hidden neuron $h_j$ to recorded neuron $r$, assign a factor $\hat{J}_{r,h_j}(\omega)$. All factors for each path are multiplied together, and all paths are then summed over.

The graphical expansion is reminiscent of recent works expanding correlation functions of linear models of network spiking in terms of network ``motifs'' \cite{PernicePLOSCB2011, HuJStatMechThExp2013,HuPRE2014}. Computationally, this expression is practical for calculating the effective interactions in small networks involving only a few hidden neurons (as in the next section), but is generally unwieldy for large networks. In practice, the linear response matrix $\hat{\Gamma}_{h,h'}(\omega)$ can be calculated directly by numerical matrix inversion and an inverse Fourier transform back into the time domain. The utility of the path-length series is the intuitive understanding of the origin of contributions to the effective coupling filters and our ability to analytically analyze the strength of contributions from each path. For example, one immediate insight the path decomposition offers is that neurons only develop effective interactions between one another if there is a path by which one neuron can send a signal to the other.

\subsection*{Feedforward inhibition and degeneracy of hidden networks in small circuits}
\label{sec:ffwi}

\subsubsection*{Effective interactions \& emergent timescales in a small circuit}

\begin{figure*}
 \centering
  \includegraphics[width=0.9\textwidth]{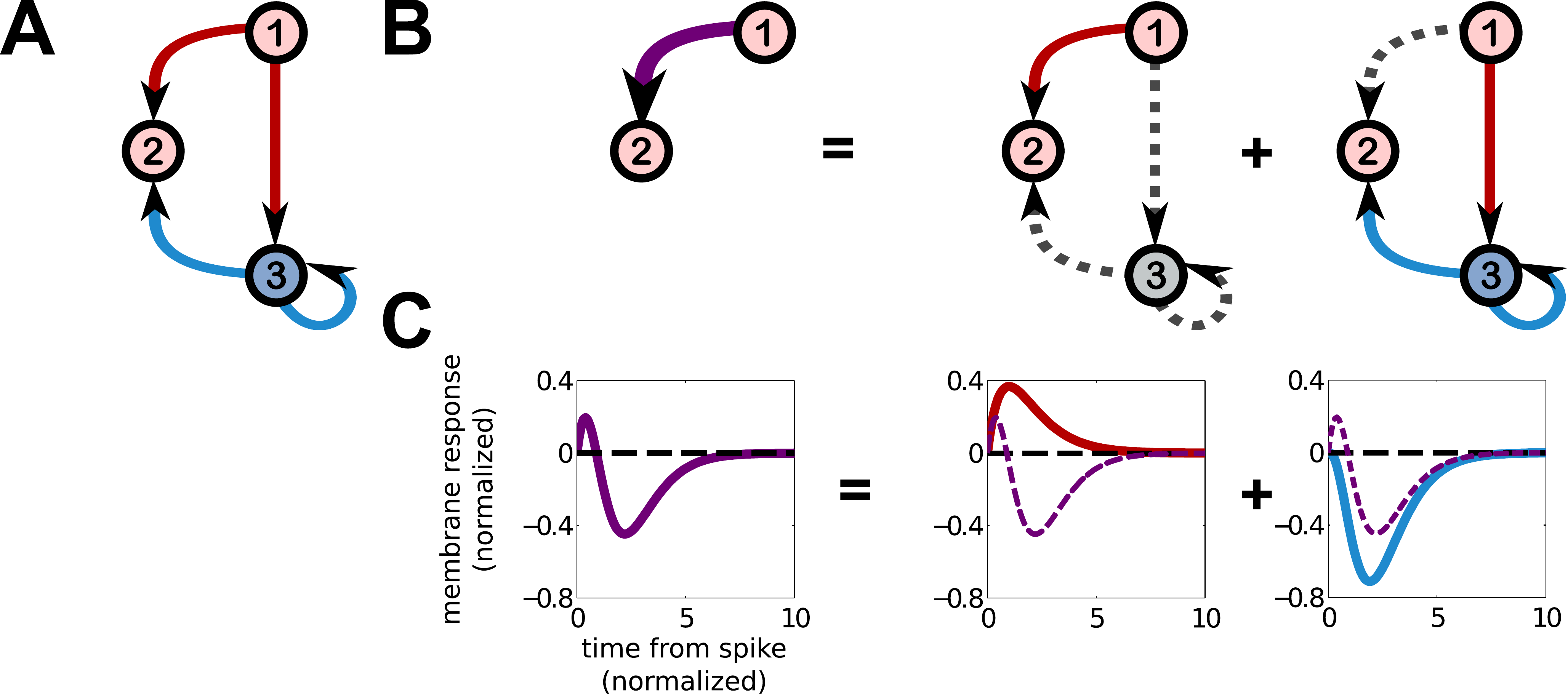}
  \caption{\textbf{3 neuron feedforward inhibition circuit.} \textbf{A}: A 3-neuron circuit displaying feedforward inhibition. Neuron 1 provides excitatory input to neurons 2 and 3, while neuron 3 provides inhibitory input to neuron 2. Neuron 3 also has a self-history coupling, denoted by an autaptic loop, which implements a refractory period in this circuit model. \textbf{B}: Leftmost, the effective interaction from neuron 1 to 2 when neuron 3 is unobserved. Subsequent plots decompose this interaction into contributions from neuron 1's direct input to neuron 2, and its indirect input through neuron 3. The indirect input through neuron 3 also takes account of neuron 3's self-history interaction. \textbf{C.} Leftmost, the effective interaction (membrane response) from neuron 1 to 2, subsequently decomposed into contributions from the direct interaction and the indirect interaction from 1 to 2.}
  \label{fig:ffwi_3neuron}
\end{figure*}

To build intuition for our result and compare to a well-known circuit phenomenon, we apply our Eq.~(\ref{eqn:Jeffgeneral}) to a $3$-neuron feedforward inhibition circuit, like that studied by Pouille and Scanziani \cite{PouilleScience2001}. Feedforward inhibition can sharpen the temporal precision of neural coding by narrowing the ``window of opportunity'' in which a neuron is likely to fire. For example, in the circuit shown in Fig.~\ref{fig:ffwi_3neuron}A, excitatory neuron 1 projects to both neurons 2 and 3, and 3 projects to 2. Neuron 1 drives both 2 and 3 to fire more, while neuron 3 is inhibitory and will counteract the drive neuron 2 receives from 1. The window of opportunity can be understood by looking at the effective interaction between neurons 1 and 2, treating neuron 3 as hidden. We use our path expansion (Fig.~\ref{fig:Jeffeqn}) to quickly write down the effective interaction we expect to measure in the frequency domain,
\begin{equation}
\hat{J}^{\rm eff}_{2,1}(\omega) = \hat{J}_{2,1}(\omega) + \frac{\hat{J}_{2,3}(\omega) \gamma_3 \hat{J}_{3,1}(\omega)}{1-\gamma_3 \hat{J}_{3,3}(\omega)}.
\label{eqn:ffwiJeff}
\end{equation}
The corresponding true synaptic interactions and resulting effective interaction are shown in Fig.~\ref{fig:ffwi_3neuron}B. The effective interaction matches qualitatively the observed membrane changes measured by Pouille and Scanziani \cite{PouilleScience2001}, and shows a narrow window after neuron 2 receives a spike in which the change in membrane potential is depolarized and neuron 2 is more likely to fire. Following this brief window, the membrane potential is hyperpolarized and the cell is less likely to fire until it receives more excitatory input. 

The effective interaction from neuron 1 to 2 in this simple circuit also displays several features that emerge in more complex circuits. Firstly, although the true interactions are either excitatory (positive) or inhibitory (negative), the effective interaction has a mixed character, being initially excitatory (due to excitatory inputs from neuron 1 arriving first through the monosynaptic pathway), but then becoming inhibitory (due to inhibitory input arriving from the disynaptic pathway). 

Secondly, emergent timescales develop due to reverberations between hidden neurons with bi-directional connections, represented as loops between neurons in our circuit schematics (e.g., between neurons $3$ and $4$ in Fig.~\ref{fig:ffwi_4neuron}). This includes self-history interactions such as refractoriness, schematically represented by loops like the $3 \rightarrow 3$ loop shown in Fig.~\ref{fig:ffwi_3neuron}, corresponding to the factor $1/(1-\gamma_3 \hat{J}_{3,3}(\omega))$). In the particular example shown in Fig.~\ref{fig:ffwi_3neuron}, in which we use a self-history interaction $J_{33}(\tau) = \mathcal J_{33} \beta_{33}\exp(-\beta_{33}\tau)$, a new timescale $\beta_{33}^{-1}(1-\gamma_3 \mathcal J_{33})^{-1}$ develops. Other choices of interactions can generate more complicated emergent timescales and temporal dynamics, including oscillations. For example, in the $4$-neuron circuit discussed below (Fig.~\ref{fig:ffwi_4neuron}), the choice $J_{3,4}(\tau) = J_{4,3}(\tau) = -|\mathcal J| \alpha^2 \tau e^{-\alpha \tau}$ yields effective interactions with new decay and oscillatory timescales equal to $(\alpha (1-\lambda_0 |\mathcal J|))^{-1}$ and $(\alpha \lambda_0 |\mathcal J|)^{-1}$. In the larger networks we consider in the next section, inter-neuron interactions must scale with network size in order to maintain network stability. Because emergent timescales depend on the synaptic strengths of hidden neurons, we typically expect emergent timescales generated by loops between hidden neurons to be negligible in large random networks. However, because the magnitudes of the self-history interaction strengths need not scale with network size, they may generate emergent timescales large enough to be detected.

It is worth noting explicitly that only the interaction from neuron 1 to 2 has been modified by the presence of the hidden neuron 3, for the particular wiring diagram shown in Fig.~\ref{fig:ffwi_3neuron}. The self-history interactions of both neurons 1 and 2, as well as the interaction from neuron 2 to 1 (zero in this case) are unmodified. The reason the hidden neuron did not modify these interactions is that the only link neuron 3 makes is from 1 to 2. There is no path by which neuron 1 can send a signal back to itself, hence its self-interaction is unmodified, nor is there a path that neuron 2 can send signals to neuron 3 or on to neuron 1, and hence neuron 2's self-history interaction and its interaction to neuron 1 are unmodified.

\subsubsection*{Degeneracy of hidden networks giving rise to effective interactions}

\begin{figure*}
 \centering
  \includegraphics[scale=0.2]{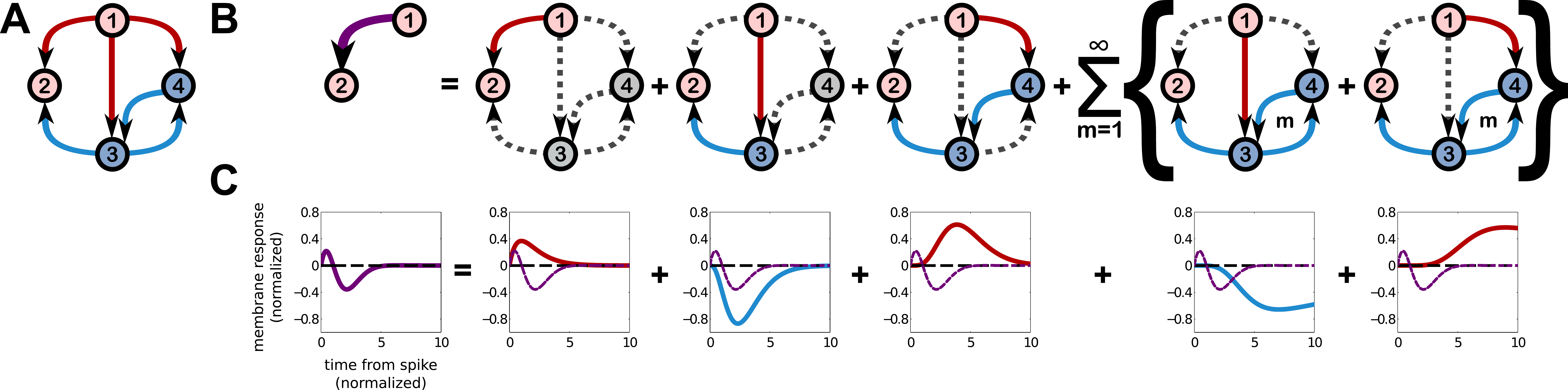}
  \caption{\textbf{Different complete circuits may underly similar effective circuits.} \textbf{A}: A circuit very similar to that in Fig.~\ref{fig:ffwi_3neuron}, except that neuron 1 also provides excitatory input to neuron 4, which in turn provides inhibitory input to neuron 3. The self-history coupling of neuron 3 to itself has also been removed in this example. \textbf{B}: Leftmost, the effective interaction from neuron 1 to 2, which is qualitatively and quantitatively similar to the effective interaction shown in Fig.~\ref{fig:ffwi_3neuron}. Subsequent plots indicate each path through the circuit that neuron 1 can send a signal to neuron 2 through the hidden neurons 3 and 4. \textbf{C.} Leftmost, the effective interaction from neuron 1 to 2. Subsequent plots decompose this interaction into contributions from the paths shown above in \textbf{B}.}
  \label{fig:ffwi_4neuron}
\end{figure*}

It is well known that different networks may produce the same observed circuit phenomena \cite{Prinz2004,GutenkunstPLOSCB2007,ApgarMolBioSyst2010,GutierrezNeuron2013,FisherNeuron2013,MarderDevNeurobio2017}. To illustrate that our approach may be used to identify degenerate solutions in which more than one network underlies observed effective interactions, we construct a $4$-neuron circuit that produces a quantitatively similar effective interaction between the recorded neurons $1$ and $2$, shown in Fig.~\ref{fig:ffwi_4neuron}. Specifically, in this circuit we have removed neuron 3's self-history interaction and introduced a second inhibitory hidden network that receives excitatory input from neuron 1 and provides inhibitory input to neuron 3. By tuning the interaction strengths we are able to produce the desired effective interaction. This demonstrates that intrinsic neural properties such as refractoriness can trade off against inputs from other hidden neurons, making it difficult to distinguish the two cases from one another (or from a potentially infinity of other circuits that could have produced this interaction; for example, a qualitatively similar interaction is produced in the $N=1000$ network in which only three neurons are recorded, shown below in Fig.~\ref{fig:Jeffs}). Statistical inference methods may favor one of the possible underlying choices of complete network consistent with a measured set of effective interactions, suggesting there may be some sense of a ``best'' solution. However, the particular ``best'' network will depend on many factors, including the amount and fidelity of data recorded, regularization choices, and how well the fitted model generalizes to new data (i.e., how ``close'' the fitted model is to the generative model). Potentially, if these conditions were met, with enough data the slight quantitative differences between the effective interactions produced by different hidden networks (including higher order effective interactions, which we assume to be negligible here; see SI), could help distinguish different hidden networks. However, the amount of data required to perform this discrimination and validate the result may be impractically large\cite{GoldenfeldBook1992,ApgarMolBioSyst2010,MachtaScience2013,CaycoGajicFrontiersCompNeuro2015}. It is thus worth studying the structure of the observed effective interactions directly in search of possible signatures that elucidate the statistical properties of the complete network.

\subsection*{Strongly coupled large networks}

Constructing networks that produce particular effective interactions is tractable for small circuits, but much more difficult for larger circuits composed of many circuit motifs. Not only can combinations of different circuit motifs interact in unexpected ways, one must also take care to ensure the resulting network is both active and stable---i.e., that firing will neither die out nor skyrocket to the maximum rate. Stability in networks is often implemented by either building networks with classical (or ``weak'') synapses whose strength scales inversely with the number of inputs they receive, assumed here to be proportional to network size, and hence $\mathcal J_{i,j} \sim 1/N$, or by building balanced networks in which excitatory and inhibitory synaptic strengths balance out, on average, and scale as $\mathcal J_{i,j} \sim 1/\sqrt{N}$ \cite{vanVreeswijkScience1996,BarralNatNeuro2016}. In both cases the synapses tend to be small in value in large networks, but are compensated for by large numbers of incoming connections. In the case of $1/N$ scaling, neurons are driven primarily by the mean of their inputs, while in ``strong'' balanced $1/\sqrt{N}$ networks neurons are driven primarily by fluctuations in their inputs. 

Our goal is to understand how the interplay between the presence of hidden neurons and different synaptic scaling or network architectures shapes effective interactions. Previous work has studied the hidden-neuron problem in the weak coupling limit \cite{NykampSIAMJAM2005,NykampMathBiosci2007,NykampSIAMJAM2007,NykampPRE2008} using a different approach to relate inferred synaptic parameters to true parameters; here we use our approach to study the $1/\sqrt{N}$ strong coupling limit, theoretically predicted to be an important feature that supports computations in networks in a balanced regime \cite{vanVreeswijkScience1996,LitwinKumarNatNeuro2012,RosenbaumPRX2014,DeneveNatNeuro2016}. Moreover, experiments in cultured neural tissue have been found to be more consistent with the $1/\sqrt{N}$ scaling than $1/N$\cite{BarralNatNeuro2016}, indicating that it may have intrinsic physiological importance.

We analytically determine how significantly effective interaction strengths are skewed away from the true interaction strengths as a function of both the number of observed neurons and typical synaptic strength. We consider several simple networks ubiquitous in neural modeling: first, an \ERtext (ER) network with ``mixed synapses'' (i.e., a neuron may have both positive and negative synaptic weights), a balanced ER network with Dale's law imposed (a neuron's synapses are all the same sign), and a Watts-Strogatz (WS) small world network with mixed synapses. Each network has $N$ neurons and connection sparsity $p$ (only $100p\%$ of connections are non-zero). Connections in ER networks are chosen randomly and independently, while connections in the WS network are determined by randomly rewiring a fraction $\beta$ of the connections in a $(pN)^{\rm th}$-nearest-neighbor ring network. In each network $N_{\rm rec}$ neurons are recorded randomly. 

For simplicity we take the baselines of all neurons to be equal, $\mu_i = \mu_0$ (such that in the absence of synaptic input the probability that a neuron fires in a short time window $\Delta t$ is $\lambda_0 \Delta t \exp(\mu_0))$. We choose the rate nonlinearity to be exponential, $\phi(x) = e^x$; this is the ``canonical'' choice of nonlinearity often used when fitting this model to data \cite{Simoncelli2004,Paninski2004,Pillow2005,Pillow2008,SuppInfo}. We will further assume $\exp(\mu_0) \ll 1$, so that we may use this as a small control parameter. For $i \neq j$, the non-zero synaptic weights between neurons $\mathcal J_{i,j}$ are independently drawn from a normal distribution with zero mean and standard deviation $J_0/(p N)^{2a}$, where $J_0$ controls the overall strength of the weights and $a = 1$ or $1/2$, corresponding to ``weak'' and ``strong'' coupling. For simplicity, we do not consider intrinsic self-coupling effects in this part of the analysis, i.e., we take $\mathcal J_{i,i} = 0$ for all neurons $i$. For the Dale's law network, the overall distribution of synaptic weights follows the same normal distribution as the mixed synapse networks, but the signs of the weights correspond to whether the pre-synaptic neuron is excitatory or inhibitory. Neurons are randomly chosen to be excitatory and inhibitory, the average number of each type being equal so that the network is balanced. Numerical values of all parameters are given in Table~\ref{tab:params} in the Methods.

We seek to assess how the presence of hidden neurons can shape measured network interactions. We first focus on the typical strength of the effective interactions as a function of both the fraction of neurons recorded, $f = N_{\rm rec}/N$, and the strength of the synaptic weights $J_0$. We quantify the strength of the effective interactions by defining the effective synaptic weights $\mathcal J^{\rm eff}_{r,r'} \equiv \int_0^\infty d\tau~J^{\rm eff}_{r,r'}(\tau) = \hat{J}^{\rm eff}_{r,r'}(\omega = 0)$; c.f. $\mathcal J_{r,r'} = \int_0^\infty d\tau~J_{r,r'}(\tau)$ for the true synaptic weights. We then study the sample statistics of the difference, $\mathcal J^{\rm eff}_{r,r'} - \mathcal J_{r,r'}$, averaged across both subsets of recorded neurons and network instantiations, to estimate the typical contribution of hidden neurons to the measured interactions. The mean of the synaptic weights is near zero (because the weights are normally distributed with zero mean in the mixed synapse networks and due to balance of excitatory and inhibitory neurons in the Dale's law network), so we focus on the root-mean-square of $\mathcal J^{\rm eff}_{r,r'} - \mathcal J_{r,r'}$. This measure is a conservative estimate of changes in strength, as $J^{\rm eff}_{r,r'}(\tau)$ may have both positive and negative components that partially cancel when integrated over time, unlike $J_{r,r'}(\tau)$. An alternative measure we could have chosen that avoids potential cancellations is $\int_0^\infty d\tau~|J^{\rm eff}_{r,r'}(\tau) - J_{r,r'}(\tau)|$, i.e., the integrated absolute difference between effective and true interactions. However, this will depend on our specific choices of waveform $g(\tau)$, whereas $\mathcal J^{\rm eff}_{r,r'} - \mathcal J_{r,r'}$ does not due to our normalization $\int_0^\infty d\tau~g(\tau) = 1$. As $\left|\int d\tau~f(\tau) \right| \leq \int d\tau~\left|f(\tau)\right|$, for any $f(\tau)$, we can consider our choice of $\mathcal J^{\rm eff}_{r,r'} - \mathcal J_{r,r'}$ as a lower bound on the strength that would be quantified by $\int_0^\infty d\tau~|J^{\rm eff}_{r,r'}(\tau) - J_{r,r'}(\tau)|$.

\begin{figure}[h!]
 \centering
 \includegraphics[scale=0.4]{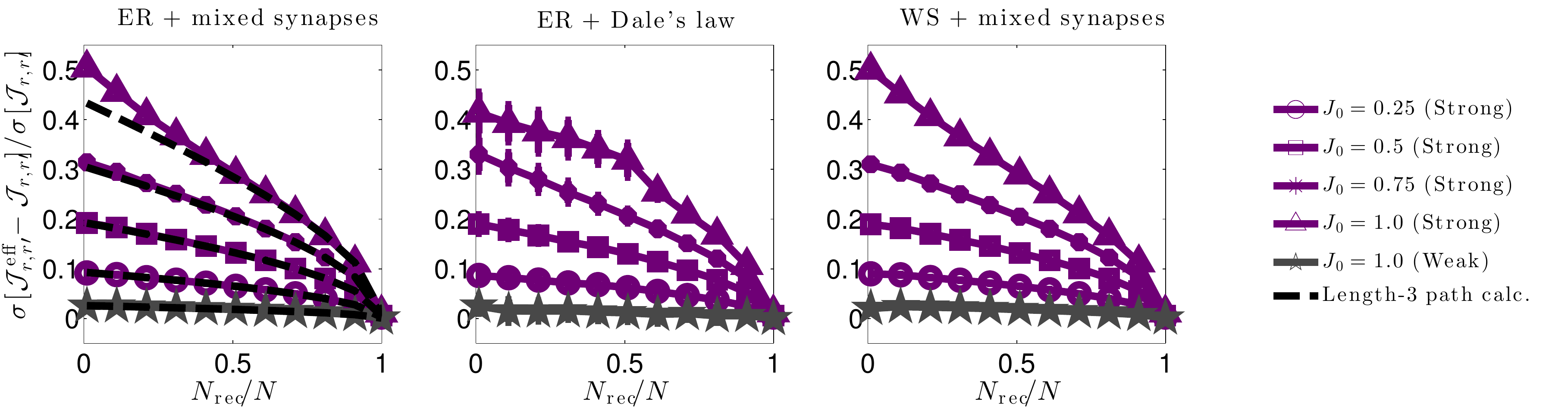}
  \caption{\textbf{Relative changes in interaction strength due to hidden neurons for three network types.} We quantify relative changes in interaction strength between effective ($\mathcal J^{\rm eff}_{r,r'}$) and true ($\mathcal J_{r,r'}$) interactions by the (sample) root-square-mean deviation, $\sigma[\mathcal J^{\rm eff}_{r,r'} - \mathcal J_{r,r'}]$, normalized by the true synaptic weight (sample) standard deviation $\sigma[\mathcal J_{r,r'}]$. We do so for three (sparse) network types: \textbf{Left.} An \ERtext (ER) network with ``mixed synapses'' (i.e., Dale's law not imposed) with normally distributed synaptic weights. \textbf{Middle.} An ER network with Dale's law imposed, (i.e., each neuron's outgoing synaptic weights all have the same sign). \textbf{Right.} A Watts-Strogatz (WS) small world network with $30\%$ rewired connections and mixed synapses. All network types yield qualitatively similar results. In each plot solid lines are numerical estimates of the sample standard deviation of the difference between effective coupling weights $\mathcal J^{\rm eff}_{r,r'}$ and true coupling weights $\mathcal J_{r,r'}$ between neurons $r \neq r'$, normalized by the standard deviation of $\mathcal J_{r,r'}$. These estimates account for all paths through hidden neurons. Purple lines correspond to synaptic weights with standard deviation $J_0/\sqrt{pN}$ (strong coupling), while grey lines correspond to synaptic weights with standard deviation $J_0/pN$ (weak coupling). For weak $1/N$ coupling (grey), the ratio of standard deviations is $\mathcal O(1/\sqrt{N})$. For strong $1/\sqrt{N}$ coupling (purple) the ratio is $\mathcal O(1)$ and grows in strength as the fraction of recorded neurons $N_{\rm rec}/N$ decreases or the typical synaptic strength $J_0$ increases. The dashed black lines in the left plot show theoretical estimates accounting only for hidden paths of length-$3$ connecting recorded neurons (Eq.~(\ref{eqn:varJeff}). Deviations from the length-$3$ prediction at small $f$ and large $J_0$ indicate that contributions from circuit paths involving many hidden neurons are significant in these regimes.}
  \label{fig:Janalysis}
\end{figure}

Numerical evaluations of the population statistics for all three network types are shown as solid curves in Fig.~(\ref{fig:Janalysis}), for both strong coupling and weak coupling. All three networks yield qualitatively similar results. The vertical axes measure the root-mean-square deviations between the statistically expected true synaptic $\mathcal J_{r,r'}$ and the corresponding effective synaptic weight $\mathcal J^{\rm eff}_{r,r'}$, normalized by the true root mean square of $\mathcal J_{r,r'}$. Thus, a ratio of 0.5 corresponds to a $50\%$ root-mean difference in effective versus true synaptic strength. We measure these ratios as a function of both the fraction of neurons recorded (horizontal axis) and the parameter $J_0$ (labeled curves).  

There are two striking effects.  First, deviations are nearly negligible ($\mathcal O(1/\sqrt{pN})$) for $1/N$ scaling of connections.  Thus, for large networks with synapses that scale with the system size, vast numbers of hidden neurons combine to have negligible effect on effective couplings.  This is in marked contrast to the case when coupling is strong ($1/\sqrt{N}$ scaling), when hidden neurons have a pronounced impact ($\mathcal O(1)$).  This is particularly the case when $f \ll 1$---the typical experimental case in which the hidden neurons outnumber observed ones by orders of magnitude---or when $J_0 \lesssim 1.0$, when typical deviations become half the magnitude of the true couplings themselves (upper blue line). For $J_0 \gtrsim 1.0$, the network activity is unstable for an exponential nonlinearity. 

To gain analytical insight into these numerical results, we calculate the standard deviation $\sigma[\mathcal J^{\rm eff}_{r,r'} - \mathcal J_{r,r'}]$, normalized by $\sigma[\mathcal J_{r,r'}]$, for contributions from paths up to length-$3$, focusing on the case of the ER network with mixed synapses (the Dale's law and WS networks are more complicated, as the moments of the synaptic weights depend on the identity of the neurons). For strong $1/\sqrt{N}$ coupling we find
%\begin{linenomath*}
\begin{align}
\frac{\sigma[\mathcal J_{r,r'}^{\rm eff}-\mathcal J_{r,r'}]}{\sigma[\mathcal J_{r,r'}]} &\approx \lambda_0 J_0 e^{\mu_0} \sqrt{1-f} \nonumber \\
& ~~~~\times\left(1 + \frac{3}{2}(\lambda_0 J_0 e^{\mu_0})^2 (1-f)\right),
\label{eqn:varJeff}
\end{align}
%\end{linenomath*}
corresponding to the black dashed curves in Fig.~\ref{fig:Janalysis} left.  Eq.~(\ref{eqn:varJeff}) is a truncation of a series in powers of $\lambda_0 J_0 e^{\mu_0}\sqrt{1-f}$, where $f = N_{\rm rec}/N$ is the fraction of recorded neurons. The most important feature of this series is the fact that it only depends on the \emph{fraction} of recorded neurons $f$, not the absolute number, $N$. Contributions from long paths remain finite, even as $N \rightarrow \infty$.  In contrast, the corresponding expression for $\sigma[\mathcal J_{r,r'}^{\rm eff}-\mathcal J_{r,r'}]/\sigma[\mathcal J_{r,r'}]$ in the case of weak $1/N$ coupling is a series is in powers of $\lambda_0 J_0 e^{\mu_0}\sqrt{(1-f)/(pN)}$, so that contributions from long paths are negligible in large networks $N \gg 1$. (See \cite{SuppInfo} for derivation and results for $N=100$.) Deviations of Eq.~(\ref{eqn:varJeff}) from the numerical solutions in Fig.~\ref{fig:Janalysis} indicate that contributions from truncated terms are not negligible when $f \ll 1$. As these terms correspond to paths of length-$4$ or more, this shows that long chains through the network contribute significantly to shaping effective interactions.

\begin{figure}[h!]
 \centering
 \includegraphics[width=0.9\textwidth]{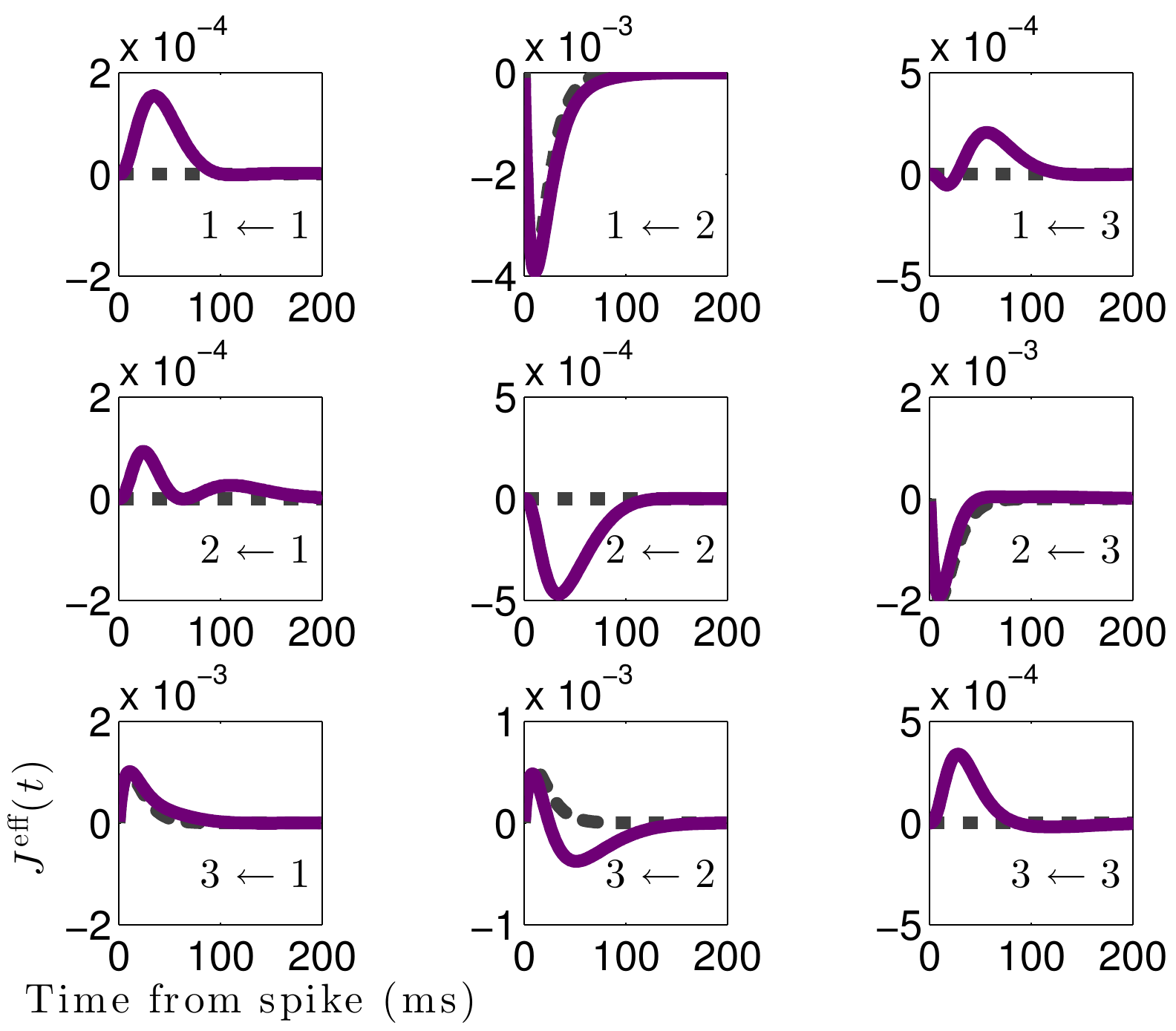}
  \caption{\textbf{Effective interactions between recorded neurons differ qualitatively from true interactions.} Effective interactions $J^{\rm eff}_{r,r'}(t)$ (solid purple) versus true coupling filters (dashed black) for $N_{\rm rec} = 3$ recorded neurons in a network of $N = 1000$ total neurons. Inset labels $i \leftarrow j$ indicate the interaction is from neuron $j$ to $i$, for $i,j \in \{1,2,3\}$. The simulated network has an \ERtext connectivity with sparsity $p = 0.2$ and normally distributed non-zero weights with zero mean and standard deviation $1/\sqrt{pN}$. Although the network is sparse, the effective interactions are not: non-zero effective interactions develop where no direct connection exists. The effective interactions can differ \emph{qualitatively} from the true interactions, as evidenced by the biphasic $3 \leftarrow 2$ effective interaction, whereas the true $3 \leftarrow 2$ is purely excitatory.}
  \label{fig:Jeffs}
\end{figure}

The above analysis demonstrates that the strength of the effective interactions can deviate from that of the true direct interactions by as much as $50\%$. However, changes in strength do not give us the full picture---we must also investigate how the temporal dynamics of the effective interactions change. To illustrate how hidden units can skew temporal dynamics, in Fig.~\ref{fig:Jeffs} we plot the effective vs. true interactions between $N_{\rm rec} = 3$ neurons in an $N = 1000$ neuron network. Because the three network types considered in Fig.~\ref{fig:Janalysis} yield qualitatively similar results, we focus on the \ERtext network with mixed synapses. 

Four of the true interactions between neurons shown in Fig.~\ref{fig:Jeffs} are non-zero ($J^{\rm eff}_{1,2}(t)$, $J^{\rm eff}_{3,2}(t)$, $J^{\rm eff}_{3,1}(t)$, and $J^{\rm eff}_{2,3}(t)$). Of these, three exhibit only slight differences between the true and effective interactions: $J^{\rm eff}_{1,2}(t)$ and $J^{\rm eff}_{3,1}(t)$ have slightly longer decay timescales than their true counterparts, while $J^{\rm eff}_{2,3}(t)$ has a slightly shorter timescale, indicating the contribution of the hidden network to these interactions was either small or cancelled out. However, the interaction $J^{\rm eff}_{3,2}(t)$ differs significantly from the true interaction, becoming initially excitatory but switching to inhibitory after a short time, as in our earlier example case of feedforward inhibition. This indicates that neuron $2$ must drive a cascade of neurons that ultimately provide inhibitory input to neuron $3$.

Contrasting the true and effective interactions shown in Fig.~\ref{fig:Jeffs} highlights many of the ways in which hidden neurons skew the temporal properties of measured interactions. An immediately obvious difference is that although the true synaptic connections in the network are sparse, the effective interactions are not. This is a generic feature of the effective interaction matrix, as in order for an effective interaction from a neuron $r'$ to $r$ to be identically zero there cannot be any paths through the network by which $r'$ can send a signal to $r$. In a random network the probability that there are no paths connecting two nodes tends to zero as the network size $N$ grows large. Note that this includes paths by which each neuron can send a signal back to itself, hence the neurons developed effective self-interactions, even though the true self-interactions are zero in these particular simulations. 

%%%%%%%%%%%%%%%%%%%%%%%%%%%%%%%
% DISCUSSION 
%%%%%%%%%%%%%%%%%%%%%%%%%%%%%%%

\section*{Discussion}
\label{sec:discussion}

We have derived a quantitative relationship between ``ground-truth'' synaptic interactions and the effective interactions (interpreted here as post-synaptic membrane responses) that unobserved neurons generate between subsets of observed neurons. This relationship, Eq.~(\ref{eqn:Jeffgeneral}) and Fig.~\ref{fig:Jeffeqn}, provides a foundation for studying how different network architectures and neural properties shape the effective interactions between subsets of observed neurons. Our approach can be also be used to study higher order effective interactions between 3 or more neurons, and can be systematically extended to account for corrections to our mean-field approximations and investigate effective noise generated by hidden neurons (using field theoretic techniques from \cite{OckerPLOSCB2017}, see SI), as well as time-dependent external drives or steady-states.

Here, as first explorations, we focused on the effective interactions corresponding to linear membrane responses. We first demonstrated that our approach applied to small feedforward inhibitory circuits yields effective interactions that capture the role of inhibition in shortening the time window for spiking, and are qualitatively similar to experimentally observed measurements \cite{PouilleScience2001}. Moreover, we used this example to demonstrate explicitly that different hidden networks can give rise to the same effective interactions between neurons. We then showed that the influence of hidden neurons can remain significant even in large networks in which the typical synaptic strengths scale with network size. In particular, when the synaptic weights scale as $1/\sqrt{N}$, the relative influence of hidden neurons depends only on the fraction of neurons recorded. Together with theoretical and experimental evidence for this scaling in cortical slices \cite{vanVreeswijkScience1996,LitwinKumarNatNeuro2012,RosenbaumPRX2014,BarralNatNeuro2016,DeneveNatNeuro2016}, this suggests that neural interactions inferred from cortical activity data may differ markedly from the true interactions and connectivity. 

The issue of degeneracy in complex biological systems and networks has been discussed at length in the literature, in the context of both inherent degeneracies---multiple different network architectures can produce the same qualitative behaviors \cite{Prinz2004,GutierrezNeuron2013,FisherNeuron2013,MarderDevNeurobio2017}, as well as degeneracies in our model descriptions---many models may reproduce experimental observations, demanding sometimes arbitrary criteria for selecting one model over another. All have implications for how successfully one can infer unobserved network properties. One kind of model degeneracy, ``sloppiness'' \cite{GutenkunstPLOSCB2007,MachtaScience2013}, describes models in which the behavior of the model is sensitive to changes in only a relatively small number of directions in parameter space. Many models of biological systems have been shown to be sloppy \cite{GutenkunstPLOSCB2007}; this could account for experimentally observed networks that are quite different in composition but produce remarkably similar behaviors. Sloppiness suggests that rather than trying to infer all properties of a hidden network, there may be certain parameter combinations that are much more important to the overall network operation, and could potentially be inferred from subsampled observations. 

Another perspective on model degeneracy comes from the concepts of ``universality'' that occur in random matrix theory \cite{ErdosRussianMathSurv2011,AhmadianPhysRevE2015} and  Renormalization Group methods of statistical physics \cite{GoldenfeldBook1992}. Many bulk properties of matrices (e.g., the distribution of eigenvalues) whose entrees are combinations random variables, such as our $\mathcal J^{\rm eff}_{r,r'}$ , are universal in that they depend on only a few key details of the distribution that individual elements are drawn from \cite{TaoCommContMath2008}. Similarly, one of the central results of the Renormalization Group shows that models with drastically disparate features may yield the same coarse-grained model structure when many degrees of freedom are averaged out, as in our case of approximately averaging out hidden neurons. Different distributions (in the case of random matrix theory) or different models (in the case of the Renormalization group) that yield the same bulk properties or coarse-grained models are said to be in the same ``universality class.'' Measuring particular quantities under a range of experimental conditions (e.g., different stimuli) may be able to reveal which universality class an experimental system belongs to and eliminate models belonging to other universality classes as candidate generating models of the data, but these measurements cannot distinguish between models within a universality class. As our network of subsampled neurons can be thought of as a model in which the hidden network has been approximately averaged over, this means we can potentially use Eq.~(\ref{eqn:Jeffgeneral}) to rule out sets of models of the hidden network that are inconsistent with measured sets of effective interactions $J^{\rm eff}_{r,r'}(t)$ (e.g., hidden networks with given network \emph{statistics}), even if we are unable to uniquely pin down the true hidden network (i.e., the exact or even approximate \emph{configuration} of network parameters drawn those statistical distributions).

The fact that many different hidden networks may yield the same set of effective interactions suggests that the effective interactions themselves may yield direct insight into a circuit's functions. For instance, many circuits consist of principal neurons that transmit the results of circuit computation to downstream circuitry, but often do not make direct connections with one another, instead interacting through (predominantly inhibitory) intermediaries called interneurons. From the point of view of a downstream circuit, the principal neurons are ``recorded'' and the interneurons are ``hidden.'' A potential reason for this general arrangement is that direct synaptic interactions alone are insufficient to produce the membrane responses required to perform the circuit's computations, and the network of interneurons reshapes the membrane responses of projection neurons into effective interactions that can perform the desired computations---it may thus be that the effective interactions should be of primary interest, not necessarily the (possibly degenerate choices of) physiological synaptic interactions. For example, in the feedforward inhibitory circuits of Figs.~\ref{fig:ffwi_3neuron} and \ref{fig:ffwi_4neuron}, the roles of the hidden inhibitory neurons may simply be to act as interneurons that reshape the interaction between the excitatory projection neurons $1$ and $2$, and the choice of which particular circuit motif is implemented in a real network is determined by other physiological constraints, not only computational requirements. 

One of the greatest achievements in systems neuroscience would be the ability to perform targeted modifications to a large neural circuit and \emph{selectively} alter its suite of computations. This would have powerful applications for both studying a circuit's native computations, but also repurposing circuits or repairing damaged circuitry (due to, e.g., disease). If the computational roles of circuits are indeed most sensitive to the effective interactions between principal neurons, this suggests we can exploit potential degeneracies in the interneuron architecture and intrinsic properties to find \emph{some} circuit that achieves a desired computation, even if it is not a physiologically natural circuit. Our main result relating effective and true interactions, Eq.~(\ref{eqn:Jeffgeneral}), provides a foundation for future work investigating how to identify sets of circuits that perform a desired set of computations. We have shown in this work that it can be done for small circuits (Figs.~\ref{fig:ffwi_3neuron} and \ref{fig:ffwi_4neuron}), and that the effective interactions in large random networks can be significantly skewed away from the true interactions when synaptic weights scale as $1/\sqrt{N}$, as observed in experiments \cite{BarralNatNeuro2016}. This holds promise for identifying principled approaches to tuning or controlling neural interactions, such as by using neuromodulators to adjust interneuron properties or inserting artificial or synthetic circuit implants into neural tissue to act as ``hidden'' neurons. If successful, this could contribute to the long term goal of using such interventions to aid in reshaping the effective synaptic interactions between diseased neurons, and thereby restore healthy circuit behaviors. 

%
%%%%%%%%%%%%%%%%%%%%%%%%%%%%%%%%
%% METHODS 
%%%%%%%%%%%%%%%%%%%%%%%%%%%%%%%%

\section*{Methods}
\label{sec:methods}

\subsection*{Model definition and details}
\label{sec:modeldefn}

The firing rate of a neuron $i$ in the full network is given by
\begin{equation}
\lambda_i(t) = \lambda_0 \phi\left(\mu_i + \mu^{\rm ext}_i(t) + \sum_j \int_{-\infty}^\infty dt' J_{ij}(t-t') \dot{n}_j(t') \right),
\label{eqn:modeldefn}
\end{equation}
where $\lambda_0$ is a characteristic rate, $\phi(x) \geq 0$ is a nonlinear function, $\mu_i$ (potentially a function of some external stimulus $\theta$) is a time-independent tonic drive that sets the baseline firing rate of the neuron in the absence of input from other neurons, $\mu^{\rm ext}_i(t)$ is an external input current, and $J_{ij}(t-t')$ is a coupling filter that filters spikes $\dot{n}_j(t')$ fired by presynaptic neuron $j$ at time $t'$, incident on post-synaptic neuron $i$. We will take $\mu^{\rm ext}_i(t) = 0$ for simplicity in this work, focusing on the activity of the network due to the tonic drives $\mu_i$ (which could be still be interpreted as external tonic inputs, so the activity of the network need not be interpreted as spontaneous activity). 

While we need not attach a mechanistic interpretation to these filters, a convenient interpretation is that the nonlinear Hawkes model approximates the stochastic dynamics of a leaky integrate-and-fire network model driven by noisy inputs  \cite{GerstnerBook,OstojicPLOSCB2011}. In fact, the nonlinear Hawkes model is equivalent to a current-based integrate-and-fire model in which the deterministic spiking rule (a spike fires when a neuron's membrane potential reaches a threshold value $V_{\rm th}$) is replaced by a stochastic spiking rule (the higher a neuron's membrane potential, the higher the probability a neuron will fire a spike). (It can also be mapped directly to a conductance-based in special cases \cite{LatimerNIPS2014}). For completeness, we present the mapping from a leaky integrate-and-fire model with stochastic spiking to Eq.~(\ref{eqn:modeldefn}) in the Supporting Information (SI). 

\subsection*{Derivation of effective baselines and coupling filters}
\label{sec:effectivederivations}

To study how hidden neurons affect the inferred properties of recorded neurons, we partition the network into ``recorded'' neurons, labeled by indices $r$ (with sub- or superscripts to differentiate different recorded neurons, e.g., $r$ and $r'$ or $r_1$ and $r_2$) and ``hidden'' neurons labeled by indices $h$ (with sub- or superscripts). The rates of these two groups are thus
$$\lambda_r(t) = \lambda_0 \phi\left(\mu_r + \sum_{r'} J_{r,r'} \ast \dot{n}_{r'} + \sum_h J_{r,h} \ast \dot{n}_h \right),$$ 
$$\lambda_h(t) = \lambda_0 \phi\left(\mu_h + \sum_{r} J_{h,r} \ast \dot{n}_{r} + \sum_{h'} J_{h,h'} \ast \dot{n}_{h'} \right).$$
To simplify notation, we write $J_{i,j} \ast \dot{n}_j = \int_{-\infty}^\infty dt'~J_{i,j}(t-t')\dot{n}_j(t')$. If we seek to describe the firing of the recorded neurons only in terms of their own spiking history, input from hidden neurons effectively acts like noise with some mean amount of input. We thus begin by splitting the hidden input to the recorded neurons up into two terms, the mean plus fluctuations around the mean:
$$\sum_h J_{r,h} \ast \dot{n}_h(t) = \sum_h J_{r,h} \ast \mathbb{E}\left[ \dot{n}_h(t) | \left\{\dot{n}_r \right\} \right] + \xi_r(t),$$
where $\mathbb{E}\left[ \dot{n}_h(t) | \left\{\dot{n}_r \right\} \right]$ denotes the mean activity of the hidden neurons conditioned on the activity of the recorded units, and $\xi_r(t)$ are the fluctuations, i.e., $\xi_r(t) \equiv \sum_h J_{r,h} \ast (\dot{n}_h - \mathbb{E}\left[ \dot{n}_h(t) | \left\{\dot{n}_r \right\} \right])$. Note that $\xi_r(t)$ is also conditional on the activity of the recorded units.

By construction, the mean of the fluctuations is identically zero, while the cross-correlations can be expressed as
$$\mathbb{E}\left[ \xi_r(t) \xi_{r'}(t') \right] = \int_{-\infty}^\infty dt_1 dt_2 \sum_{h_1,h_2} J_{r,h_1}(t-t_1) J_{r',h_2}(t'-t_2) C_{h_1,h_2}(t_1,t_2),$$
where $C_{h_1,h_2}(t_1,t_2)$ is the cross-correlation between hidden neurons $h_1$ and $h_2$ (conditioned on the spiking of recorded neurons). If the autocorrelation of the fluctuations ($r=r'$) is small compared to the mean input to the recorded neurons ($\sum_h J_{r,h} \ast \mathbb{E}\left[ \dot{n}_h(t) | \left\{\dot{n}_r \right\} \right]$), or if $J_{r,h}$ is small, then we may neglect these fluctuations and focus only on the effects that the mean input has on the recorded subnetwork. At the level of the mean field theory approximation we make in this work, the spike-train correlations are zero. One can calculate corrections to mean field theory (see SI) to estimate the size of this noise, however, even when this noise is non-negligible it can be treated as a separate input to the recorded neurons, and hence will not alter the form of the effective couplings between neurons. Averaging out the effective noise, however, will generate new interactions between neurons; we leave investigation of this issue for future work.

In order to calculate how hidden input shapes the activity of recorded neurons, we need to calculate the mean $\mathbb{E}\left[\dot{n}_h | \left\{\dot{n}_r \right\}\right]$. This mean input is difficult to calculate in general, especially when conditioned on the activity of the recorded neurons. In principle, the mean can be calculated as
$$\mathbb{E}\left[ \dot{n}_h | \left\{\dot{n}_r\right\} \right] = \mathbb{E}\left[ \left. \lambda_0 \phi\left(\mu_h + \sum_{r} J_{h,r} \ast \dot{n}_{r} + \sum_{h'} J_{h,h'} \ast \dot{n}_{h'}\right)\right| \left\{\dot{n}_r\right\} \right].$$
This is not a tractable calculation. Taylor series expanding the nonlinearity $\phi(x)$ reveals that the mean will depend on \emph{all} higher cumulants of the hidden unit spike trains, which cannot in general be calculated explicitly. Instead, we again appeal to the fact that in a large, sufficiently connected network, we expect fluctuations to be small, as long as the network is not near a critical point. In this case, we may make a mean field approximation, which amounts to solving the self-consistent equation
\begin{equation}
\mathbb{E}\left[ \dot{n}_{h} | \left\{\dot{n}_r \right\}\right] = \lambda_0 \phi\left(\mu_h + \sum_r J_{h,r} \ast \dot{n}_r + \sum_{h'} J_{h,h'} \ast \mathbb{E}\left[ \dot{n}_{h'} | \left\{\dot{n}_r\right\}\right] \right).
\label{eqn:MethodsMFThidden0}
\end{equation}
In general, this equation must be solved numerically. Unfortunately, the conditional dependence on the activity of the recorded neurons presents a problem, as in principle we must solve this equation for \emph{all possible patterns of recorded unit activity}. Instead, we note that the mean hidden neuron firing rate is a \emph{functional} of the filtered recorded input $I_h(t) \equiv \sum_r J_{h,r} \ast \dot{n}_r$, so we can expand it as a functional Taylor series around some reference filtered activity $I^0_h(t) = \sum_r J_{h,r} \ast \dot{n}^0_r$,
\begin{align*}
\mathbb{E}\left[ \dot{n}_{h}(t) | \left\{I_h(t)\right\} \right] &= \mathbb{E}\left[ \dot{n}_{h} | \left\{I^0_h(t)\right\}\right]\\
& + \int dt_1 \sum_{h_1} \frac{\delta \mathbb{E}\left[ \dot{n}_{h}(t) | \left\{I^0_h(t)\right\}\right]}{\delta I_{h_1}(t_1)} (I_{h_1}(t_1) - I^0_{h_1}(t_1)) \\
& + \frac{1}{2} \int dt_1 dt_2 \sum_{h_1,h_2}\frac{\delta^2 \mathbb{E}\left[ \dot{n}_{h}(t) | \left\{I^0_h(t)\right\}\right]}{\delta I_{h_2}(t_2)\delta I_{h_1}(t_1)} (I_{h_2}(t_2) - I^0_{h_2}(t_2))(I_{h_1}(t_1) - I^0_{h_1}(t_1))\\ 
&+ \dots
\end{align*}
Within our mean field approximation, the Taylor coefficients are simply the response functions of the network --- i.e., the zeroth order coefficient is the mean firing rate of the neurons in the reference state $I^0_h(t)$, the first order coefficient is the linear response function of the network, the second order coefficient is a nonlinear response function, and so on.

There are two natural choices for the reference state $I^0_h(t)$. The first is simply the state of zero recorded unit activity, while the second is the mean activity of the recorded neurons. The zero-activity case conforms to the choice of nonlinear Hawkes models used in practice. Choosing the mean activity as the reference state may be more appropriate if the recorded neurons have high firing rates, but requires adjusting the form of the nonlinear Hawkes model so that firing rates are modulated by filtering the \emph{deviations} of spikes from the mean firing rate, rather than filtering the spikes themselves. Here, we focus on the zero-activity reference state. We present the formulation for the mean field reference state in the SI.

For the zero-activity reference state $I^0_h(t) = 0$, the conditional mean is 
\begin{align*}
\mathbb{E}\left[ \dot{n}_{h}(t) | \left\{I_h(t)\right\} \right] &= \mathbb{E}\left[ \dot{n}_{h} | 0 \right] + \int dt_1 \sum_{h_1} \frac{\delta \mathbb{E}\left[ \dot{n}_{h}(t) | 0 \right]}{\delta I_{h_1}(t_1)} I_{h_1}(t_1)\\
& + \frac{1}{2} \int dt_1 dt_2 \sum_{h_1,h_2}\frac{\delta^2 \mathbb{E}\left[ \dot{n}_{h}(t) |0\right]}{\delta I_{h_2}(t_2)\delta I_{h_1}(t_1)} I_{h_2}(t_2)I_{h_1}(t_1) + \dots.
\end{align*}
The mean inputs $\mathbb{E}\left[ \dot{n}_{h} | 0 \right]$ are the mean field approximations to the firing rates of the hidden neurons in the absence of the recorded neurons. Defining $\nu_h \equiv \mathbb{E}\left[ \dot{n}_{h} | 0 \right]$, these firing rates are given by
$$\nu_h = \lambda_0 \phi\left(\mu_h + \sum_{h'} \mathcal J_{h,h'} \nu_{h'} \right);$$
in writing this equation we have assumed that the steady-state mean field firing rates will be time-independent, and hence the convolution $J_{h,h'} \ast \nu_{h'} = \mathcal J_{h,h'}\nu_{h'}$, where $\mathcal J_{h,h'} = \int_0^\infty dt~J_{h,h'}(t)$. This assumption will generally be valid for at least some parameter regime of the network, but there can be cases where it breaks down, such as if the nonlinearity $\phi(x)$ is bounded, in which case a transition to chaotic firing rates $\nu_h(t)$ may exist (c.f. \cite{SompolinskyPRL1988}). The mean field equations for the $\nu_h$ are a system of transcendental equations that in general cannot be solved exactly. In practice we will solve the equations numerically, but we can develop a series expansion for the solutions (see below).

The next term in the series expansion is the linear response function of the hidden unit network, $\Gamma_{h,h'}(t-t') \equiv \frac{\delta \mathbb{E}\left[ \dot{n}_{h}(t) | 0 \right]}{\delta I_{h'}(t')},$ given by the solution to the integral equation
$$\Gamma_{h,h'}(t-t') = \gamma_h \left(\delta_{h,h'}\delta(t-t') + \sum_{h''} \int_0^\infty dt'' J_{h,h''}(t-t'') \Gamma_{h'',h'}(t''-t')\right).$$
The ``gain'' $\gamma_h$ is defined by
$$\gamma_h \equiv \lambda_0 \phi'\left(\mu_h + \sum_{h'}\mathcal J_{h,h'} \nu_{h'}\right),$$
where $\phi'(x)$ is the derivative of the nonlinearity with respect to its argument.

For time-independent drives $\mu_r$ and steady states $\nu_h$ (and hence $\gamma_h$), we may solve for $\Gamma_{h,h'}(t-t')$ by first converting to the frequency domain and then performing a matrix inverse:
$$\hat{\Gamma}_{h,h'}(\omega) = \left[\mathbb{I} - \hat{\mathbf{V}}(\omega) \right]^{-1}_{h,h'} \gamma_{h'},$$
where $\hat{V}_{h,h'}(\omega) = \gamma_h J_{h,h'}(\omega)$. 

If the zero and first order Taylor series coefficients in our expansion of $\mathbb{E}[\dot{n}_{h}(t)|\{\dot{n}_r\}]$ are the dominant terms---i.e., if we may neglect higher order terms in this expansion---then we may approximate the instantaneous firing rates of the recorded neurons by
$$\lambda_r(t) \approx \lambda_0 \phi\left(\mu^{\rm eff}_r + \sum_{r'} J^{\rm eff}_{r,r'} \ast \dot{n}_{r'}(t) \right),$$
where
$$\mu^{\rm eff}_r = \mu_r + \sum_h \mathcal J_{r,h} \nu_h$$
are the effective baselines of the recorded neurons and
$$\hat{J}^{\rm eff}_{r,r'}(\omega) = \hat{J}_{r,r'}(\omega) + \sum_{h,h'} \hat{J}_{r,h}(\omega) \hat{\Gamma}_{h,h'}(\omega) \hat{J}_{h',r'}(\omega)$$
are the effective coupling filters in the frequency domain, as given in the main text. In addition to neglecting the higher order spike filtering terms here, we have also neglected fluctuations around the mean input from the hidden network. These fluctuations are zero within our mean field approximation, but we could in principle calculate corrections to the mean field predictions using the techniques of \cite{OckerPLOSCB2017}; we do so to estimate the size of the effective noise correlations in the SI. 

In the main text, we decompose our expression for $\hat{J}^{\rm eff}_{r,r'}(\omega)$ into contributions from all paths that a signal can travel from neuron $r'$ to $r$. To arrive at this interpretation, we note that we can expand $\hat{\Gamma}_{h,h'}(\omega)$ in a series over paths through the hidden network. To start, we note that if $||\mathbf{\hat{V}}(\omega)|| < 1$ for some matrix norm $|| \cdot ||$, then the matrix $\left[\mathbb{I} - \mathbf{V}(\omega)\right]^{-1}$ admits a convergent series expansion \cite{MatrixAnalysisHornJohnson}
$$ \left[\mathbb{I} - \mathbf{\hat{V}}(\omega)\right]^{-1} = \sum_{\ell = 0} \mathbf{\hat{V}}(\omega)^\ell,$$
where $\mathbf{\hat{V}}(\omega)^\ell$ is a matrix product and $\mathbf{\hat{V}}(\omega)^0 \equiv \mathbb{I}$. We can write an element of the matrix product out as
$$\left[\mathbf{\hat{V}}(\omega)^\ell\right]_{h,h'} = \sum_{h_1,\dots,h_\ell} \hat{V}_{h,h_1}(\omega) \hat{V}_{h_1,h_2}(\omega) \dots \hat{V}_{h_{\ell-1},h_\ell}(\omega) \hat{V}_{h_\ell,h'}(\omega);$$
inserting $\hat{V}_{h_i,h_j}(\omega) = \gamma_{h_i} \hat{J}_{h_i,h_j}(\omega)$ yields
$$\left[\mathbf{\hat{V}}(\omega)^\ell\right]_{h,h'} = \sum_{h_1,\dots,h_\ell} \gamma_{h} \hat{J}_{h,h_1}(\omega) \gamma_{h_1} \hat{J}_{h_1,h_2}(\omega) \dots \gamma_{h_{\ell-1}}\hat{J}_{h_{\ell-1},h_\ell}(\omega) \gamma_{h_\ell} \hat{J}_{h_\ell,h'}(\omega).$$
This expression can be interpreted in terms of summing over paths through network of hidden neurons that join two observed neurons: the $\hat{J}_{h_i,h_j}(\omega)$ are represented by edges from neuron $h_j$ to $h_i$, and the $\gamma_{h_i}$ are represented by the nodes. In this expansion, we allow edges from one neuron back to itself, meaning we include paths in which signals loop back around to the same neuron arbitrarily many times before the signal is propagated further. However, such loops can be easily factored, contributing a factor $\sum_{m=0}^\infty (\gamma_h \hat{J}_{h,h}(\omega) )^m= 1/(1-\gamma_h \hat{J}_{h,h}(\omega))$. We thus remove the need to consider self-loops in our rules for calculating effective coupling filters by assigning a factor $\gamma_h/(1-\gamma_h J_{h,h}(\omega))$ to each node, as discussed in the main text and depicted in Fig.~\ref{fig:Jeffeqn}. (The contribution of the self-feedback loops can be derived rigorously; see the SI for the full derivation).

Although we have worked here in the frequency domain, our formalism does adapt straightforwardly to handle time-dependent inputs; however, among the consequences of this explicit time-dependence are that the mean field rates $\nu_h(t)$ are not only time-dependent, but solutions of a system of nonlinear integral equations, and hence more challenging to solve. Furthermore, quantities like the linear response of the hidden network, $\Gamma_{h,h'}(t,t')$, will depend on both absolute times $t$ and $t'$, rather than just their difference, $t-t'$, and hence we must also (numerically) solve for $\Gamma_{h,h'}(t,t')$ directly in the time domain. We leave these challenges for future work. 

\subsection*{Model network architectures}
\label{sec:networkarch}

Our main result, Eq.~(\ref{eqn:Jeffgeneral}), is valid for general network architectures with arbitrary weighted synaptic connections, so long as the hidden subset of the network has stable dynamics when the recorded neurons are removed. An example for which our method must be modified would be a network in which all or the majority of the hidden neurons are excitatory, as the hidden network is unlikely to be stable when the inhibitory recorded neurons are disconnected. Similarly, we find that synaptic weight distributions with undefined moments will generally cause the network activity to be unstable. For example, $\mathcal J_{i,j}$ drawn from a Cauchy distribution generally yield unstable network dynamics unless the weights are scaled inversely with a large power of the network size $N$. 

\subsubsection*{Specific networks---common features}

The specific network architectures we study in the main text share several features in common: all are sparse networks with sparsity $p$ (i.e., only a fraction $p$ of connections are non-zero) and non-zero synaptic weight strengths drawn independently from a random distribution with zero population mean and population standard deviation $J_0/(pN)^a$; the overall standard deviation of weights, accounting for the expected $1-p$ fraction of zero weights is $\sqrt{p}J_0/(pN)^a$. The parameter $a$ determines whether the synaptic strengths are ``strong'' ($a= 1/2$) or ``weak'' ($a=1$). In most of our analytical results we only need the mean and variances of the weights, so we do not need to specify the exact distribution. In simulations, we use a normal distribution. The reason for scaling the weights as $1/(pN)^a$, as opposed to just $1/N^a$, is that the mean incoming degree of connections is $p(N-1) \approx pN$ for large networks; this scaling thus controls for the typical magnitude of incoming synaptic currents.

For strongly coupled networks, the combined effect of sparsity and synaptic weight distribution yields an overall standard deviation of $\sqrt{p} J_0/\sqrt{pN} = J_0/\sqrt{N}$. Because the sparsity parameter $p$ cancels out, it does not matter if we consider $p$ to be fixed or $k_0 = pN$ to be fixed---both cases are equivalent. However, this is not the case if we scale $\mathcal J_{i,j}$ by $1/k_0$, as the overall standard deviation would then be $\sqrt{p}J_0/k_0$, which only corresponds to the weak-coupling limit if $p$ is fixed. If $k_0$ is fixed, the standard deviation would scale as $1/\sqrt{N}$. 

It is worth noting that the determination of ``weak'' versus ``strong'' coupling depends not only on the power of $N$ with which synaptic weights scale, but also on the network architecture and correlation structure of the weights $\mathcal J_{i,j}$. For example, for an all-to-all connected matrix with symmetric rank-1 synaptic weights of the form $\mathcal J_{i,j} = \zeta_i \zeta_j$, where the $\zeta_i$ are independently distributed normal random variates, the standard deviation of \emph{each} $\zeta$ must scale as $1/\sqrt{N}$ in order for hidden paths to generate $\mathcal O(1)$ contributions to effective interactions, such that $\mathcal J_{i,j}$ scales as $1/N$ but the coupling is still strong.

\subsubsection*{Specific networks---differences in architecture and synaptic constraints}

Beyond the common features outlined above, we perform our analysis of the distribution of effective synaptic interaction strengths for three network architectures commonly studied in network models. These architectures are not intended to be realistic representations of neuronal network structures, but to capture basic features of network architecture and therefore give insight into the basic features of the effective interaction networks.\\
~~~~

\noindent \emph{\ERtext + mixed synapses}---The first network we consider (and the one we perform most of our later analyses on as well) is an \ERtext random network architecture with ``mixed synapses.'' That is, each connection between neurons is chosen randomly with probably $p$. By ``mixed synapses'' we mean that each neuron's outgoing synaptic weights are chosen completely independently. i.e., in this network there are no excitatory or inhibitory neurons; each neuron make make both excitatory and inhibitory connections. The corresponding analysis is shown in Fig.~\ref{fig:Janalysis}A.\\
~~~~

\noindent \emph{\ERtext + Dale's law imposed}---Real neurons appear to split into separate excitatory and inhibitory classes, a dichotomy know as ``Dale's law'' (or alternatively, ``Dale's principle'' to highlight that it is not really a law of nature). Neurons in a network that obeys this law will have coupling filters $J_{i,j}(t)$ that are strictly positive for excitatory neurons and strictly negative for inhibitory neurons. This constraint complicates analytic calculations slightly, as the moments of the synaptic weights now depend on the identity of the neuron, and more care must be taken in calculating expected values or population averages. We instead impose this numerically to generate the results shown in Fig.~\ref{fig:Janalysis}B. The trends are the same as in the network with mixed synapses, with the resulting ratios being slightly reduced. 

As a technical point, because our analysis requires calculation of the mean field firing rates of the hidden network in absence of the recorded neurons, random sampling of the network may, by chance, yield hidden networks with an imbalance of excitatory neurons, for which the mean field firing rates of the hidden network may diverge for our choice of exponential nonlinearity. This is the origin of the relatively larger error bars in Fig.~\ref{fig:Janalysis}B: less random samplings for which the hidden network was stable were available to perform the computation. One way this artifact can be prevented is by choosing a nonlinearity that saturates, such as $\phi(x) = c/(1+\exp(-x))$, which prevents the mean-field firing rates from diverging and yields stable network activity (see Fig.~\ref{fig:JanalysisDLERsigmoid}). Another is to choose a different reference state of network activity around which we perform our expansion of $\mathbb{E}[\dot{n}_h|\{\dot{n}_r\}]$, such as the mean field state discussed in the SI. \\
~~~~

\noindent \emph{Watts-Strogatz network + mixed synapses}---Finally, although \ERtext networks are relatively easy to analyze analytically, and are ubiquitous in many influential computational and theoretical studies, real world networks typically have more structure.  Therefore, we also consider a network architecture with more structure, a Watts-Strogatz (small world) network. A Watts-Strogatz network is generated by starting with a $K$-nearest neighbor network (such that fraction of non-zero connections each neuron makes is $p = K/(N-1)$) and rewiring a fraction $\beta$ of those connections. The limit $\beta = 0$ remains a $K$-nearest neighbor network, while $\beta \rightarrow 1$ yields an \ERtext network. We generated the adjacency matrices of the Watts-Strogatz networks using code available in \cite{SongPRE2014}. Here we consider only a Watts-Strogatz network with mixed synapses; a network with spatial structure and Dale's law with become sensitive to both the distribution of excitatory and inhibitory neurons in the network as well as the way in which the neurons are sampled, an investigation we leave for future work. The results for the Watts-Strogatz network with mixed synapses are shown in Fig.~\ref{fig:Janalysis}C, and are qualitatively similar to the \ERtext netowrk with mixed synapses.

Because all three network types we considered yield qualitatively similar results, for the remainder of our analyses, we focus on the \ERtext + mixed synapses network for simplicity in both simulations and analytical calculations.

Parameter values used to generate our networks are given in Table~\ref{tab:params}.

\begin{table}[htp]
\caption{Network connectivity parameter values for Figs.~\ref{fig:Janalysis}-\ref{fig:JanalysisDLERsigmoid},~\ref{fig:MFTvalidation}-\ref{fig:linearapproxvalidationNhid500}. See individual captions for other figures.}
\begin{center}
\begin{tabular}{|c|c|}
\hline
Number of neurons $N$ & 1000 \\ \hline
Number of hidden neurons $N_{\rm hid}$ & $\{1,90,190,290,390,490,590,690,790,890,990\}$ \\ \hline
Number of recorded neurons $N_{\rm rec}$ & $N-N_{\rm hid}$ \\ \hline
Baselines $\mu_i$ & -1.0, $\forall i$ \\ \hline
Sparsity $p$ & $0.2$ \\ \hline
Coupling weights $\mathcal J_{ij}~(i \neq j)$ & $\mathcal N(0,J_0^2/(pN)^{2a})$ \\ \hline
Self-coupling weights $\mathcal J_{ii}$ & $0$ \\ \hline
Coupling regime $a$ & $1$ (weak coupling) or $1/2$ (strong coupling) \\ \hline
Rewiring probability $\beta$ (Watts-Strogatz only) & 0.3 \\ \hline
Characteristic synaptic weight $J_0$ & $\{0.25, 0.5, 0.75, 1.0\}$ \\ \hline
Firing frequency $\lambda_0$ & 1.0 \\ \hline
\end{tabular}
\end{center}
\label{tab:params}
\end{table}%

\subsection*{Choice of nonlinearity $\phi(x)$}
\label{sec:choiceofnonlin}

The nonlinear function $\phi(x)$ sets the instantaneous firing rate for the neurons in our model. Our main analytical results (e.g., Eq.~(\ref{eqn:Jeffgeneral}) hold for arbitrary choice of $\phi(x)$. Where specific choices are required in order to perform simulations, we used $\phi(x) = \mbox{max}(x,0)$ for the results presented in Figs.~\ref{fig:ffwi_3neuron} and \ref{fig:ffwi_4neuron} and $\phi(x) = \exp(x)$ otherwise. The rectified linear choice is convenient for small networks, as high-order derivatives are zero, which eliminates corresponding high-order ``loop corrections'' to mean field theory \cite{OckerPLOSCB2017}. The exponential function is the ``canonical'' choice of nonlinearity for the nonlinear Hawkes process \cite{Simoncelli2004,Paninski2004,Pillow2005,Pillow2008}. The exponential has particularly nice theoretical properties, but is also convenient for fitting the nonlinear Hawkes model to data, as the log-likelihood function of the model simplifies considerably and is  convex (though some similar families of nonlinearities also yield convex log-likelihood functions). 

An important property that both choices of nonlinearity possess is that they are unbounded. This property is necessary to \emph{guarantee} that a neuron spikes given enough input. A bounded nonlinearity imposes a maximum firing rate, and neurons cannot be forced to spike reliably by providing a large bolus of input. The downside of an unbounded nonlinearity is that it is possible for the average firing rates to diverge, and the network never reaches a steady state. For example, in a purely excitatory network (all $\mathcal J_{i,j} \geq 0$) with an exponential nonlinearity, neural firing will run away without a sufficiently strong self-refractory coupling to suppress the firing rate. This will not occur with a bounded nonlinearity, as excitation can only drive neurons to fire at some maximum but finite rate.

This can be a problem in simulations of networks obeying Dale's law. For unbounded nonlinearities, the mean field theory for the hidden network occasionally does not exist due to an imbalance of excitatory and inhibitory neurons caused by our random selection of recorded of neurons. However, the Dale's law network is stable if the nonlinearity is bounded. We demonstrate this below in Figs.~\ref{fig:JanalysissignedERsigmoid} and \ref{fig:JanalysisDLERsigmoid}, comparing simulations of the effective interaction statistics in \ERtext networks with and without Dale's law for a sigmoidal nonlinearity $\phi(x) = 2/(1+e^{-x})$.

\begin{figure}[h!]
 \centering
 \includegraphics[width=0.55\textwidth]{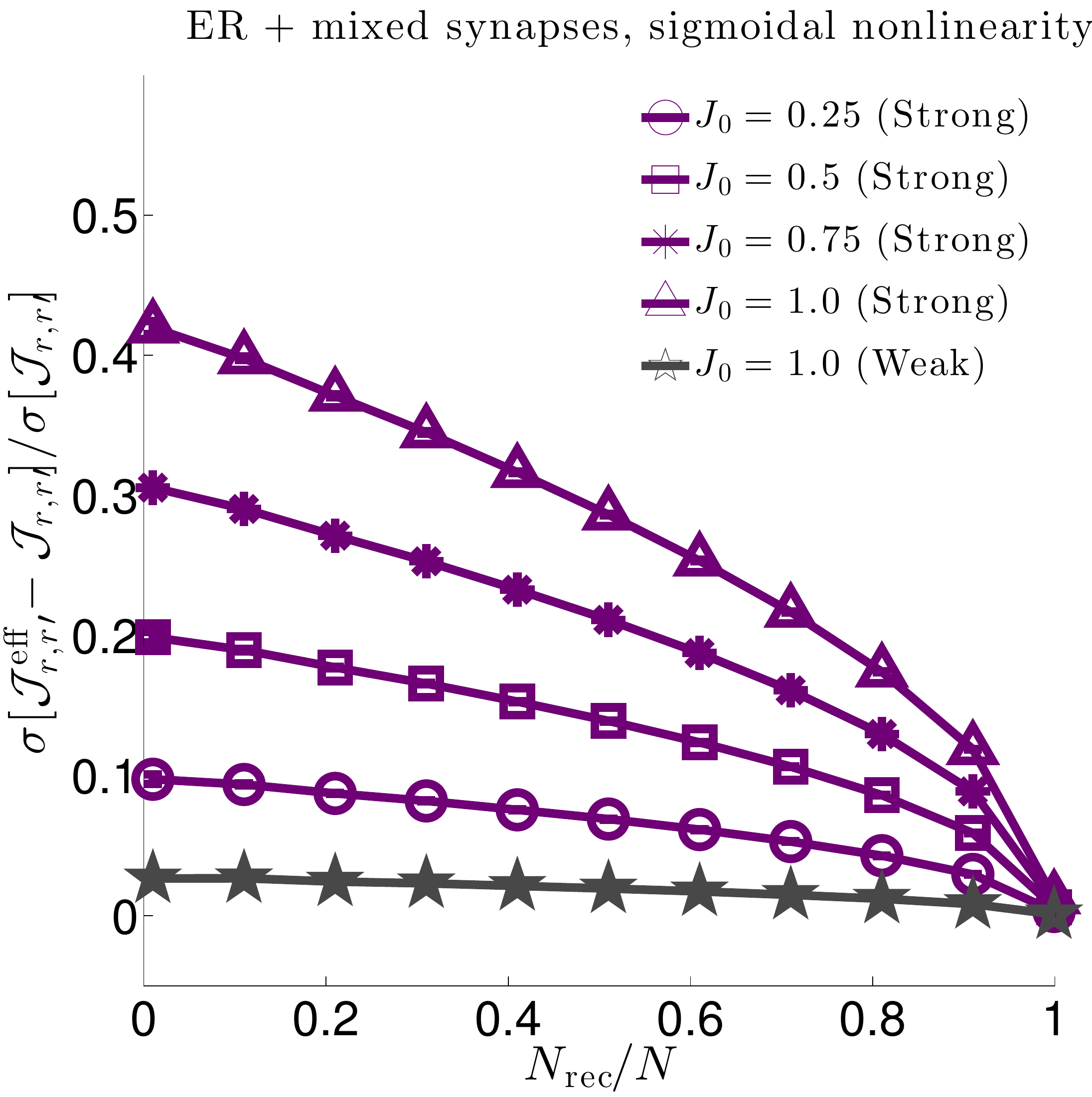}
  \caption{Same as Fig.~\ref{fig:Janalysis}A in main text, but for a sigmoidal nonlinearity $\phi(x) = 2/(1+e^{-x})$.}
  \label{fig:JanalysissignedERsigmoid}
\end{figure}

\begin{figure}[h!]
 \centering
 \includegraphics[width=0.55\textwidth]{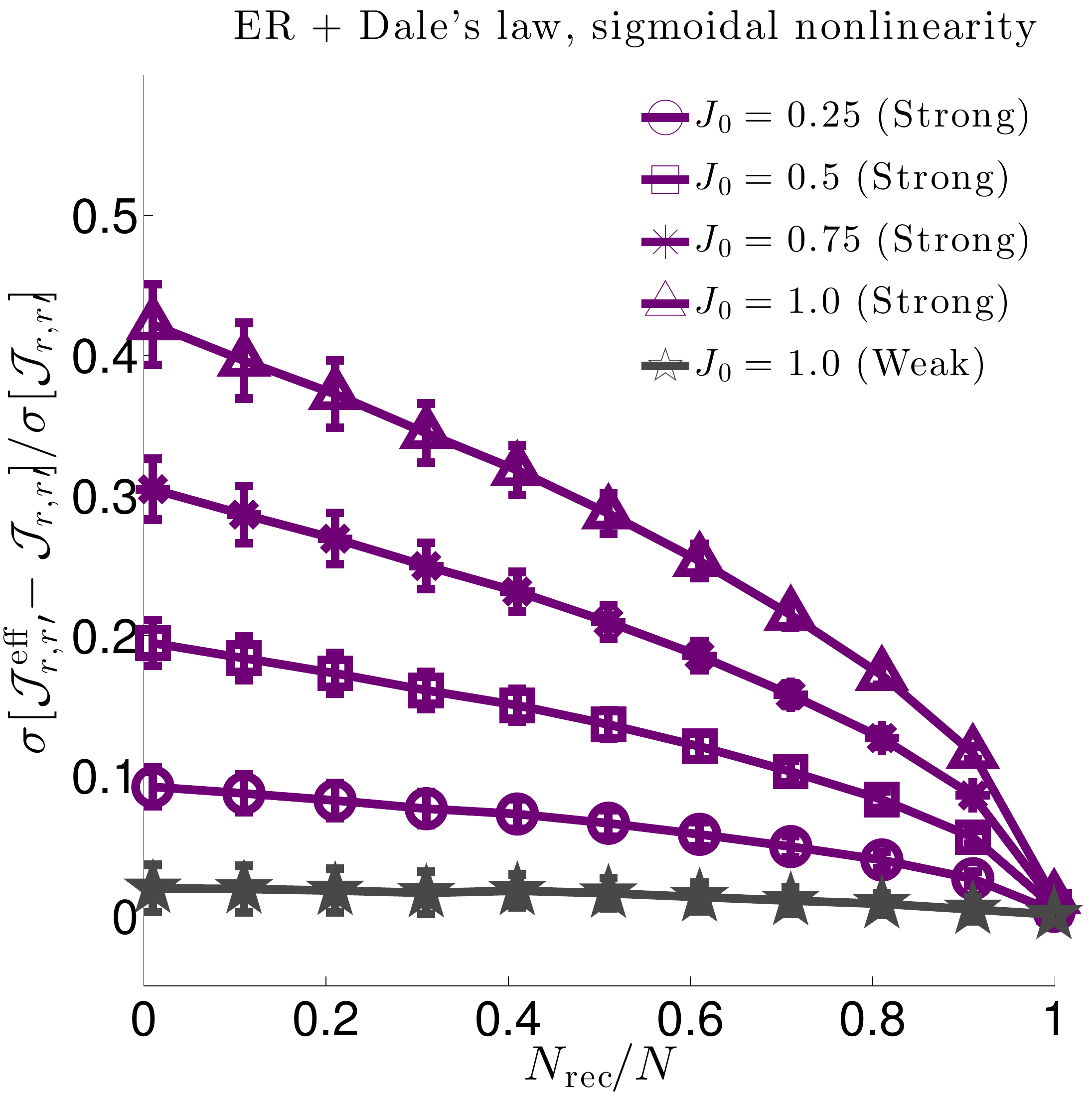}
  \caption{Same as Fig.~\ref{fig:Janalysis}B in the main text, but for a sigmoidal nonlinearity $\phi(x) = 2/(1+e^{-x})$. Because the sigmoid is bounded the mean field solution cannot diverge, yielding better results.}
  \label{fig:JanalysisDLERsigmoid}
\end{figure}

Another consequence of unbounded nonlinearities is that the mean firing rates are either finite or they diverge. Bounded nonlinearities, on the other hand, may allow for the possibility of a transition to chaotic dynamics in the mean-field firing rate dynamics (cf. the results of the \cite{SompolinskyPRL1988}).

\subsection*{Specific choices of network properties used to generate figures}
\label{sec:networkchoices}

\subsubsection*{Feedforward-inhibitiory circuit model details}

\noindent \emph{3 neuron circuit (Fig.~\ref{fig:ffwi_3neuron})}

Using our graphical rules (Fig.~\ref{fig:Jeffeqn}), we calculated the effective interaction from neuron $1$ to $2$ for the circuit shown in Fig.~\ref{fig:ffwi_3neuron}A, giving Eq.~\ref{eqn:ffwiJeff}. In principle, our mean field approximation would not be expected to hold for such a small circuit; in particular, loop corrections \cite{OckerPLOSCB2017} to our calculation of the rate $\nu_3$ and associated gain $\gamma_3$ might be significant. However, as loop corrections depend on derivatives of the nonlinearity $\phi(x)$, we can minimize these errors by choosing $\phi(x) = \mbox{max}(x,0)$, for which $\phi'(x) = \Theta(x)$, the Heaviside step function. Accordingly, we can solve for $\nu_3 = \lambda_0 \mu_3/(1-\lambda_0 \mathcal J_{33})$ and $\gamma_3 = \lambda_0$ for this particular network.

To generate the plots shown in Fig.~\ref{fig:ffwi_3neuron}C, we take the inter-neuron couplings to have the form $J_{i,j}(\tau) = \mathcal J_{i,j} \alpha_{i,j}^2 \tau e^{-\alpha_{i,j} \tau}$ and the self-history couplings to have the form $J_{i,i}(\tau) = \mathcal J_{i,i}  \beta_{i,i} e^{-\beta_{i,i} \tau}$. 

Using Mathematica to perform the inverse Fourier transform, we obtain an explicit expression for the effective interaction,
\begin{align*}
J^{\rm eff}_{2,1}(\tau) &= \mathcal J_{21} \alpha_{21}^2 \tau e^{-\alpha_{21} \tau} \\
&~~~~+ \mathcal J_{23}\mathcal J_{31} \alpha_{23}^2 \alpha_{31}^2 \times \\
&~~~~~~~~ \Bigg[ \frac{\beta_{33} \mathcal J_{33}}{(\alpha_{23}-\beta_{33}(1-\lambda_0 \mathcal J_{33}))^2(\alpha_{31}-\beta_{33}(1-\lambda_0 \mathcal J_{33}))^2}e^{-\beta_{33}(1-\lambda_0 \mathcal J_{33})\tau} \\
&~~~~~~~~~~~~ +\frac{(-2\alpha_{31}^2 + \beta_{33} \alpha_{31}(4-\lambda_0\mathcal J_{33}) - 2\beta_{33}^2(1-\lambda_0 \mathcal J_{33}) - \beta_{33}\mathcal J_{33}\alpha_{23} )}{(\alpha_{23}-\alpha_{31})^2(\alpha_{31}-\beta_{33}(1-\lambda_0 \mathcal J_{33}))^2}e^{-\alpha_{31}\tau} \\
&~~~~~~~~~~~~ +\frac{(-2\alpha_{23}^2 + \beta_{33} \alpha_{23}(4-\lambda_0\mathcal J_{33}) - 2\beta_{33}^2(1-\lambda_0 \mathcal J_{33}) - \beta_{33}\mathcal J_{33}\alpha_{31} )}{(\alpha_{31}-\alpha_{23})^2(\alpha_{23}-\beta_{33}(1-\lambda_0 \mathcal J_{33}))^2}e^{-\alpha_{23}\tau} \\
&~~~~~~~~~~~~ + \frac{\alpha_{23}-\beta_{33}}{(\alpha_{23}-\alpha_{31})^2(\alpha_{23}-\beta_{33}(1-\lambda_0 \mathcal J_{33}))^2}\tau e^{-\alpha_{23}\tau} \\
&~~~~~~~~~~~~ + \frac{\alpha_{31}-\beta_{33}}{(\alpha_{31}-\alpha_{23})^2(\alpha_{31}-\beta_{33}(1-\lambda_0 \mathcal J_{33}))^2}\tau e^{-\alpha_{31}\tau} \Bigg].
\end{align*}
In order for the inverse Fourier transform to converge and result in a causal function, we require that $1 - \lambda_0 \mathcal J_{33} > 0$.

Parameter values used to generate the plots in Fig.~\ref{fig:ffwi_3neuron}C are given in Table~\ref{tab:ffwi_3neuron_params}.

\begin{table}[htp]
\caption{Parameter values for Fig.~\ref{fig:ffwi_3neuron}C. Setting $\lambda_0 = 1.0$ simply sets the units of frequency and time to be measured relative to $\lambda_0$ (e.g., the value $\alpha_{31} = 1.8$ really means $\alpha_{31} = 1.8\lambda_0$ and $\mathcal J_{31} = 2.0$ really means $\mathcal J_{31} = 2.0/\lambda_0$).}
\begin{center}
\begin{tabular}{|c|c|}
\hline
Parameter & value \\ \hline
$\lambda_0$ & 1.0 \\ \hline
$\mathcal J_{21}$ & $1.0$ \\ \hline
$\mathcal J_{23}$ & $-2.0$ \\ \hline
$\mathcal J_{31}$ & $2.0$ \\ \hline
$\mathcal J_{33}$ & $-0.9$ \\ \hline
$\alpha_{21} = \alpha_{23} = \beta_{33}$ & $1.0$ \\ \hline
$\alpha_{31}$ & $1.8$ \\ \hline
\end{tabular}
\end{center}
\label{tab:ffwi_3neuron_params}
\end{table}%

\noindent \emph{4 neuron circuit (Fig.~\ref{fig:ffwi_4neuron})}

Like for the 3-neuron circuit, we can use our graphical rules (Fig.~\ref{fig:Jeffeqn}) to calculate the effective interaction for our 4-neuron circuit (Fig.~\ref{fig:ffwi_4neuron}A) in the frequency domain:
\begin{align*}
\hat{J}^{\rm eff}_{21}(\omega)-\hat{J}_{21}(\omega) &= \hat{J}_{23}(\omega)\left[\sum_{m=0}^\infty \left(\gamma_3 \hat{J}_{34}(\omega) \gamma_4 \hat{J}_{43}(\omega)\right)^m\right]\hat{J}_{31}(\omega) \\
&~~~~+ \hat{J}_{23}(\omega)\left[\sum_{m=0}^\infty \left(\gamma_3 \hat{J}_{34}(\omega) \gamma_4 \hat{J}_{43}(\omega)\right)^m\right]\gamma_3\hat{J}_{34}(\omega)\gamma_4\hat{J}_{41}(\omega)\\
&= \hat{J}_{23}(\omega)\frac{1}{1 - \gamma_3 \hat{J}_{34}(\omega) \gamma_4 \hat{J}_{43}(\omega)}\hat{J}_{31}(\omega) \\
&~~~~+ \hat{J}_{23}(\omega)\frac{1}{1 - \gamma_3 \hat{J}_{34}(\omega) \gamma_4 \hat{J}_{43}(\omega)}\gamma_3\hat{J}_{34}(\omega)\gamma_4\hat{J}_{41}(\omega)\\
&= \hat{J}_{23}(\omega)\hat{J}_{31}(\omega) + \hat{J}_{23}(\omega)\gamma_3\hat{J}_{34}(\omega)\gamma_4\hat{J}_{41}(\omega) \\
&~~~~+ \frac{\hat{J}_{23}(\omega)\gamma_3 \hat{J}_{34}(\omega) \gamma_4 \hat{J}_{43}(\omega)\hat{J}_{31}(\omega)}{1 - \gamma_3 \hat{J}_{34}(\omega) \gamma_4 \hat{J}_{43}(\omega)} + \frac{\hat{J}_{23}(\omega)\gamma_3 \hat{J}_{34}(\omega) \gamma_4 \hat{J}_{43}(\omega)\gamma_3\hat{J}_{34}(\omega)\gamma_4\hat{J}_{41}(\omega)}{1 - \gamma_3 \hat{J}_{34}(\omega) \gamma_4 \hat{J}_{43}(\omega)};
\end{align*}
in going to the last equality we have separated the terms out into contributions from each of the paths, in order, shown in Fig.~\ref{fig:ffwi_4neuron}B.

To generate the plots in Fig.~\ref{fig:ffwi_4neuron}C, we choose $\phi(x) = \mbox{max}(x,0)$, which gives $\gamma_i = \lambda_0$, as in Fig.~\ref{fig:ffwi_3neuron}C, and interaction filters $J_{2,1}(\tau) = \mathcal J_{21} \alpha_{21}^2 \tau e^{-\alpha \tau}$ for the direct interaction and $J_{i,j}(\tau) = \mathcal J_{ij} \alpha^2 \tau e^{-\alpha \tau}$ for all other interactions shown---i.e., all other interactions have the same decay time $\alpha^{-1}$ for simplicity.

Inverting the Fourier transform using Mathematica yields
\begin{align*}
J^{\rm eff}_{2,1}(\tau) &= \mathcal J_{21}\alpha^2 \tau e^{-\alpha \tau} - \frac{\mathcal J_{23} \mathcal J_{31}\alpha e^{-\alpha \tau}}{2|\mathcal J_{34}|^{3/4}|\mathcal J_{43}|^{3/4}}\left(\sin(\alpha(|\mathcal J_{34}| |\mathcal J_{43}|)^{1/4}\tau) - \sinh(\alpha(|\mathcal J_{34}| |\mathcal J_{43}|)^{1/4}\tau)\right) \\
& + \frac{\mathcal J_{23} \mathcal J_{41}\alpha e^{-\alpha \tau}}{2|\mathcal J_{34}|^{1/4} |\mathcal J_{43}|^{5/4}}\left(2\alpha (|\mathcal J_{34}| |\mathcal J_{43}|)^{1/4}\tau - \sin(\alpha(|\mathcal J_{34}| |\mathcal J_{43}|)^{1/4}\tau) - \sinh(\alpha(|\mathcal J_{34}| |\mathcal J_{43}|)^{1/4}\tau)\right)
\end{align*}
In order for this result to converge, we require $|\mathcal J_{34}||\mathcal J_{43}| < 1$. Splitting this result up into the contributions to each plot in Fig.~\ref{fig:ffwi_4neuron}C, using the specific parameter choices $\lambda_0 = 1$ and $\mathcal J_{34} = \mathcal J_{43} \equiv \mathcal J$, gives
\begin{align*}
2 \leftarrow 3 \leftarrow 1: & \frac{1}{6}\alpha^4 \mathcal J_{23}\mathcal J_{31} \tau^3 e^{-\alpha \tau},\\
2 \leftarrow 3\leftarrow 4 \leftarrow 1: & -\frac{1}{120}\alpha^6 |\mathcal J| \mathcal J_{23} \mathcal J_{41} \tau^5 e^{-\alpha \tau},\\
2 \leftarrow 3 \leftrightarrow 4 \leftarrow 3 \leftarrow 1: & \frac{\alpha \mathcal J_{23} \mathcal J_{31}\left(\cosh(\alpha \tau) - \sinh(\alpha\tau)\right)(-2\alpha^3|\mathcal J|^{3/2} \tau^3 - 6\sin(\alpha\sqrt{|\mathcal J|}\tau) + 6\sinh(\alpha\sqrt{|\mathcal J|}\tau)}{12|\mathcal J|^{3/2}},\\
2 \leftarrow 3 \leftrightarrow 4 \leftarrow 1: & \frac{\alpha e^{-\alpha \tau} \mathcal J_{23} \mathcal J_{41} (\alpha\sqrt{|\mathcal J|} \tau (120 + \alpha^4 \mathcal J^2 \tau^4) - 60\sin(\alpha\sqrt{J}\tau) - 60\sinh(\alpha\sqrt{|\mathcal J|}\tau))}{120|\mathcal J|^{3/2}}.
\end{align*}
Parameter values used to generate the plots in Fig.~\ref{fig:ffwi_4neuron}C are given in Table~\ref{tab:ffwi_4neuron_params}.

\begin{table}[htp]
\caption{Parameter values for Fig.~\ref{fig:ffwi_4neuron}C. Setting $\lambda_0 = 1.0$ simply sets the units of frequency and time to be measured relative to $\lambda_0$ (e.g., the value $\alpha = 1.294$ really means $\alpha = 1.294\lambda_0$ and $\mathcal J_{23} = -3.0$ really means $\mathcal J_{23} = -3.0/\lambda_0$).}
\begin{center}
\begin{tabular}{|c|c|}
\hline
Parameter & value \\ \hline
$\lambda_0$ & 1.0 \\ \hline
$\mathcal J_{21} = \mathcal J_{31} = \mathcal J_{41}$ & $1.0$ \\ \hline
$\mathcal J_{23}$ & $-3.0$ \\ \hline
$\mathcal J_{34} = \mathcal J_{43} = \mathcal J$ & $-0.9$ \\ \hline
$\alpha_{21}$ & $1.0$ \\ \hline
$\alpha$ & $1.294$ \\ \hline
\end{tabular}
\end{center}
\label{tab:ffwi_4neuron_params}
\end{table}%

\subsubsection*{Large networks}

To generate the results in Fig.~\ref{fig:Jeffs} in the main text, we choose the coupling filters to be $J_{i,j}(t) = \mathcal J_{i,j} \alpha^2 t e^{-\alpha t}$, for $i \neq j$, which has Fourier transform
$$\hat{J}_{i,j}(\omega) = \frac{\mathcal J_{i,j} \alpha^2}{(\alpha + i \omega)^2},$$
using the Fourier convention
$$\hat{f}(\omega) = \int_{-\infty}^\infty dt~e^{-i\omega t} f(t).$$
The weight matrix $\mathbf{\mathcal J}$ is generated as described in ``Model network architectures,'' choosing $J_0 = 1.0$. We partition this network up into recorded and hidden subsets. For a network of $N$ neurons, we choose neurons $1$ to $N_{\rm rec}$ to be recorded, and the remainder to be hidden, hence we define (using an index notation starting at 1; indices should be subtracted by 1 for 0-based index counting)
$$\mathcal J^{\rm RR} = \mathcal J[1:N_{\rm rec},1:N_{\rm rec}],$$
$$\mathcal J^{\rm RH} = \mathcal J[1:N_{\rm rec},(N_{\rm rec}+1):N],$$
$$\mathcal J^{\rm HR} = \mathcal J[(N_{\rm rec}+1):N,1:N_{\rm rec}],$$
and
$$\mathcal J^{\rm HH} = \mathcal J[(N_{\rm rec}+1):N,(N_{\rm rec}+1):N].$$

We numerically calculate the linear response matrix $\mathbf{\hat{\Gamma}}(\omega)$ by evaluating
$$\mathbf{\hat{\Gamma}}(\omega) = \left[\mathbb{I} - \mathbf{\hat{V}}^{\rm HH}(\omega) \right]^{-1} \mbox{diag}(\vec{\gamma}),$$ 
where $\hat{V}^{\rm HH}_{h,h'}(\omega) = \gamma_h \mathcal J_{h,h'}(\omega)$ and \mbox{diag}($\vec{\gamma})$ is an $N_{\rm hid} \times N_{\rm hid}$ diagonal matrix with elements $\gamma_h$. 

The effective coupling filter in the frequency domain can then be evaluated pointwise at a desired set of frequencies $\omega$ by matrix multiplication,
$$\mathbf{\hat{J}}^{\rm eff}(\omega) =  \frac{\alpha^2}{(\alpha + i\omega)^2} \mathcal J^{\rm RR}+ \left(\frac{\alpha^2}{(\alpha + i\omega)^2} \right)^2 \mathcal J^{\rm RH} \mathbf{\hat{\Gamma}}(\omega) \mathcal J^{\rm HR}.$$
We then return to the time domain by inverse Fourier transforming the result, achieved by treating $\mathbf{\hat{J}}^{\rm eff}(\omega)$ as an $N_{\rm rec} \times N_{\rm rec} \times N_{\rm freq}$ array (where $N_{\rm freq}$ is the number of frequencies at which we evaluate the effective coupling) and multiplying along the frequency dimension by an $N_{\rm freq} \times N_{\rm time}$ matrix $\mathbf{E}$ with elements $E_{\omega,t} = \exp(i \omega t) \Delta t/(2\pi)$, for $N_{\rm time}$ sufficiently small time bins of size $\delta t = 0.1/\alpha$, for $\alpha = 10$, as listed in Table~\ref{tab:simparams}.

To generate Fig.~\ref{fig:Janalysis}, we focus on the zero-frequency component of $\hat{J}^{\rm eff}(\omega)$, which is also equal to the time integral of $\mathbf{J}^{\rm eff}(t)$. As in the main text, we label the elements of this component $\mathcal J^{\rm eff}_{r,r'} = \hat{J}^{\rm eff}_{r,r'}(\omega=0)$, which is equal to
$$\mathcal J^{\rm eff}_{r,r'} = \mathcal J_{r,r'} + \sum_{h,h'} \mathcal J_{r,h} \hat{\Gamma}_{h,h'}(0) \mathcal J_{h',r'}.$$
We do not need to simulate the full network to study the statistics of $\mathcal J^{\rm eff}_{r,r'}$. We only need to generate samples of the matrix $\mathcal J$ and evaluate $\mathbf{\hat{\Gamma}}(0)$. This is where the choice of an \ERtext network that is not restricted to obey Dale's law becomes convenient. Because the weights $\mathcal J_{i,j}$ are \emph{i.i.d.} and the sign of the weight is random, population averages will be equivalent to expected values. i.e., the sample mean
$$ \tilde{\mathcal J}_{\rm mean} = \frac{1}{N_{\rm rec}(N_{\rm rec}-1)} \sum_{r\neq r'} \mathcal J^{\rm eff}_{r,r'}$$
and sample variance
$$\tilde{\mathcal J}_{\rm var} = \frac{1}{N_{\rm rec}(N_{\rm rec}-1) - 1} \sum_{r\neq r'} \left(\mathcal J^{\rm eff}_{r,r'} - \tilde{\mathcal J}_{\rm mean}\right)^2$$
will tend to the expected values $\mathbb{E}[\mathcal J^{\rm eff}_{r,r'}]$ and $\mbox{var}[\mathcal J^{\rm eff}_{r,r'}]$ for large networks. We have explicitly removed the diagonal elements from these averages because these elements will have slightly different statistics from the off-diagonal elements due to the fact that all ground-truth self-couplings are set to zero, $\mathcal J_{r,r} = 0$. This allows us to compare the population variance, plotted in Fig.~\ref{fig:Janalysis} (after normalization by the population variance of the true off-diagonal weights), to the expected variance calculated analytically below.

The error bars in Fig.~\ref{fig:Janalysis} are generated by first drawing a single sample of true weights $\mathcal J$, and then taking $100$ random subsets of $N_{\rm rec} = \left\{10, 110, 210, 310, 410, 510, 610, 710, 810, 910, 999 \right\}$ recorded neurons. For this analysis, random subsets were generated by permuting the indices of the full weight matrix $\mathcal J$ and taking the last $N_{\rm rec}$ neurons to be recorded. For each random subset of the network we calculate the population statistics. The standard error of, for example, the population variance $\tilde{\mathcal J}_{\rm var}$ across subsets gives an estimate of the error. However, if we only use a single sample of the network architecture and weights $\mathcal J_{i,j}$, this estimate may depend on the particular instantiation of the network. To average over the effects of global network architecture, we draw a total of $10$ network architecture samples, and average a second time over these samples to obtain our final estimates of the population variance of $\mathcal J^{\rm eff}_{r,r'}$. We note that for an \ERtext network with mixed synapses, this second stage of averaging is probabilistically unnecessary: for a large enough network random subsets of a single large network are statistically identical to random subsets drawn from several samples of full \ERtext networks. However, this will not be true for networks with more structure, such as the Watts-Strogatz or Dale's law networks we also considered, for which the second stage of averaging over the global network architecture is necessary to average over network configurations.

\subsection*{Series approximation for the mean field firing rates for the case of exponential nonlinearity $\phi(x)=e^x$}
\label{sec:seriesapprox}

The mean field firing rates for the hidden neurons are given by
$$\nu_h = \lambda_0 \exp\left(\mu_h + \sum_{h'} \mathcal J_{h,h'} \nu_{h'} \right),$$
where we focus specifically on the case of exponential nonlinearity $\phi(x) = \exp(x)$. For this choice of nonlinearity, $\gamma_h = \nu_h$, so we do not need to calculate a separate series for the gains.

This system of transcendental equations generally cannot be solved analytically. However, for small $\exp(\mu_h) \ll 1$ we can derive, recursively, a series expansion for the firing rates. We first consider the case of $\mu_h = \mu_0$ for all hidden neurons $h$. Let $\epsilon = \exp(\mu_0)$. We may then write
$$\nu_h = \lambda_0 \epsilon \sum_{\ell = 0}^\infty a^{(\ell)}_h (\lambda_0 \epsilon)^\ell.$$
Plugging this into the mean field equation,
\begin{align*}
\sum_{\ell = 0}^\infty a^{(\ell)}_h (\lambda_0 \epsilon)^\ell &= \exp\left(\sum_{h'} \mathcal J_{h,h'} \sum_{\ell = 0}^\infty a^{(\ell)}_{h'} (\lambda_0 \epsilon)^{\ell+1} \right)\\
&= 1+\sum_{m=1}^\infty \frac{1}{m!} \left(\sum_{h'} \mathcal J_{h,h'} \sum_{\ell = 0}^\infty a^{(\ell)}_{h'} (\lambda_0 \epsilon)^{\ell+1} \right)^m\\
&= 1+\sum_{m=1}^\infty \frac{1}{m!} \sum_{\ell_1,\dots,\ell_m,h'_1,\dots,h'_m} \mathcal J_{h,h'_1}a^{(\ell_1)}_{h'_1} \dots \mathcal J_{h,h'_m}a^{(\ell_m)}_{h'_m} (\lambda_0 \epsilon)^{\ell_1 + \dots + \ell_m+m} \\
&= 1+ \sum_{\ell=1}^\infty \left\{\sum_{m=1}^\infty \frac{1}{m!} \sum_{\ell_1,\dots,\ell_m,h'_1,\dots,h'_m} \hspace{-0.5cm}\mathcal J_{h,h'_1}a^{(\ell_1)}_{h'_1} \dots \mathcal J_{h,h'_m}a^{(\ell_m)}_{h'_m} \delta_{\ell,\ell_1 + \dots + \ell_m+m}\right\} (\lambda_0 \epsilon)^{\ell}.
\end{align*}
Thus, matching powers of $\lambda_0 \epsilon$ on the left and right hand sides, we find $a^{(0)}_h = 1$ and
$$a^{(\ell)}_h = \sum_{m=1}^\infty \frac{1}{m!} \sum_{\ell_1,\dots,\ell_m,h'_1,\dots,h'_m} \mathcal J_{h,h'_1}a^{(\ell_1)}_{h'_1} \dots \mathcal J_{h,h'_m}a^{(\ell_m)}_{h'_m} \delta_{\ell,\ell_1 + \dots + \ell_m+m}$$
for $\ell > 0$.

For $\ell = 1$, the sum in $m$ truncates at $m=1$ (as $\delta_{\ell,\ell_1+\dots+\ell_m+m}$ is zero for $m > \ell$, as all indices are positive). Thus,
\begin{align*}
a^{(1)}_h &= \sum_{h_1'} \mathcal J_{h,h'_1},\\
a^{(2)}_h &= \sum_{h_1',h_2'}\left\{ \mathcal J_{h,h_1'}\mathcal J_{h_1',h_2'} + \frac{1}{2} \mathcal J_{h,h'_1}\mathcal J_{h,h'_2}\right\},\\
a^{(3)}_h &= \sum_{h_1',h_2',h_3'} \Big\{\mathcal J_{h,h_1'}\mathcal J_{h_1',h_2'}\mathcal J_{h_2',h_3'} + \frac{1}{2} \mathcal \mathcal J_{h,h_1'}\mathcal J_{h_1',h_2'}\mathcal J_{h_1',h_3'} + \mathcal J_{h,h_1'}\mathcal J_{h,h'_2} \mathcal J_{h'_2,h_3'}\\
& \hspace{7.5cm} + \frac{1}{3!} \mathcal J_{h,h_1'} \mathcal J_{h,h_2'}\mathcal J_{h,h_3'}\Big\}.
\end{align*}
With this we have calculated the firing rates to $\mathcal O(\epsilon^4)$. 

The analysis can be straightforwardly extended to the case of heterogeneous $\mu_h$, though it becomes more tedious to compute terms in the (now multivariate) series. Assuming $\epsilon_h \equiv \exp(\mu_h) \ll 1$ for all $h$, to $\mathcal O(\epsilon^3)$ we find
$$\nu_h = \lambda_0 \epsilon_h\left( 1 + \sum_{h'}  \mathcal J_{h,h'} \lambda_0\epsilon_{h'} + \sum_{h_1',h_2'} \left\{\mathcal J_{h,h_1'}\mathcal J_{h_1',h_2'} + \frac{1}{2} \mathcal J_{h,h_1'}\mathcal J_{h,h_2'} \right\}\lambda_0 \epsilon_{h_1'} \lambda_0 \epsilon_{h_2'}+\dots\right).$$
\subsection*{Variance of the effective coupling to second order in $N_{\rm rec}/N$ \& fourth order in $\lambda_0J_0e^{\mu_0}$ (exponential nonlinearity)}
\label{sec:varJeff}

To estimate the strength of the hidden paths, we would like to calculate the variance of the effective coupling $\mathcal J_{r,r'}^{\rm eff}$ and compare its strength to the variance of the direct couplings $\mathcal J_{r,r'}$, where $\mathcal J^{\rm eff}_{r,r'} \equiv \int_0^\infty dt~J^{\rm eff}_{r,r'}(t)$ and $\mathcal J_{r,r'} \equiv \int_0^\infty dt~J^{\rm eff}_{r,r'}(t)$, as in the main text.

We assume that the synaptic weights $\mathcal J_{i,j}$ are independently and identically distributed with zero mean and variance $\mbox{var}(\mathcal J) = p \frac{J_0^2}{(pN)^{2a}}$ for $i \neq j$, where $a = 1$ corresponds to weak coupling and $a = 1/2$ corresponds to strong coupling. We assume no self-couplings, $\mathcal J_{i,i} = 0$ for all neurons $i$. The overall factor of $p$ in $\mbox{var}[\mathcal J]$ comes from the sparsity of the network. For example, for normally distributed non-zero weights with variance $J_0^2/N^{2a}$, the total probability for every connection in the network is
$$\rho_{ER \times J}(\mathcal J) = (1-p) \delta(\mathcal J) + p \frac{\exp\left(-\frac{N^{2a}}{2}\frac{\mathcal J^2}{J_0^2}\right)}{\sqrt{2\pi J_0^2/N^{2a}}}.$$

Because the $\mathcal J_{i,j}$ are \emph{i.i.d.}, the mean of $\mathcal J^{\rm eff}_{r,r'}$:
\begin{align*}
\overline{\mathcal J^{\rm eff}_{r,r'}} &= \overline{\mathcal J_{r,r'}} + \sum_{h,h'} \overline{\mathcal J_{r,h} \hat{\Gamma} _{h,h'} \mathcal J_{h',r'}} \\
&= 0  + \sum_{h,h'} \overline{\mathcal J_{r,h}} ~ \overline{ \hat{\Gamma}_{h,h'}} ~ \overline{\mathcal J_{h',r'}}\\
&= 0,
\end{align*}
where we used the fact that $\hat{\Gamma}_{h,h'} \equiv \hat{\Gamma}_{h,h'}(0)$ depends only on the hidden neuron couplings $\mathcal J_{h,h'}$, which are independent of the couplings to the recorded neurons, $\mathcal J_{r,h}$ and $\mathcal J_{h',r'}$. This holds for any pair of neurons $(r,r')$, including $r =r '$ because of the assumption of no self-coupling. 

The variance of $\mathcal J^{\rm eff}_{r,r'}$ is thus equal to the mean of its square, for $r \neq r'$,
\begin{align*}
\mbox{var}\left[\mathcal J^{\rm eff}_{r,r'}\right] &= \overline{\left(\mathcal J^{\rm eff}_{r,r'} \right)^2} \\
&= \overline{\left(\mathcal J_{r,r'}\right)^2} + \overline{\left(\sum_{h,h'} \mathcal J_{r,h} \hat{\Gamma}_{h,h'} \mathcal J_{h',r'} \right)^2}\\
&= \mbox{var}[\mathcal J] + \sum_{h_1,h_1',h_2,h_2'} \overline{\mathcal J_{r,h_1} \hat{\Gamma}_{h_1,h_1'} \mathcal J_{h_1',r'}\mathcal J_{r,h_2} \Gamma_{h_2,h_2'} \mathcal J_{h_2',r'}} \\
&= \mbox{var}[\mathcal J]  + \sum_{h,h'} \overline{\mathcal J_{r,h}^2} ~  \overline{\hat{\Gamma}_{h,h'}^2} ~ \overline{\mathcal J_{h',r'}^2} \\
&= \mbox{var}[\mathcal J]  + \mbox{var}[\mathcal J]^2 \sum_{h,h'} \overline{\hat{\Gamma}_{h,h'}^2}
\end{align*}
In this derivation, we used the fact that $\overline{\mathcal J_{r,h_1} \mathcal J_{r,h_2}} = \overline{\mathcal J^2_{r,h_1}} \delta_{h_1,h_2}$ due to the fact that the synaptic weights are uncorrelated. We now need to compute $\overline{\hat{\Gamma}^2_{h,h'}}$. This is intractable in general, so we will resort to calculating this in a series expansion in powers of $\epsilon \equiv \exp(\mu_0)$ for the exponential nonlinearity model. Our result will also turn out to be an expansion in powers of $J_0$ and $1-f \equiv N_{\rm hid}/N$. 

The lowest order approximation is obtained by the approximation $\nu_h \approx \lambda_0 \epsilon$ and $\Gamma_{h,h'} \approx \nu_h \delta_{h,h'}$, yielding
\begin{align}
\frac{\mbox{var}\left[\mathcal J^{\rm eff}_{r,r'}\right]}{\mbox{var}[\mathcal J]} &= 1 + (\lambda_0 \epsilon)^2 N_{\rm hid} \mbox{var}[\mathcal J]\nonumber \\
&= 1 + (\lambda_0J_0 \epsilon)^2 (1-f)  \frac{1}{(pN)^{2a-1}}.
\label{eqn:varJeff_1storder}
\end{align}
This result varies linearly with $f$, while numerical evaluation of the variance shows obvious curvature for $f \ll 1$ and $J_0 \lesssim 1$, so we need to go to higher order. This becomes tedious very quickly, so we will only work to $\mathcal O(\epsilon^4)$ (it turns out $\mathcal O(\epsilon^3)$ corrections vanish).

We calculate $\overline{\hat{\Gamma}^2_{h,h'}}$ using a recursive strategy, though we could also use the path-length series expression for $\hat{\Gamma}_{h,h'}(\omega)$, keeping terms up to fourth order in $\epsilon$. We begin with the expression
$$\hat{\Gamma}_{h,h'} = \nu_h \delta_{h,h'} + \sum_{h''} \nu_h \mathcal J_{h,h''} \hat{\Gamma}_{h'',h'}$$
and plug it into itself until we obtain an expression to a desired order in $\epsilon$. In doing so, we note that $\nu_h \sim \mathcal O(\epsilon)$, so we will first work to fourth order in $\nu_h$, and then plug in the series for $\nu_h$ in powers of $\epsilon$.

We begin with
\begin{align*}
\hat{\Gamma}^2_{h,h'} &= \nu^2_h \delta_{h,h'} + 2\delta_{h,h'} \sum_{h''} \nu^2_h \mathcal J_{h,h''} \hat{\Gamma}_{h'',h'} + \left( \sum_{h''} \nu_h \mathcal J_{h,h''} \hat{\Gamma}_{h'',h'}\right)^2 \\
&= \nu^2_h \delta_{h,h'} + 2\delta_{h,h'} \sum_{h''} \nu^2_h \mathcal J_{h,h''} \hat{\Gamma}_{h'',h'} + \sum_{h_1,h_2} \nu^2_h \mathcal J_{h,h_1}\mathcal J_{h,h_2}  \hat{\Gamma}_{h_1,h'} \hat{\Gamma}_{h_2,h'} \\
&\approx \nu^2_h \delta_{h,h'} + 2\delta_{h,h'} \sum_{h''} \nu^2_h \mathcal J_{h,h''}\left\{\nu_{h''} \delta_{h'',h'} + \sum_{h_2} \nu_{h''} \mathcal J_{h'',h_2} \nu_{h_2} \delta_{h_2,h'} \right\} \\
& \hspace{2.5cm}+ \sum_{h_1,h_2} \nu^2_h \nu^2_{h'}\mathcal J_{h,h_1}\mathcal J_{h,h_2}   \delta_{h_1,h'} \delta_{h_2,h'}\\
&= \nu^2_h \delta_{h,h'} + 2\delta_{h,h'}  \left\{\nu^2_h \nu_{h'}\mathcal J_{h,h'} + \sum_{h''} \nu^2_h \nu_{h''} \mathcal J_{h,h''}\mathcal J_{h'',h'} \nu_{h'}  \right\} + \nu^2_h \nu^2_{h'}\mathcal J^2_{h,h'}\\
&= \left\{\nu^2_h  + 2\nu^2_h \nu_{h'}\mathcal J_{h,h'} + 2\sum_{h''} \nu^2_h\nu_{h''} \mathcal J_{h,h''} \mathcal J_{h'',h'} \nu_{h'}  \right\}\delta_{h,h'} + \nu^2_h \nu^2_{h'}\mathcal J^2_{h,h'}\\
&= \left\{\nu^2_h  + 2\sum_{h''} \nu^3_h\nu_{h''} \mathcal J_{h,h''} \mathcal J_{h'',h}  \right\}\delta_{h,h'} + \nu^2_h \nu^2_{h'}\mathcal J^2_{h,h'}
\end{align*}
The third order term $\nu_h^3\mathcal J_{h,h'} \delta_{h,h'}$ vanished because we assume no self-couplings. We have obtained $\hat{\Gamma}^2_{h,h'}$ to fourth order in $\nu_h$; now we need to plug in the series expression for $\nu_h$ to obtain the series in powers of $\lambda_0 \epsilon$. We will do this order by order in $\nu_h$. The easiest terms are the fourth order terms, as
$$\nu_h^2 \nu_{h'}^2 \approx (\lambda_0 \epsilon)^4~\mbox{and}~\nu_h^3 \nu_{h''} \approx (\lambda_0 \epsilon)^4.$$
The second order term is
\begin{align*}
\nu^2_h &\approx (\lambda_0\epsilon)^2 \left(1 + \sum_{h_1} \mathcal J_{h,h_1} \lambda_0 \epsilon + \sum_{h_1,h_2} a^{(2)}_{h,h_1,h_2} (\lambda_0 \epsilon)^2 \right) \\
& \hspace{2.0cm} \times  \left(1 + \sum_{h'_1} \mathcal J_{h,h'_1} \lambda_0 \epsilon + \sum_{h'_1,h'_2} a^{(2)}_{h,h'_1,h'_2} (\lambda_0 \epsilon)^2 \right)\\
&\approx (\lambda_0\epsilon)^2\left(1 + 2\left( \sum_{h_1} \mathcal J_{h,h_1} \lambda_0 \epsilon + \sum_{h_1,h_2} a^{(2)}_{h,h_1,h_2} (\lambda_0 \epsilon)^2\right) + \left(\sum_{h_1} \mathcal J_{h,h_1} \lambda_0 \epsilon \right)^2 \right)\\
&= (\lambda_0\epsilon)^2\left(1 + 2 \sum_{h_1} \mathcal J_{h,h_1} \lambda_0 \epsilon + \sum_{h_1,h_2} \left\{ 2a^{(2)}_{h,h_1,h_2} + \mathcal J_{h,h_1} \mathcal J_{h,h_2}\right\} (\lambda_0 \epsilon)^2\right),
\end{align*}
where $a^{(2)}_{h,h_1,h_2} = \mathcal J_{h,h_1} \mathcal J_{h_1,h_2} + \frac{1}{2} \mathcal J_{h,h_1} \mathcal J_{h,h_2}$. We need the average $\overline{\nu_h^2}$. The third-order term will vanish upon averaging, and 
$$\overline{2a^{(2)}_{h,h_1,h_2} + \mathcal J_{h,h_1} \mathcal J_{h,h_2}} = \overline{2\mathcal J_{h,h_1} \mathcal J_{h_1,h_2} + 2 \mathcal J_{h,h_1} \mathcal J_{h,h_2}}  = 2\mbox{var}[\mathcal J]\delta_{h_1,h_2}(1-\delta_{h,h_1}),$$
using the fact that synaptic weights are independent (giving the $\delta_{h_1,h_2}$ factor) and self-couplings are zero (giving the $1 - \delta_{h,h_1}$ factor). We thus obtain
$$\overline{\nu^2_h} = (\lambda_0\epsilon)^2+ 2(\lambda_0\epsilon)^4 (N_{\rm hid}-1) \mbox{var}[\mathcal J].$$

The first fourth order term in $\hat{\Gamma}^2_{h,h'}$, $2\sum_{h''} \nu^3_h\nu_{h''} \mathcal J_{h,h''} \mathcal J_{h'',h} \delta_{h,h'}$, will vanish upon averaging because matching indices requires $h'' = h = h'$ and we assume no self-couplings. The second fourth order term is $\mathcal J^2_{h,h'}$, which averages to $\mbox{var}[\mathcal J](1-\delta_{h,h'})$, where the factor of $(1-\delta_{h,h'})$ again accounts for the fact that this term does not contribute when $h = h'$ due to no self-couplings.
We thus arrive at
\begin{align*}
\overline{\hat{\Gamma}^2_{h,h'}} &= \left((\lambda_0\epsilon)^2+ 2(\lambda_0\epsilon)^4 (N_{\rm hid}-1) \mbox{var}[\mathcal J] \right) \delta_{h,h'} + (\lambda_0 \epsilon)^4 \mbox{var}[\mathcal J](1-\delta_{h,h'})\\
&= \left((\lambda_0\epsilon)^2+ (\lambda_0\epsilon)^4 (2N_{\rm hid}-3) \mbox{var}[\mathcal J]\right) \delta_{h,h'} + (\lambda_0 \epsilon)^4 \mbox{var}[\mathcal J];
\end{align*}
Putting everything together,
\begin{align*}
\frac{\mbox{var}\left[\mathcal J^{\rm eff}_{r,r'}\right]}{\mbox{var}[\mathcal J]} &= 1 + \mbox{var}[\mathcal J]\sum_{h,h'}\overline{\hat{\Gamma}^2_{h,h'}} \\
&= 1 + \mbox{var}[\mathcal J]\left[\sum_h \left\{(\lambda_0\epsilon)^2+ (\lambda_0\epsilon)^4 (2N_{\rm hid}-3) \mbox{var}[\mathcal J] \right\} + \sum_{h,h'} (\lambda_0 \epsilon)^4 \mbox{var}[\mathcal J] \right]\\
&= 1 + \mbox{var}[\mathcal J]\left[N_{\rm hid} \left\{(\lambda_0\epsilon)^2+ (\lambda_0\epsilon)^4 (2N_{\rm hid}-3) \mbox{var}[\mathcal J] \right\} + N_{\rm hid}^2 (\lambda_0 \epsilon)^4 \mbox{var}[\mathcal J] \right] \\
&= 1 + N_{\rm hid} \mbox{var}[\mathcal J]\left[(\lambda_0\epsilon)^2+ (\lambda_0\epsilon)^4 \left(2N_{\rm hid} - 3\right) \mbox{var}[\mathcal J]  + N_{\rm hid} (\lambda_0 \epsilon)^4 \mbox{var}[\mathcal J] \right]\\
&= 1 + N_{\rm hid} \mbox{var}[\mathcal J]\left[(\lambda_0\epsilon)^2+ (\lambda_0\epsilon)^4 \left(3 - \frac{3}{N_{\rm hid}}\right) N_{\rm hid}\mbox{var}[\mathcal J]\right]
\end{align*}
For weak coupling, this tends to $1$ in the $N \gg 1$ limit, as $N_{\rm hid}\mbox{var}[\mathcal J] = (1-f)J_0^2/N \rightarrow 0$, for fixed fraction of observed neurons $f = N_{\rm rec}/N$. For strong coupling, $N_{\rm hid} \mbox{var}[\mathcal J] = (1-f) J_0^2$, which is constant as $N \rightarrow \infty$, and hence
\begin{equation}
\frac{\mbox{var}\left[\mathcal J^{\rm eff}_{r,r'}\right]}{\mbox{var}[\mathcal J]} = 1 + (\lambda_0 J_0 \epsilon)^2 (1-f) + 3 (\lambda_0 J_0 \epsilon)^4 (1-f)^2 + o\left((\lambda_0 J_0 \epsilon)^4 (1-f)^2\right),
\label{eqn:varJeff_2ndorder}
\end{equation}
where we have used little-$o$ notation to denote that there are higher order terms dominated by $(\lambda_0 J_0 \epsilon)^4 (1-f)^2$. With this expression, we have improved on our approximation of the relative variance of the effective coupling to the true coupling; however, the neglected higher order terms still become significant as $f \rightarrow 0$ and $J_0 \rightarrow 1$, indicating that hidden paths have a significant impact when synaptic strengths are moderately strong and only a small fraction of the neurons have been observed.

Because the synaptic weights $\mathcal J_{i,j}$ are independent, we may rewrite Eq.~(\ref{eqn:varJeff_2ndorder}) as
$$\frac{\mbox{var}\left[\mathcal J^{\rm eff}_{r,r'} - \mathcal J_{r,r'}\right]}{\mbox{var}[\mathcal J]} \approx (\lambda_0 J_0 \epsilon)^2 (1-f) + 3(\lambda_0 J_0 \epsilon)^4 (1-f)^2;$$
or, in terms of the ratio of standard deviations,
$$\frac{\sigma\left[\mathcal J^{\rm eff}_{r,r'} - \mathcal J_{r,r'}\right]}{\sigma[\mathcal J]} \approx (\lambda_0 J_0 \epsilon) \sqrt{1-f}\left(1 + \frac{3}{2}(\lambda_0 J_0 \epsilon)^2 (1-f)\right),$$
where we used the approximation $\sqrt{1+x} \approx 1 + x/2$ for $x$ small.

\begin{figure}[h!]
 \centering
 \includegraphics[width=0.55\textwidth]{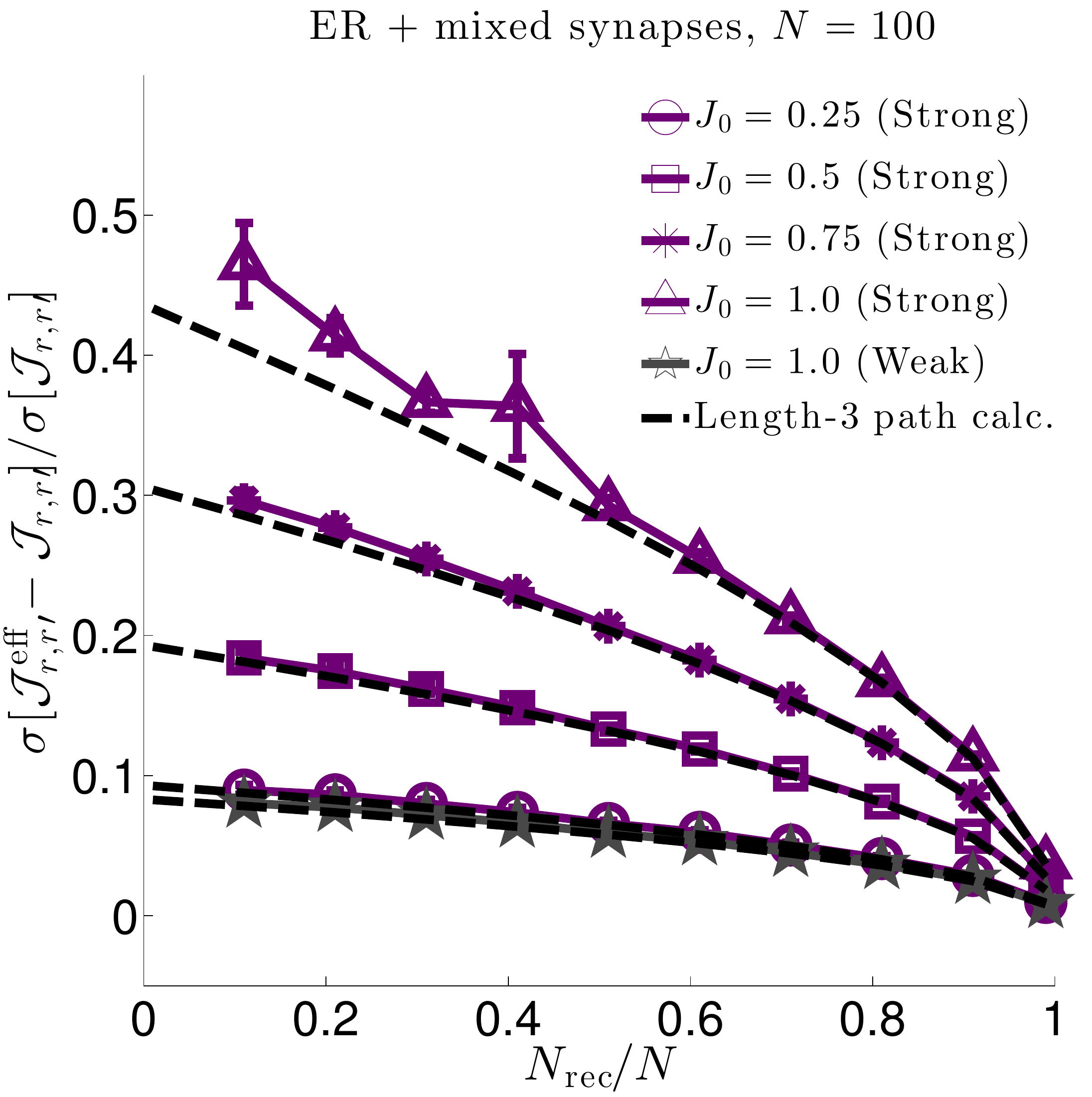}
  \caption{Same as Fig.~\ref{fig:Janalysis}, but for $N = 100$ neurons and $N_{\rm rec} = \{1, 11, 21, 31, 41, 51, 61, 71, 81, 91, 99\}$ recorded neurons. Because we plot the relative deviations of the coupling strength against the fraction of observed neurons, the curves for the strongly coupled case are the same as for $N=1000$, as expected, while the weakly coupled case yields stronger deviations.}
  \label{fig:JanalysisN100}
\end{figure}

In the main text, we plotted results for $N=1000$ total neurons (Fig.~\ref{fig:Janalysis}A). For strongly coupled networks, the results should only depend on the fraction of observed neurons, $f = N_{\rm rec}/N$, while for weak coupling the results do depend on the absolute number $N$. To demonstrate this, in Fig.~\ref{fig:JanalysisN100} we remake Fig.~\ref{fig:Janalysis} for $N=100$ neurons. We see that the strongly coupled results have not been significantly altered, whereas the weakly coupled results yield stronger deviations (as the deviations are $\mathcal O(1/\sqrt{N})$).

\newpage
\section*{Acknowledgments}
We thank Gabe Ocker for providing the nonlinear Hawkes network simulation code that we modified to perform the full network simulations in this work, Tyler Kekona for work on an early version of a related project, and Ben Lansdell and Christof Koch for helpful feedback. Support provided by the Sackler Scholar Program in Integrative Biophysics (BAWB), CRCNS grant DMS-1208027 (ESB, FR), NIH grant EY11850 and the Howard Hughes Medical Institute (http://www.hhmi.org)  (FR). This work was partially based on work supported by the Center for Brains, Minds and Machines (CBMM), funded by NSF STC award CCF-1231216 (MAB). MAB and ESB wish to thank the Allen Institute for Brain Science founders, Paul G. Allen and Jody Allen, for their vision, encouragement, and support.

%%%%%%%%%%%%%%%%%%%%%%%%%%%%%%%%
%% SUPPORTING INFO (here temporarily)
%%%%%%%%%%%%%%%%%%%%%%%%%%%%%%%%

\section*{Supporting Information}

\subsection*{Mapping a leaky integrate-and-fire network with stochastic spiking to the nonlinear Hawkes model}

As claimed in the Methods, we now show explicitly how to map a current-based leaky-integrate and fire network model with stochastic spiking rules on to the nonlinear Hawkes model we use in this work. Suppose each neuron's membrane potential obeys the differential equation
\begin{equation}
\tau_m \frac{dV_i}{dt} = -(V_i-\mathcal E_L) + \mathcal E^{\rm syn}_i(t) +\mathcal E^{\rm ext}_i(t),
\label{eqn:currentLIF}
\end{equation}
where $\tau_m$ is the membrane time constant of the neuron, $\mathcal E_L$ is its reversal potential, $\mathcal E^{\rm ext}_i(t)$ is an external current input (converted to a voltage by dividing by the membrane resistance), and
$$\mathcal E^{\rm syn}_i(t) = \sum_{j} \int_{-\infty}^t dt'~\tilde{J}_{ij}(t-t') \dot{n}_j(t')$$
are the synaptic currents flowing into neuron $i$ from presynaptic neurons $j$, where $\dot{n}_j(t')$ is the spike train from presynaptic neuron $j$ at time $t'$ and $\tilde{J}_{ij}(t-t')$ is a spike filter. For notational convenience we also include the self-history coupling $\tilde{J}_{i,i}(t-t')$ in this term, though it has a physiologically different origin, representing refractory effects that reset a neuron's voltage after it spikes, rather than having a hard reset. Similarly, rather than having a hard firing threshold, we assume that neurons spike stochastically with an instantaneous rate
$$\lambda_i(t) = \lambda_0 \phi\left(\frac{V_i(t)-\mathcal E_{\rm th}}{\mathcal E_{\rm s}} \right),$$
where $\lambda_0$ sets a baseline firing rate, $\phi(\cdot) \geq 0$ is a nonlinear function of the membrane voltage, $\mathcal E_{\rm th}$ is a ``soft'' threshold value, $\mathcal E_{\rm s}$ sets the steepness of the nonlinearity, and $V_i(t)$ is the membrane voltage given in Eq.~(\ref{eqn:currentLIF}). We term this the instantaneous firing rate because it is equal to the the trial-averaged spike trains $\dot{n}_i(t)$, conditioned on the inputs to the neuron. The value $\mathcal E_{\rm th}$ is a soft-threshold because while it is likely the neuron will fire when $V_i(t)$ reaches $\mathcal E_{\rm th}$, it is possible the neuron will fire at higher or lower voltage. In this work we assume that the probability that the number of spikes neuron $i$ fires in a small time window $\Delta t$ around time $t$, given its input history, is
$$\dot{n}_i(t) \Delta t \sim \mbox{Poiss}\left[\lambda_i(t) \Delta t \right];$$
however, we could have chosen many other point processes with instantaneous rates $\lambda_i(t)$. Note that while spiking is stochastic, a neuron is guaranteed to fire at times when its instantaneous rate $\lambda_i(t)$ diverges, so there is some sense of deterministic output retained. 

We can formally solve the membrane equation (\ref{eqn:currentLIF}), giving
$$V_i(t) = \mathcal E_L + \int_{-\infty}^t dt''~\frac{e^{-(t-t'')/\tau_m}}{\tau_m} \left[\mathcal E_{\rm ext}(t'') + \sum_j \int_{-\infty}^{t''} dt'~\tilde{J}_{ij}(t''-t') \dot{n}_j(t') \right].$$
We now define 
$$\mu_i = \frac{\mathcal E_L-\mathcal E_{\rm th}}{\mathcal E_{\rm s}},$$
$$\mu_i^{\rm ext}(t) = \frac{\int_{-\infty}^t dt''~\frac{e^{-(t-t'')/\tau_m}}{\tau_m}\mathcal E^{\rm ext}_i(t'')}{\mathcal E_{\rm s}},$$
and 
$$J_{ij}(t-t') = \frac{\frac{e^{-(t-t')/\tau_m}}{\tau_m}\int_0^{t-t'} dy~ \frac{e^{y/\tau_m}}{\tau_m} \tilde{J}_{ij}(y)}{\mathcal E_{\rm s}};$$
we arrive at this last definition by changing integration order 
\begin{align*}
&\int_{-\infty}^t dt''~\frac{e^{-(t-t'')/\tau_m}}{\tau_m} \int_{-\infty}^{t''} dt'~\tilde{J}_{ij}(t''-t') \dot{n}_j(t') \\
= &\int_{-\infty}^t dt'~\left\{  \int_{t'}^{t} dt''~\frac{e^{-(t-t'')/\tau_m}}{\tau_m} \tilde{J}_{ij}(t''-t')\right\} \dot{n}_j(t')
\end{align*} 
and then changing variables to $y = t-t''$.
With these definitions,
$$\lambda_i(t) = \lambda_0 \phi\left(\mu_i + \mu_i^{\rm ext}(t) + \sum_j \int_{-\infty}^t dt'~J_{ij}(t-t') \dot{n}_j(t') \right);$$
i.e., we have shown the the soft-threshold leaky integrate-and-fire model is equivalent to a nonlinear Hawkes model, Eq.~(\ref{eqn:modeldefn}). Because the argument of the rate is now expressed entirely in terms of the spiking of the neurons, and not the membrane voltage, we need only simulate the spiking activity of the network; i.e., we do not need to keep track of the membrane voltages and can simply use Eq.~(\ref{eqn:modeldefn}). 

Lastly, we note that membrane potential dynamics are more appropriately described by changes to a neuron's membrane conductance, rather than current inputs \cite{GerstnerBook}. If we insert conductance-based synaptic inputs, such as
$$\mathcal E^{\rm syn}_i(t) = -(V_i(t)-\mathcal E^S_i)\sum_{j} \int_{-\infty}^t dt'~\tilde{J}^{\rm cond}_{ij}(t-t') \dot{n}_j(t'),$$
into Eq.~(\ref{eqn:currentLIF}), the voltage equation is still formally solvable, but the rates $\lambda_i(t)$ will no longer be of the form of Eq.~(\ref{eqn:modeldefn}), except in approximate limits or if special conditions are met \cite{LatimerNIPS2014}. We leave a more detailed investigation of conductance-based models---including those with nonlinear voltage dependence---for future work.

\subsection*{Complete derivation of the contribution of self-cycles to nodes in Fig.~\ref{fig:Jeffeqn}}

In the Methods section of the full text, we heuristically argued that loops from a neuron back to itself in the series expansion of $\hat{\Gamma}_{h,h'}(\omega) = [\mathbb{I}-\mathbf{\hat{V}}(\omega)]^{-1}_{h,h'}$ could be explicitly summed into a factor $1/(1-\gamma_{h}\hat{J}_{h,h}(\omega))$ contributed by each node $h$. This factor can be derived directly; we do so here. 

Let us decompose the matrix $\mathbf{\hat{V}}(\omega)$ in a diagonal and off-diagonal piece, $\mathbf{\hat{V}}(\omega) = \mathbf{\hat{V}}_{\rm diag}(\omega) + \mathbf{\hat{V}}_{\rm off}(\omega)$. Then,
\begin{align*}
\left[\mathbb{I} - \mathbf{\hat{V}}(\omega) \right]^{-1} &= \left[\mathbb{I} - \mathbf{\hat{V}}_{\rm diag}(\omega) - \mathbf{\hat{V}}_{\rm off}(\omega) \right]^{-1}\\
&= \left[\left(\mathbb{I} - \mathbf{\hat{V}}_{\rm diag}(\omega)\right)\left(\mathbb{I} - \left(\mathbb{I} - \mathbf{\hat{V}}_{\rm diag}(\omega)\right)^{-1}\mathbf{\hat{V}}_{\rm off}(\omega)\right) \right]^{-1}\\
&= \left[\mathbb{I} - \left(\mathbb{I} - \mathbf{\hat{V}}_{\rm diag}(\omega)\right)^{-1}\mathbf{\hat{V}}_{\rm off}(\omega) \right]^{-1}\left[\mathbb{I} - \mathbf{\hat{V}}_{\rm diag}(\omega)\right]^{-1}
\end{align*}
We assumed that $\mathbb{I} - \mathbf{\hat{V}}_{\rm diag}(\omega)$ is invertible, which requires that there is no element for which $1 - \gamma_h \hat{J}_{h,h}(\omega) = 0$ for all $\omega$. Assuming this is the case, the inverse of the matrix is trivial to calculate, as it is diagonal:
$$\left[\mathbb{I} - \mathbf{\hat{V}}_{\rm diag}(\omega) \right]^{-1}_{h,h'} = \frac{1}{1-\gamma_h J_{h,h}(\omega)}\delta_{h,h'}.$$
The matrix $\left[\mathbb{I} - \left(\mathbb{I} - \mathbf{\hat{V}}_{\rm diag}(\omega)\right)^{-1}\mathbf{\hat{V}}_{\rm off}(\omega) \right]^{-1}$ can be expressed as a series, as before:
\begin{align*}
& \left[\mathbb{I} - \left(\mathbb{I} - \mathbf{\hat{V}}_{\rm diag}(\omega)\right)^{-1}\mathbf{\hat{V}}_{\rm off}(\omega) \right]^{-1}_{h,h''}\\
 = & \sum_{\ell = 0}  \left[\left[\left(\mathbb{I} - \mathbf{\hat{V}}_{\rm diag}(\omega)\right)^{-1}\mathbf{\hat{V}}_{\rm off}(\omega)\right]^\ell\right]_{h,h''}\\
= & \sum_{\ell = 0} \sum_{h_1,\dots,h_\ell} \left[\left(\mathbb{I} - \mathbf{\hat{V}}_{\rm diag}(\omega)\right)^{-1}\mathbf{\hat{V}}_{\rm off}(\omega)\right]_{h,h_1} \dots\left[\left(\mathbb{I} - \mathbf{\hat{V}}_{\rm diag}(\omega)\right)^{-1}\mathbf{\hat{V}}_{\rm off}(\omega)\right]_{h_\ell,h''} \\
= & \sum_{\ell = 0} \sum_{h_1, \dots, h_\ell; h_i \neq h_{i+1}} \frac{\gamma_{h}}{1-\gamma_{h} \hat{J}_{h,h}(\omega)}\hat{J}_{h,h_1}(\omega) \dots \frac{\gamma_{h_\ell}}{1-\gamma_{h_\ell} \hat{J}_{h_\ell,h_\ell}(\omega)}J_{h_\ell,h''}(\omega)
\end{align*}
Hence, inserting the contribution from the factor $\left[\mathbb{I} - \mathbf{\hat{V}}_{\rm diag}(\omega) \right]^{-1}$ that we pulled out, and the factor $\gamma_{h'}$ that left-multiplies $\left[\mathbb{I} - \mathbf{\hat{V}}(\omega)\right]^{-1}$ to give $\hat{\Gamma}_{h,h'}(\omega)$, we have
\begin{align*}
\hat{\Gamma}_{h,h'}(\omega) &= \sum_{\ell = 0} \sum_{h_1, \dots , h_\ell; h_{i} \neq h_{i+1}} \frac{\gamma_{h}\hat{J}_{h,h_1}(\omega)}{1-\gamma_{h}\hat{J}_{h,h}(\omega)} \dots \frac{\gamma_{h_\ell}\hat{J}_{h_\ell,h'}(\omega)}{1-\gamma_{h_\ell} \hat{J}_{h_\ell,h_\ell}(\omega)} \frac{\gamma_{h'}}{1-\gamma_{h'}\hat{J}_{h,h'}(\omega)}
\end{align*}
This is the same as our previous expression, with $\gamma_h \rightarrow \gamma_h/(1-\gamma_h \hat{J}_{h,h}(\omega))$ and restricting the sum over hidden units such that self-loops are removed ($h_i \neq h_{i+1}$), proving the result described informally above. We note again that this puts restrictions on the allowed size of self-interactions, as the zeros of $1 - \gamma_h \hat{J}_{h,h}(\omega)$ must be in the upper-half plane of the complex $\omega$ plane in order for the filters to be causal and physically meaningful (given our Fourier sign-convention $\hat{f}(\omega) = \int_{-\infty}^\infty dt~e^{-i\omega t} f(t)$).

The complete expression for the correction term $\sum_{h,h'} \hat{J}_{r,h}(\omega) \hat{\Gamma}_{h,h'}(\omega) \hat{J}_{h',r'}(\omega)$ is thus
\begin{align*}
&\sum_{h,h'} \hat{J}_{r,h}(\omega) \hat{\Gamma}_{h,h'}(\omega) \hat{J}_{h',r'}(\omega) = \\
&\sum_{\ell = 0} \sum_{h, h_1, \dots , h_\ell, h'; h_{i} \neq h_{i+1}} \hat{J}_{r,h}(\omega)\frac{\gamma_{h}\hat{J}_{h,h_1}(\omega)}{1-\gamma_{h}\hat{J}_{h,h}(\omega)} \dots \frac{\gamma_{h_\ell}\hat{J}_{h_\ell,h'}(\omega)}{1-\gamma_{h_\ell} \hat{J}_{h_\ell,h_\ell}(\omega)} \frac{\gamma_{h'}}{1-\gamma_{h'}\hat{J}_{h,h'}(\omega)}\hat{J}_{h',r'}(\omega).
\end{align*}
This is the exact mathematical expression underlying the graphical rules given in Fig.~\ref{fig:Jeffeqn}. 

\subsection*{Second order nonlinear response function}
\label{sec:secondnonlinresp}

Higher order terms in the series expansion represent nonlinear response functions. We do not focus on these terms in this work, assuming instead that we can truncate this series expansion at linear order. We will, however, estimate the error incurred by this truncation by calculating the second order response function, which we label $\Gamma^{(2)}_{h,h_1,h_2}(t,t_1,t_2)$. Rather than differentiate our formal solution for the linear response, we differentiate the implicit form, yielding an integral equation
\begin{align*}
\Gamma^{(2)}_{h,h_1,h_2}(t,t_1,t_2) &\equiv \left.\frac{\delta^2 \mathbb{E}[ \dot{n}_{h}| \left\{I_h(t)\right\}]}{\delta I_{h_2}(t_2) \delta I_{h_1}(t_1)}\right|_{I_h = 0} \\
&= \gamma^{(2)}_h \left[ \delta_{h,h_2} \delta(t-t_2) + \sum_{h'} J_{h,h'} \ast \Gamma_{h',h_2} \right] \left[ \delta_{h-h_1} \delta(t-t_1) + \sum_{h'} J_{h,h'} \ast \Gamma_{h',h_1} \right] \\
& ~~~~~~~~ + \gamma_h \left[\sum_{h'} \int_{-\infty}^\infty dt'~J_{h,h'}(t-t') \Gamma^{(2)}_{h,h_1,h_2}(t',t_1,t_2) \right]
\end{align*} 
where we have defined
$$\gamma^{(2)}_h \equiv \lambda_0 \phi''\left(\mu_h + \sum_{h'} \mathcal J_{h,h'} \nu_{h'} \right).$$
$\gamma_h$ without the superscript is the gain defined previously,  $\gamma_h = \lambda_0 \phi'(\mu_h + \sum_{h'} \mathcal J_{h,h'} \nu_{h'})$.
Rearranging,
\begin{align*}
&\int dt' \sum_{h'} \Bigg[\delta_{h,h'} \delta(t-t') - \gamma_{h} J_{h,h'}(t-t') \Bigg]\Gamma^{(2)}_{h',h_1,h_2}(t',t_1,t_2)\\
 &~~~~~~~~~~~~= \gamma^{(2)}_h \left[ \delta_{h-h_2} \delta(t-t_2) + \sum_{h'} J_{h,h'} \ast \Gamma_{h',h_2} \right] \left[ \delta_{h-h_1} \delta(t-t_1) + \sum_{h'} J_{h,h'} \ast \Gamma_{h',h_1} \right].
\end{align*} 
Inverting the operator on the left hand side yields the input linear response function (when introducing the factor of $1=\gamma_{h'}/\gamma_{h'}$ on the right hand side), giving the solution
\begin{align*}
\Gamma^{(2)}_{h,h_1,h_2}(t,t_1,t_2) &= \int_{-\infty}^\infty dt' \sum_{h'} \Gamma_{h,h'}(t-t') \frac{\gamma^{(2)}_{h'}}{\gamma_{h'}} \left[ \delta_{h',h_2} \delta(t'-t_2) + \sum_{h''} \int_{-\infty}^\infty dt''~J_{h',h''}(t'-t'') \Gamma_{h'',h_2}(t''-t_2) \right] \\
& \hspace{4.5cm} \times  \left[ \delta_{h',h_1} \delta(t'-t_1) + \sum_{h''} \int_{-\infty}^\infty dt''~J_{h',h''}(t'-t'') \Gamma_{h'',h_1}(t''-t_1)\right]
\end{align*}
Because $\Gamma_{h,h'}(t-t')$ is proportional to $\gamma_h$, the second order nonlinear response function is proportional to $\gamma^{(2)}_h$. For an exponential nonlinearity, $\gamma^{(2)}_h = \gamma_h = \nu_h$, and the second order response function is of the same order of the linear response (but the overall contribution to network statistics is not of the same order; see below). For a rectified linear nonlinearity (as in Figs.~\ref{fig:ffwi_3neuron} and \ref{fig:ffwi_4neuron}), $\gamma^{(2)}_h = 0$ and the second-order response vanishes.

The effective quadratic interaction from two recorded neurons $r_1'$ and $r_2'$ to neuron $r$ is thus
$$\int dt'_1 dt'_2 \sum_{r_1',r_2'} J^{(2)}_{r,r'_1,r'_2}(t,t'_1,t'_2)\dot{n}_{r_1'}(t'_1) \dot{n}_{r_2'}(t'_2),$$
where we have defined the quadratic spike interaction $J^{(2)}_{r,r'_1,r'_2}(t,t'_1,t'_2)$ to be
\begin{equation}
J^{(2)}_{r,r'_1,r'_2}(t,t'_1,t'_2) = \int dt' dt_1 dt_2 \sum_{h,h_1,h_2} J_{r,h}(t-t')\Gamma^{(2)}_{h,h_1,h_2}(t',t_1,t_2) J_{h_1,r_1'}(t_1-t'_1) J_{h_2,r_2'}(t_2-t'_2)
\label{eqn:quadraticspikefilter}
\end{equation}

\subsection*{Estimating the error from neglecting higher order spike filtering (exponential nonlinearity)}
\label{sec:higherspikeerror}

In the main text we calculate corrections to the baseline and linear spike filters, neglecting higher-order spike filtering and fluctuations around the mean input to the recorded neurons. In the methods we validated this result numerically; here we derive an analytic estimate of the order of the error we incur by neglecting these terms. We will do so within mean field theory (meaning the noise fluctuations contribute zero error as they do not contribute to the mean field approximation). Specifically, we will assume that the quadratic spike filtering term is small, and calculate the corresponding correction to our mean field approximation of the firing rates when this term is completely neglected. If we take as our approximation of the recorded neuron firing instantaneous firing rates
\begin{align*}
\lambda_r(t) &\approx \lambda_0 \exp\Bigg(\mu^{\rm eff}_{r} + \sum_{r_1}\int dt_1 J^{\rm eff}_{r,r_1}(t-t_1)\dot{n}_{r_1}(t_1) \\
& \hspace{4.0cm} + b\sum_{r_1,r_2} \int dt_1 dt_2~J^{(2)}_{r,r_1,r_2}(t,t_1,t_2) \dot{n}_{r_1}(t_1) \dot{n}_{r_2}(t_2) \Bigg),
\end{align*}
then the mean field approximation of the firing rates is
$$\langle \dot{n}_r \rangle \approx \lambda_0 \exp\left(\mu^{\rm eff}_{r} + \sum_{r_1}\mathcal J^{\rm eff}_{r,r'} \langle\dot{n}_{r_1}\rangle + b\sum_{r_1,r_2} \mathcal J^{(2)}_{r,r_1,r_2}\langle \dot{n}_{r_1} \rangle\langle \dot{n}_{r_2}\rangle \right),$$
where we have used the fact that the average firing rates are independent of time, and replaced $J^{\rm eff}_{r,r'}(t-t_1)$ and $J^{(2)}_{r,r_1,r_2}(t,t_1,t_2)$ with their time integrals, denoted by $\mathcal J^{\rm eff}_{r,r'}$ and $\mathcal J^{(2)}_{r,r_1,r_2}$. The parameter $b$ is just a book-keeping parameter.

To calculate the lowest order correction to the linear filtering approximation ($b \rightarrow 0$), we write $\langle \dot{n}_r\rangle = \nu^{\rm sub}_r + b \tilde{\nu}_r$, treating $b$ formally as a small parameter. The linear firing rate $\nu^{\rm sub}_r$ is given by
$$\nu^{\rm sub}_r = \lambda_0 \exp\left(\mu^{\rm eff}_r + \sum_{r'} \mathcal J^{\rm eff}_{r,r'} \nu^{\rm sub}_{r'} \right).$$
For the quadratically-modified firing rates, keeping terms only to linear order in $b$,
\begin{align*}
\nu^{\rm sub}_r + b \tilde{\nu}_r &\approx \lambda_0  \exp\left(\mu^{\rm eff}_{r} + \sum_{r_1}\mathcal J^{\rm eff}_{r,r'}\nu^{\rm sub}_{r'} + b\sum_{r'} \mathcal J^{\rm eff}_{r,r'} \tilde{\nu}_{r'} + b\sum_{r_1,r_2}\mathcal J^{(2)}_{r,r_1,r_2}\nu^{\rm sub}_{r_1}\nu^{\rm sub}_{r_2}\right) \\
& = \nu^{\rm sub}_r \exp\left( b\sum_{r'} \mathcal J^{\rm eff}_{r,r'} \tilde{\nu}_{r'} + b\sum_{r_1,r_2}\mathcal J^{(2)}_{r,r_1,r_2}\nu^{\rm sub}_{r_1}\nu^{\rm sub}_{r_2}\right) \\
&\approx \nu^{\rm sub}_r\left\{1 + b\sum_{r'} \mathcal J^{\rm eff}_{r,r'} \tilde{\nu}_{r'} + b\sum_{r_1,r_2}\mathcal J^{(2)}_{r,r_1,r_2}\nu^{\rm sub}_{r_1}\nu^{\rm sub}_{r_2} \right\}.
\end{align*}
Collecting on $b$ and rearranging,
$$\sum_{r'} \left[ \delta_{r,r'} - \nu^{\rm sub}_r \mathcal J^{\rm eff}_{r,r'} \right] \tilde{\nu}_{r'} = \nu^{\rm sub}_r\sum_{r_1,r_2}\mathcal J^{(2)}_{r,r_1,r_2}\nu^{\rm sub}_{r_1}\nu^{\rm sub}_{r_2}.$$
Because $\nu^{\rm sub}_r \propto \exp(\mu_r^{\rm eff}) \propto \exp(\mu_r) = \epsilon_r$, the expansion parameters we have been using, the lowest order approximation for $\tilde{\nu}_r$ is
$$\tilde{\nu}_r \approx \nu^{\rm sub}_r\sum_{r_1,r_2}\mathcal J^{(2)}_{r,r_1,r_2}\nu^{\rm sub}_{r_1}\nu^{\rm sub}_{r_2}.$$
To evaluate the coefficient $\mathcal J^{(2)}_{r,r_1,r_2}$, we may use the fact $\gamma_h = \nu_h$ and to leading order $\nu_h \sim \lambda_0 \epsilon$ and $\Gamma^{(2)}_{h,h_1,h_2}(t,t_1,t_2) \approx \lambda_0 \epsilon \delta_{h,h_1} \delta_{h,h_2} \delta(t-t_1) \delta(t-t_2)$, giving
$$J^{(2)}_{r,r_1,r_2}(t,t_1,t_2) \approx \lambda_0 \epsilon \int dt' \sum_{h} J_{r,h}(t-t') J_{h,r_1'}(t_1-t') J_{h,r_2'}(t_2-t')$$
and hence
$$\mathcal J^{(2)}_{r,r_1,r_2} \equiv \int dt_1 dt_2~J^{(2)}_{r,r_1,_2}(t,t_1,t_2) \approx \lambda_0 \epsilon \sum_{h} \mathcal J_{r,h} \mathcal J_{h,r_1} \mathcal J_{h,r_2}.$$
To lowest order the error term $\tilde{\nu}_r$ is
$$\tilde{\nu_{r}} = (\lambda_0 \epsilon)^4\sum_{h,r_1,r_2} \mathcal J_{r,h} \mathcal J_{h,r_1} \mathcal J_{h,r_2}.$$
For $\mathcal J_{i,j}$ \emph{i.i.d.}, the population average should converge to the expected value, which is zero because the $\mathcal J_{i,j}$ have mean zero. We can calculate the root-mean-squared-error (RMSE) by looking at the variance:
\begin{align*}
\mbox{var}(\tilde{\nu}_r) = \mbox{var}\left(\sum_{h,r_1,r_2} J_{r,h} J_{h,r_1} J_{h,r_2}\right) &= \overline{ \left(\sum_{h,r_1,r_2} \mathcal J_{r,h} \mathcal J_{h,r_1} \mathcal J_{h,r_2}\right)^2} \\
&=  \sum_{h,r_1,r_2} \overline{ \mathcal J_{r,h}^2}~\overline{ \mathcal J_{h,r_1}^2 \mathcal J_{h,r_2}^2 }
\end{align*}
In principle, we should take care to separate out the $r_1 \neq r_2$ and $r_1 = r_2$ terms from the sum, as the latter will contribute a factor $\overline{\mathcal J_{h,r_1}^4}$, which we have not specified yet (though one could calculate for specific choices, such as the normal distribution that we use for most of our numerical analyses). However, both $\overline{\mathcal J^2_{h,r}}^2$ and $\overline{\mathcal J_{h,r_1}^4}$ will scale as $(J_0^2/(pN)^{2a})^2$, so we will neglect constant factors and simply use this scaling to arrive at the result
$$\mbox{var}(\tilde{\nu}_r) \sim (\lambda_0 \epsilon)^8 N^2_{\rm rec} N_{\rm hid} \frac{J_0^6}{(pN)^{6a}}.$$
If we take $N \rightarrow \infty$ with $N_{\rm rec} = fN$ and $N_{\rm hid} = (1-f)N$ for $f$ fixed, the RMSE scales as
$$\tilde{\nu}_{\rm RMSE} \sim (\lambda_0 \epsilon)^4 f \sqrt{1-f} \frac{J_0^3}{(pN)^{3a-3/2}}.$$
For $a=1$ (weak coupling), the error falls off quite quickly as $N^{3/2}$, while it is independent of $N$ for $a=1/2$ (strong coupling). However, the error does still scale with the fraction of observed neurons, as $f\sqrt{1-f}$. This demonstrates that the typical error that arises from neglecting the nonlinear filtering is small both when most neurons have been observed ($f \lesssim 1$) and when very few neurons have been observed ($f \gtrsim 0$). While it may at first seem surprising that the error is small when very few neurons have been observed, the result does make intuitive sense: when a very small fraction of the network is observed, we can treat the unobserved portion of the network as a ``reservoir'' or ``bath.'' Feedback from the observed neurons into the reservoir has a comparatively small effect, so we can get away with neglecting feedback between the observed and unobserved partitions of the network. However, when the number of observed neurons is comparable to the number of unobserved neurons, neither can be treated as a reservoir, and feedback between the two partitions is substantial. Neglecting the higher order spike filter terms may be inaccurate in this case.

\subsection*{Tree-level calculation of the effective noise correlations}
\label{sec:treelevel}

In our approximation of the model for the recorded neurons, we also neglected fluctuations from the mean input around the hidden neuron input. We should therefore check how strong this noise is. At the level of a mean-field approximation we may neglect it, so we will need to go to a tree-level approximation to calculate it. (The means and response functions are not modified at tree-level.)

The noise is defined by
$$\xi_r(t) = \sum_h \int_{-\infty}^\infty dt'~J_{r,h}(t-t') \Big(\dot{n}_h(t')-\mathbb{E}[ \dot{n}_h| \{\dot{n}_r\} ]\Big).$$
It has zero mean (by construction), conditioned on the activity of the recorded units --- i.e., the ``noise'' receives feedback from the recorded neurons. We can evaluate the cross-correlation function of this noise, conditioned on the recorded unit activity. This is given by
\begin{align*}
\mathbb{E}[ \xi_r(t)\xi_{r'}(t')|\{\dot{n}_r\}]^c = \sum_{h_1,h_2} \int_{-\infty}^\infty dt_1 dt_2 ~ J_{r,h_1}(t-t_1)J_{r',h_2}(t'-t_2) \mathbb{E}[ \dot{n}_{h_1}(t_1) \dot{n}_{h_2}(t_2) |\{\dot{n}_r\} ]^c,
\end{align*}
where $$\mathbb{E}[ \dot{n}_{h_1}(t_1) \dot{n}_{h_2}(t_2) |\{\dot{n}_r\} ]^c = \mathbb{E}[ \dot{n}_{h_1}(t_1) \dot{n}_{h_2}(t_2) |\{\dot{n}_r\} ] - \mathbb{E}[ \dot{n}_{h_1}(t_1)|\{\dot{n}_r\} ] \mathbb{E}[ \dot{n}_{h_2}(t_2) |\{\dot{n}_r\} ]$$ is the cross-correlation function of the spikes (the superscript $c$ denotes `cumulant' or `connected' correlation function to distinguish it from the moments without the superscript). At the level of mean field theory $$\mathbb{E}[ \dot{n}_{h_1}(t_1) \dot{n}_{h_2}(t_2) |\{\dot{n}_r\} ] \approx \mathbb{E}[ \dot{n}_{h_1}(t_1)|\{\dot{n}_r\} ] \mathbb{E}[ \dot{n}_{h_2}(t_2) |\{\dot{n}_r\} ],$$ and thus the cross-correlation function is zero. We can go beyond mean field theory and calculate the tree-level contribution to the correlation functions using the field theory diagrammatic techniques of \cite{OckerPLOSCB2017}. We will do so for the reference state of zero-recorded unit activity, $\{\dot{n}_r\} =\{0\}$, as we expect this to be the leading order contribution to the correlation function. As we are interested primarily in the typical magnitude of the noise compared to the interaction terms, we will work only to leading order in $\epsilon = \exp(\mu_0)$ for the exponential nonlinearity network. We find
\begin{align*}
\mathbb{E}[ \dot{n}_{h_1}(t_1) \dot{n}_{h_2}(t_2) |0]^c_{\rm tree} &= \int_{-\infty}^\infty dt' \sum_{h'} \Delta_{h_1,h'}(t_1-t')\Delta_{h_2,h'}(t_2-t') \nu_{h'} \\
&\approx \lambda_0 \epsilon \delta_{h_1,h_2} \delta(t_1-t_2),
\end{align*}
where $\Delta_{h,h'}(\omega) \approx \delta_{h,h'} + \mathcal O(\epsilon)$ is the linear response to perturbations to the \emph{output} of a neuron's rate. It is related to $\Gamma_{h,h'}(\omega)$ by $\Gamma_{h,h'}(\omega) = \Delta_{h,h'}(\omega) \gamma_{h'}$, where $\gamma_h = \nu_h$ for $\phi(x) = e^x$. 
The overall noise cross-correlation function is then approximately
$$\mathbb{E}[\xi_r(t)\xi_{r'}(t')|0]^c = \lambda_0 \epsilon \sum_h \int_{-\infty}^\infty dt_1~J_{r,h}(t-t_1) J_{r',h}(t'-t_1).$$
If $r \neq r'$, the expected noise cross-correlation, averaged over the synaptic weights $\mathcal J_{i,j}$, is zero. If $r = r'$, the expected value is non-zero. The expected noise auto-correlation function is then
\begin{align*}
\overline{\mathbb{E}[\xi_r(t)\xi_{r}(t')|0]^c} &= \lambda_0 \epsilon N_{\rm hid} \mbox{var}[\mathcal J] \int_{-\infty}^\infty dt_1~g(t-t_1) g(t'-t_1)\\
&= \lambda_0 \epsilon (1-f)J_0^2 \frac{1}{(pN)^{2a-1}} \int_{-\infty}^\infty dt_1~g(t-t_1) g(t'-t_1).
\end{align*}
For the specific case of $g(t) = \alpha^2 t e^{-\alpha t} \Theta(t)$, we have
$$\overline{\mathbb{E}[\xi_r(t)\xi_{r}(t')|0]^c} = \frac{1}{4}\lambda_0 \epsilon (1-f)J_0^2 \frac{1}{(pN)^{2a-1}} \alpha e^{
-\alpha|t-t'|}\Big(1 + \alpha |t-t'| \Big).$$
For weak coupling ($a = 1$), the expected autocorrelation function falls off with network size as $1/N$, while for strong coupling ($a=1/2$), it scales with the fraction of observed neurons $f$, but is independent of the absolute network size. The overall $\lambda_0 \epsilon$ scaling puts the magnitude of the autocorrelation function on par with contributions from hidden paths through a single hidden neuron that contributes a factor of $\lambda_0 \epsilon$ to the correction to the coupling filters. Based on our results shown in Fig.~\ref{fig:Janalysis}, which suggest that contributions from long paths through hidden neurons are significant when the fraction of neurons $f$ is small and $J_0 \lesssim 1$, we expect that network noise will also be significant in these regimes. This will not modify the results presented in the main paper, however. It simply means that this noise should be retained in the rate of our approximate model,
$$\lambda_r(t) \approx \lambda_0 \exp\left(\mu^{\rm eff}_r + \sum_{r'} J^{\rm eff}_{r,r'} \ast \dot{n}_{r'}(t) + \xi_r(t) \right).$$  

\subsection*{Validating the mean field approximation and linear conditional rate approximation via direct simulations of network activity (exponential nonlinearity)}
\label{sec:networksims}

The results presented in the main text are based on analytical calculations or numerical analyses using analytically derived formulas. For example, the statistics of $\mathcal J^{\rm eff}_{r,r'}$ are calculated based on our expression $\mathcal J^{\rm eff}_{r,r'} = \mathcal J_{r,r'} + \sum_{h,h'} \mathcal J_{r,h} \hat{\Gamma}_{h,h'}(0) \mathcal J_{h',r'}$, where $\hat{\Gamma}_{h,h'}(0)$ can be calculated by solving the matrix equation
$$\hat{\Gamma}_{h,h'}(0) = \delta_{h,h'} + \sum_{h''} \nu_h \mathcal J_{h,h''} \hat{\Gamma}_{h'',h'}(0).$$
Generating these results does not require a simulating the full network, so we check here that our approximations indeed agree with the results of full network simulations.

We check the validity of two main results: 1) that mean field theory is an accurate approximation for the parameters we consider, and 2) that our truncation of the conditional hidden firing rates $\mathbb{E}[\dot{n}_h(t)|\{\dot{n}_r\}]$ at linear order in $\dot{n}_r(t)$ is valid.

We first discuss some basic details of the simulation. The simulation code we use is a modification of the code used in \cite{OckerPLOSCB2017}, written by Gabe Ocker; refer to this paper for full details of the simulation.

The main changes we made are considering exponential nonlinearities and synaptic weights drawn from normal or lognormal distributions.

As in \cite{OckerPLOSCB2017} and the main text, we choose the coupling filters to follow an alpha function
$$g_j(t) = \alpha^2 t e^{-\alpha t} \Theta(t),~\forall j.$$
The Heaviside step function $\Theta(t)$ enforces causality of the filter, using the convention $\Theta(0) = 0$. All neurons have the same time constant $1/\alpha$.

To efficiently simulate this network the code computes the synaptic variable $s_j(t) = \int dt' g(t-t') \dot{n}_j(t')$ not by direct convolution but by solving the inhomogeneous system of differential equations, setting $x(t) = s(t)$ and $y(t) = \dot{s}(t)$,
\begin{align*}
\dot{x}_j(t) &= y_j(t)\\
\dot{y}_j(t) &= -2 \alpha_j y_j(t) - \alpha_j^2 x_j(t) + \alpha_j^2 \dot{n}_j(t),
\end{align*}
The instantaneous firing rates of the neurons can in this way be quickly computed in time steps of a specified size $\Delta t$. The number of spikes $n_i$ that neuron $i$ fires in the $t^{\rm th}$ time bin is drawn from a Poisson distribution with probability $(\lambda_i(t)\Delta t)^{n_i} \exp(-\lambda_i(t)\Delta t)/(n_i)!$. An initial transient period of spiking before the network achieves a steady state is discarded.

The parameters we use in our simulations of the full network are given in Table~\ref{tab:simparams}.

\begin{table}[htp]
\caption{Network activity simulation parameter values.}
\begin{center}
\begin{tabular}{|c|c|}
\hline
Network connectivity parameters & See Table~\ref{tab:params}. \\ \hline
Alpha function decay time $\tau \equiv 1/\alpha$ & 10 \\ \hline
Time bin width $\Delta t$ & $0.01\tau$ \\ \hline
Transient time window & $5\tau$ \\ \hline
Simulation stopping time & $4000\tau$ + transient \\ \hline
\end{tabular}
\end{center}
\label{tab:simparams}
\end{table}%

\subsubsection*{Validating the mean field approximation}
\label{sec:verifyingmft}

To confirm that the mean field approximation is valid, we seek to compare the empirically measured spike rates measured from simulations of the network activity to the calculated mean field rates. The empirical rates are measured as
$$\langle \dot{n}_{i} \rangle^{\rm emp} = \frac{\mbox{number of spikes emitted by neuron}~i}{\mbox{length of spike train window}},$$
calculated after discarding the initial transient period of firing, for any neuron $i$ (recorded or hidden). 

\begin{figure}[h!]
 \centering
 \includegraphics[width=\textwidth]{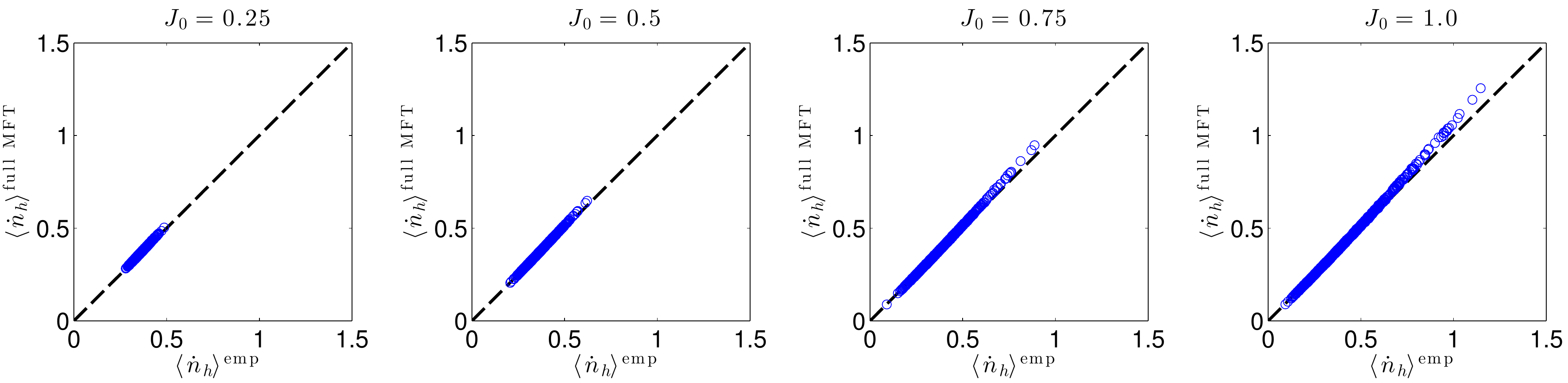}
  \caption{Empirical estimates of average neuron firing rates from simulations plotted against mean firing rates predicted by mean field theory. The fact that the data lies along the identity line demonstrates validity of the mean field theory approximation up to $J_0 = 1.0$.}
  \label{fig:MFTvalidation}
\end{figure}

The steady-state mean field firing rates are the solutions of the transcendental equation
$$\langle \dot{n}_i \rangle^{\rm full~MFT} = \lambda_0 \exp\left(\mu_i + \sum_j \mathcal J_{i,j} \langle \dot{n}_j \rangle^{\rm full~MFT} \right).$$
The only difference between this equation and the equation for $\nu_h$ is that the neuron indices are not restricted to hidden units. i.e., the $\nu_h$ are the mean field rates for the hidden neurons \emph{only} (recorded neurons removed entirely), whereas the $\langle \dot{n}_i \rangle^{\rm full~MFT}$ are the mean field rates for the entire network. If the mean field approximation is valid, the empirical rates should be approximately equal to the mean field rates, so a scatter plot of $\langle \dot{n}_i \rangle^{\rm MFT}$ versus $\langle \dot{n}_i \rangle^{\rm emp}$ should roughly lie along the identity line. We test this for a network in the strong coupling limit ($\sqrt{\mbox{var}(\mathcal J)} = J_0/\sqrt{N}$) for four values of $J_0$, $J_0 = 0.25, 0.5, 0.75$, and $1.0$. We expect $J_0 = 1.0$ to be close to the stability threshold of the model based on a linearized analysis \cite{HawkesBiometrika1971,bremaudAnnProb1996}; i.e., for $J_0 \gtrsim 1.0$ there may not be a steady state, so this may be where we expect the mean field approximation to break down. As seen in Fig.~\ref{fig:MFTvalidation}, the mean field approximation appears to hold well even up to $J_0 = 1.0$, though there are some slight deviations for neurons with large rates.

\subsubsection*{Verifying the linearized conditional mean approximation}

Having verified that the mean field approximation is valid, we now seek to check our linearized approximation of the firing rates of the hidden neurons \emph{conditioned on the activity of the recorded neurons}, $\mathbb{E}\left[\dot{n}_{h}(t) | \left\{ \dot{n}_r(t)\right\}\right]$. That is, we calculated above that
$$\mathbb{E}\left[\dot{n}_{h}(t) | \left\{ \dot{n}_r(t)\right\}\right] \approx \nu_h + \sum_{h',r} [\Gamma_{h,h'} \ast J_{h',r} \ast \dot{n}_r](t) + \dots;$$
the $\dots$ correspond to higher order spike filtering terms that we have neglected in our analyses, assuming them to be small. In an earlier calculation above, we estimated that the error incurred by neglecting higher order spike filtering is of the order $(\lambda_0 \exp(\mu_0))^4 f\sqrt{1-f}J_0^3$, but we would like to confirm the negligibility of the higher order coupling through simulations.

To do so, we compare the empirical firing rates of the designated ``hidden'' neurons obtained from simulations of the full network models with the approximation of the firing rates of the hidden neurons conditioned on the recorded neurons using the linear expansion, averaged over recorded neuron activity to give
$$\langle \dot{n}_h \rangle^{\rm approx} \approx \nu_h + \sum_{h',r} \hat{\Gamma}_{h,h'}(0) \mathcal J_{h',r} \langle \dot{n}_r\rangle^{\rm emp},$$
where as usual the zero-frequency component of the linear response function $\hat{\Gamma}_{h,h'}(0)$ of the hidden neurons is calculated in the absence of recorded neurons.

If we make a scatter plot of this against the empirical estimates of the hidden neurons, $\langle \dot{n}_h \rangle^{\rm emp}$, the data points will lie along the identity line if our approximation is valid. If the data deviates from the identity line, it indicates that the neglected higher-order filtering terms contribute substantially to the firing rates of the neurons. It is possible that the zeroth order rate approximation, $\nu_h$, would be sufficient to describe the data, so we compare the empirical rates to both $\nu_h$ and $\langle \dot{n}_h \rangle^{\rm approx}$.

\begin{figure}[h!]
 \centering
 \includegraphics[width=\textwidth]{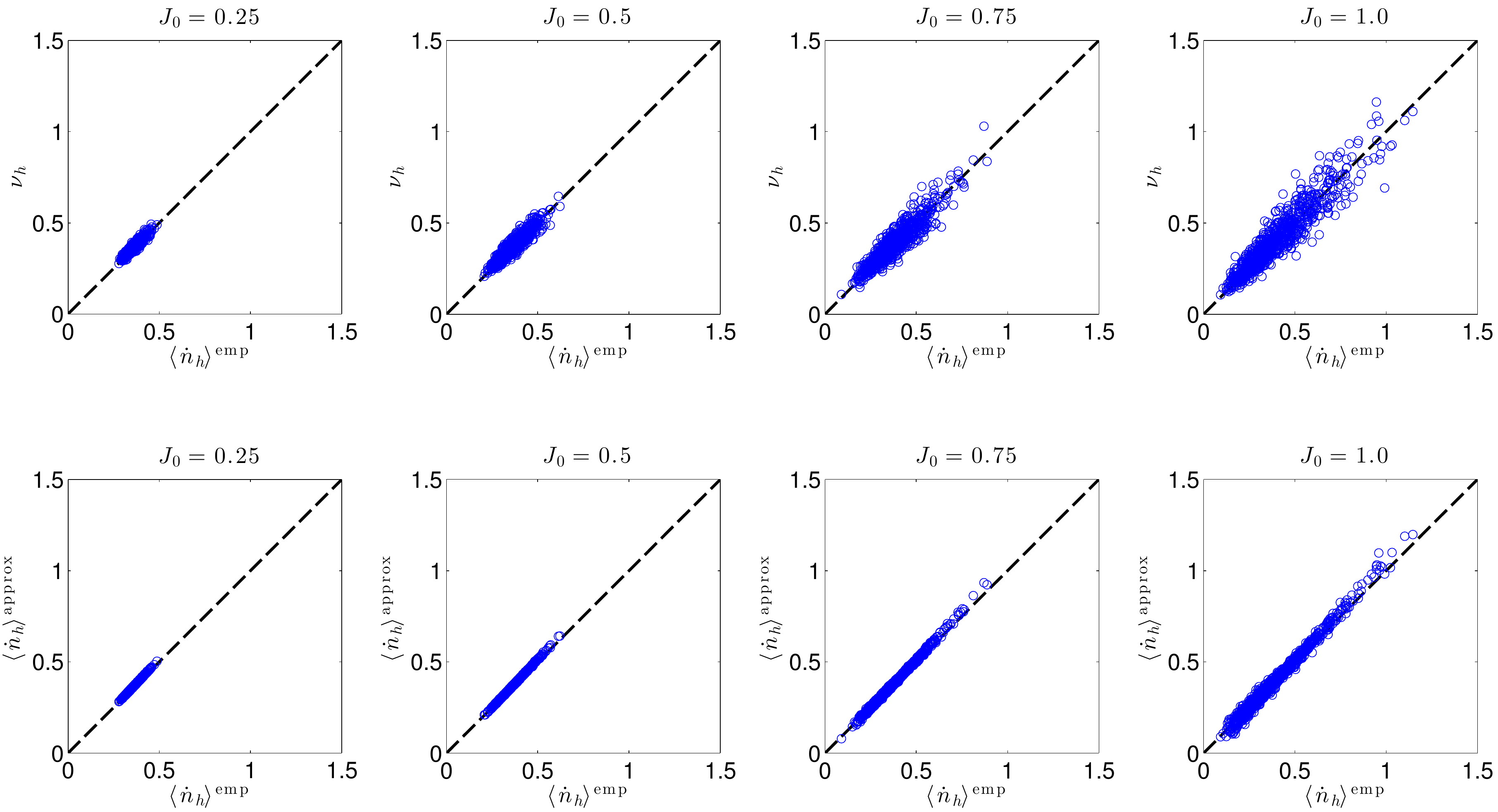}
  \caption{\textbf{Top row:} scatter plot comparing $\nu_h$, the mean field firing rates of the hidden neurons in the absence of recorded neurons, to empirically estimated firing rates in simulations of the full network, for four different values of typical synaptic strength, $J_0 = 0.25, 0.5, 0.75$, and $1.0$. The data lie along the identity line, demonstrating a strong correlation between $\nu_h$ and the empirical data. However, the spread of data around the identity line indicates that deviations of the mean firing rates away from $\nu_h$, caused by coupling to the recorded neurons, is significant. \textbf{Bottom row:} Comparison of the first order approximation of the firing rates of hidden neurons, which accounts for the effects of recorded neurons, to the empirical rates. The data lie tightly along the identity with very little dispersion, demonstrating that higher order spike filtering is unnecessary even up to $J_0 = 1.0$, for $N_{\rm rec} = 100$.}
  \label{fig:linearapproxvalidation}
\end{figure}

\begin{figure}[h!]
 \centering
 \includegraphics[width=\textwidth]{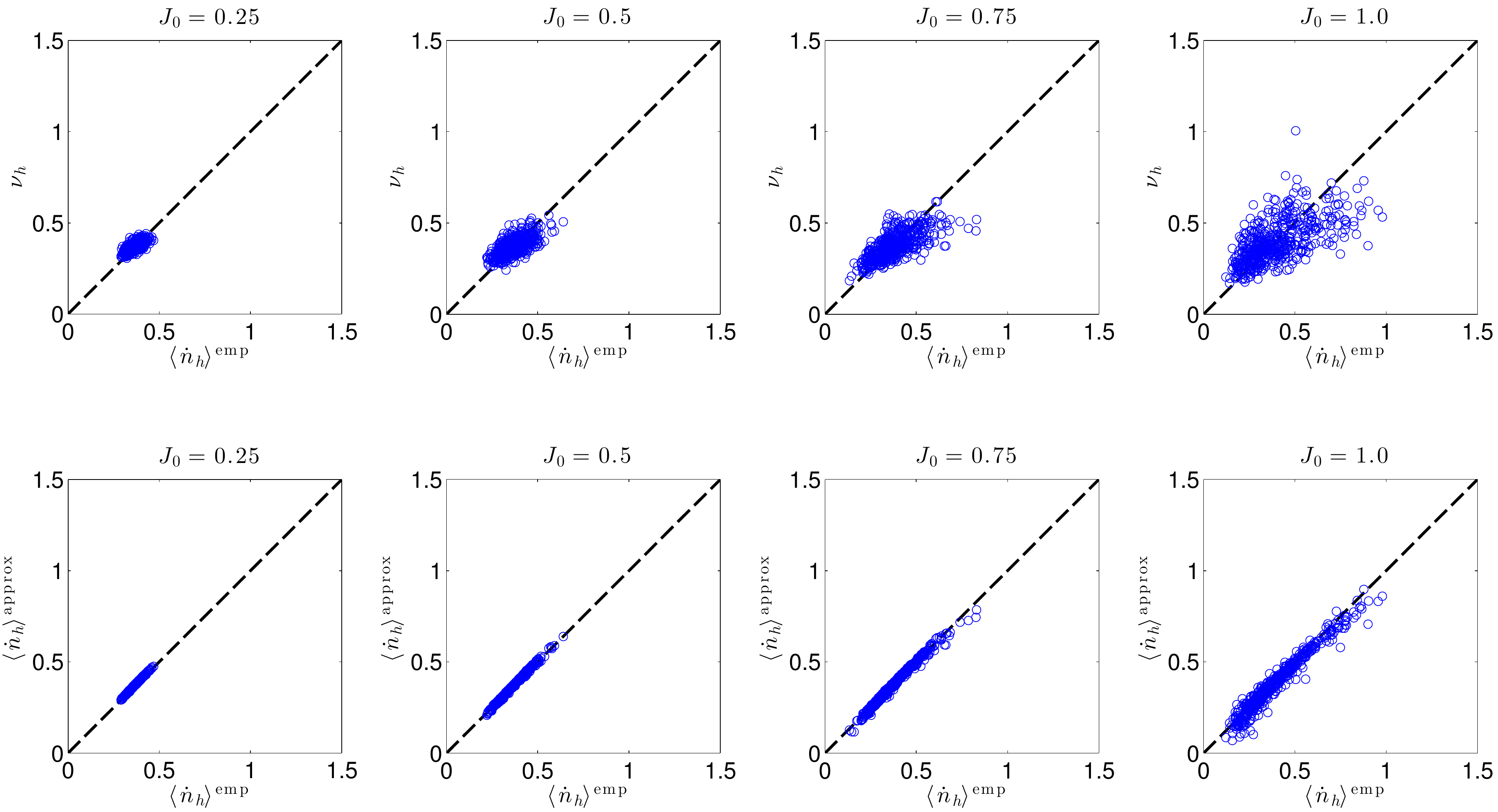}
  \caption{Same as Fig.~\ref{fig:linearapproxvalidation} but for $N_{\rm rec} = 500$ recorded neurons out of a total of $N = 1000$. Demonstrates validity of linear approximation (neglecting higher order spike filtering) up to $J_0 = 1.0$, for $N_{\rm rec} = 500$. The zeroth order approximation (top row) is quite poor, indicating the necessity of accounting for feedback from the recorded neurons. This first order approximation (bottom row) lies tightly along the identity line, indicating that even when the recorded and hidden populations are of comparable size, higher order spike filtering may not be significant. However, there appears to be some deviation for $J_0 = 1.0$, indicating that accounting for higher order spike filtering may be beneficial in this parameter regime.}
  \label{fig:linearapproxvalidationNhid500}
\end{figure}

As in the mean field approximation test, we focused on a strongly coupled network with $J_0 = 0.25, 0.5, 0.75,$ and $1.0$. In the SI we analytically estimate the error, predicting it is small for both small and large fractions of recorded neurons and largest error when $N_{\rm rec} \sim N_{\rm hid}$, so we check both $N_{\rm rec} = 100$ neurons out of $N = 1000$ neurons ($f = 0.1$) in Fig.~\ref{fig:linearapproxvalidation} and $N_{\rm rec} = 500$ neurons out of $N = 1000$ ($f = 0.5$) in Fig.~\ref{fig:linearapproxvalidationNhid500}. 

For each value of $J_0$, we present two plots: the empirical rates versus the mean field rates $\nu_h$ in the absence of recorded neurons (the zeroth order approximation; Figs.~\ref{fig:linearapproxvalidation} and \ref{fig:linearapproxvalidationNhid500}, top row), and the empirical rates versus the linear approximation $\langle \dot{n}_h\rangle^{\rm approx}$ (the first order approximation; Figs.~\ref{fig:linearapproxvalidation} and \ref{fig:linearapproxvalidationNhid500}, bottom row). We find that in both cases the data is centered around the identity line, but the spread of data grows with $J_0$ for the zeroth order approximation, while it is quite tight for the first order approximation up to $J_0 = 1.0$, validating our neglect of the higher order spike filtering terms. We also confirm that $N_{\rm rec} = 500$ offers worse agreement than $N_{\rm rec} = 100$, though the agreement between $\langle \dot{n}_h \rangle^{\rm emp}$ and $\langle \dot{n}_h \rangle^{\rm approx}$ is still not too bad. 

\subsection*{Full mean-field reference state}
\label{sec:fullmft}

For most of our analyses, we have been expanding the conditional firing rates of the hidden neurons around a reference state of zero activity of the recorded neurons. The quantities $\nu_h$, $\gamma_h$, $\hat{\Gamma}_{h,h'}(\omega)$, and so on, are thus calculated using a network in which the recorded neurons have been removed. We have demonstrated that this approximation is valid for the networks considered in this paper. However, this approximation may break down in networks in which the recorded neurons spike at high rates. In this case, we may need another reference state to expand the conditional rates around. A natural choice of reference state $\dot{n}^{(0)}_r(t)$ in this case would be the mean firing rates of the neurons. We will set up this expansion here. 

The mean firing rates of the neurons are intractable to calculate exactly, so we will estimate them by the mean field rates, an approximation that we expect to be accurate in the high-firing rate regime. 

The mean field equations for the full network are
$$\langle \dot{n}_i \rangle = \lambda_0 \phi\left(\mu_i + \sum_{j} J_{ij} \ast \langle \dot{n}_j \rangle \right).$$

We can then expand $\mathbb{E}\left[ \dot{n}_h | \left\{\dot{n}_r \right\} \right]$ around $\dot{n}_r = \langle \dot{n}_r\rangle$, truncating at linear order to obtain 
$$\mathbb{E}\left[ \dot{n}_h(t) | \left\{\dot{n}_r \right\} \right] \approx \langle \dot{n}_h \rangle + \sum_{h',r}\int_{-\infty}^\infty dt' dt''~\Gamma^{\rm full}_{h,h'}(t-t') J_{h',r}(t'-t'')(\dot{n}_r(t'') - \langle \dot{n}_r \rangle),$$
where $\Gamma^{\rm full}_{h,h'}(t-t')$ is the input linear response of the \emph{full network}, including the recorded neurons.

We can then approximate the instantaneous firing rates of the recorded neurons by
\begin{align*}
\lambda_r(t) &\approx \lambda_0 \phi\left(\left\{\mu_r + \sum_{r'} J_{r,r'} \ast \langle \dot{n}_r\rangle\right\}  + \sum_{r'} J_{r,r'} \ast (\dot{n}_r -\langle \dot{n}_r\rangle) + \sum_h J_{r,h} \ast\mathbb{E}\left[ \dot{n}_h(t) | \left\{\dot{n}_r \right\} \right] \right) \\
&\approx \lambda_0 \phi\left(\left\{\mu_r + \sum_{r'} J_{r,r'} \ast \langle \dot{n}_r\rangle + \sum_h J_{r,h} \ast \langle \dot{n}_h\rangle \right\}  + \sum_{r'} \left\{J_{r,r'} + \sum_{h,h'} J_{r,h} \ast \Gamma^{\rm full}_{h,h'} \ast J_{h',r'} \right\} \ast (\dot{n}_r -\langle \dot{n}_r\rangle) \right); 
\end{align*}
note that we introduced $0 = \sum_{r'} \langle \dot{n}_{r'}\rangle - \sum_{r'} \langle \dot{n}_{r'}\rangle$ so that we could write the instantaneous firing not as a function of filtered spike trains but as a function of filtered deviations from the mean firing rate. Importantly, although it looks like only the baseline is different from the zero-activity reference state case but the coupling is the same, the linear response function $\Gamma^{\rm full}_{h,h'}(\tau)$ is \emph{not} the same as the zero-reference state case, and hence the correction to the coupling is slightly different. The solutions look similar, but the linear response functions now incorporate the effects of the recorded units as well. In particular, $\Gamma^{\rm full}_{ij}(t-t')$ satisfies the equation
$$\int_{-\infty}^\infty dt''\left[ \delta_{ik} - \gamma^{\rm full}_i J_{ik}(t-t'')\right]\Gamma^{\rm full}_{kj}(t''-t') = \gamma^{\rm full}_i \delta_{ij}\delta(t-t'),$$
where $\gamma^{\rm full}_i$ is the gain of neuron $i$ accounting for the entire network,
$$\gamma^{\rm full}_i = \lambda_0 \phi'\left(\mu_i + \sum_{j} J_{ij} \ast \langle \dot{n}_j \rangle\right).$$
Thus, in Fourier space
\begin{align*}
\hat{\Gamma}^{\rm full}_{ij}(\omega) &= \left[\mathbb{I} - \mathbf{\hat{V}}^{\rm full}(\omega) \right]^{-1}_{ij} \langle \dot{n}_j \rangle \\
&= \sum_{\ell = 0}^\infty \left[\mathbf{\hat{V}}^{\rm full}(\omega))^\ell \right]_{ij} \langle \dot{n}_j \rangle,
\end{align*}
where $\hat{V}_{i,j}^{\rm full}(\omega) = \gamma^{\rm full}_i \hat{J}_{i,j}(\omega)$ is an $N \times N$ matrix -- i.e., it contains the couplings and firing rates of \emph{all} neurons, recorded and hidden. Hence, while this looks formally similar to the result we obtained in the zero activity state, the inclusion of recorded neurons modifies our rules for calculating contributions to the effective coupling filters. In particular, $\hat{J}^{\rm eff}_{r,r'}(\omega)- \hat{J}_{r,r'}(\omega)$ involves contributions from paths through both hidden and recorded neurons, unlike the zero-activity reference case, which involved contributions only from paths through hidden neurons. The reason for this, of course, is that the reference state depends on the entire network, not just the hidden neurons. The difference between the cases matters only at higher orders in our expansion --- paths of length $\ell = 4$ or greater. We can see this by writing out the first few terms in the path length expansion of the effective coupling,
\begin{align*}
\hat{J}^{\rm eff}_{r,r'}(\omega) &= \hat{J}_{r,r'}(\omega) + \sum_h \hat{J}_{r,h}(\omega) \gamma^{\rm full}_h \hat{J}_{h,r} (\omega) + \sum_{h,h'} \hat{J}_{r,h}(\omega) \gamma^{\rm full}_h \hat{J}_{h,h'}(\omega) \gamma^{\rm full}_{h'} \hat{J}_{h',r'}(\omega) \\
& ~~~~~~~~~~~~~~~~+ \sum_{h,h',j} \hat{J}_{r,h}(\omega) \gamma^{\rm full}_h \hat{J}_{h,j}(\omega) \gamma^{\rm full}_j \hat{J}_{j,h'}(\omega) \gamma^{\rm full}_{h'} \hat{J}_{h',r'}(\omega) + \dots;
\end{align*}
for conciseness, we have assumed zero-self coupling ($\hat{J}_{i,i}(\omega) = 0$), but this can be restored by setting $\gamma^{\rm full}_i \rightarrow  \gamma^{\rm full}_i/(1-\gamma^{\rm full}_i \hat{J}_{i,i}(\omega))$. 

We see that the first few terms of the expansion are the same as the zero-activity reference case, with the exception that the $\gamma^{\rm full}_h$ are the gains for the entire network, not just the hidden network absent the recorded neurons. It is only the $\ell = 4$ term at which contributions to the linear response functions involving paths through any neuron $j$, recorded or hidden, start to appear. Because we typically expect the amplitude of these terms to be small, we anticipate expanding around the mean field reference state will yield similar results to the expansion around the zero-activity reference state presented in the main paper.

%\nolinenumbers

%%%%%%%%%%%%%%%%%%%%%%%%%%%%%%%%
%% REFERENCES 
%%%%%%%%%%%%%%%%%%%%%%%%%%%%%%%%
%

\bibliographystyle{unsrt}
\bibliography{paper1}

\begin{thebibliography}{10}

\bibitem{BassettJNeurosci2008}
Danielle~S. Bassett, Edward Bullmore, Beth~A. Verchinski, Venkata~S. Mattay,
  Daniel~R. Weinberger, and Andreas Meyer-Lindenberg.
\newblock Hierarchical organization of human cortical networks in health and
  schizophrenia.
\newblock {\em Journal of Neuroscience}, 28(37):9239--9248, 2008.

\bibitem{KramerEpilepsyRes2008}
Mark~A. Kramer, Eric~D. Kolaczyk, and Heidi~E. Kirsch.
\newblock Emergent network topology at seizure onset in humans.
\newblock {\em Epilepsy Research}, 79(2–3):173 -- 186, 2008.

\bibitem{SupekarPLOSCB2008}
Kaustubh Supekar, Vinod Menon, Daniel Rubin, Mark Musen, and Michael~D.
  Greicius.
\newblock Network analysis of intrinsic functional brain connectivity in
  alzheimer's disease.
\newblock {\em PLOS Computational Biology}, 4(6):1--11, 06 2008.

\bibitem{LoJNeurosci2010}
Chun-Yi Lo, Pei-Ning Wang, Kun-Hsien Chou, Jinhui Wang, Yong He, and Ching-Po
  Lin.
\newblock Diffusion tensor tractography reveals abnormal topological
  organization in structural cortical networks in alzheimer{\textquoteright}s
  disease.
\newblock {\em Journal of Neuroscience}, 30(50):16876--16885, 2010.

\bibitem{ChavezPRL2010}
M.~Chavez, M.~Valencia, V.~Navarro, V.~Latora, and J.~Martinerie.
\newblock Functional modularity of background activities in normal and
  epileptic brain networks.
\newblock {\em Phys. Rev. Lett.}, 104:118701, Mar 2010.

\bibitem{DouwPLOSOne2010}
Linda Douw, Marjolein de~Groot, Edwin van Dellen, Jan~J. Heimans, Hanneke~E.
  Ronner, Cornelis~J. Stam, and Jaap~C. Reijneveld.
\newblock ‘functional connectivity’ is a sensitive predictor of epilepsy
  diagnosis after the first seizure.
\newblock {\em PLOS ONE}, 5(5):1--7, 05 2010.

\bibitem{vanDiessenNeuroImage2013}
Eric van Diessen, Judith~I. Hanemaaijer, Willem~M. Otte, Rina Zelmann, Julia
  Jacobs, Floor~E. Jansen, François Dubeau, Cornelis~J. Stam, Jean Gotman, and
  Maeike Zijlmans.
\newblock Are high frequency oscillations associated with altered network
  topology in partial epilepsy?
\newblock {\em NeuroImage}, 82:564 -- 573, 2013.

\bibitem{ReijmerNeurology2013}
Yael~D. Reijmer, Alexander Leemans, Karen Caeyenberghs, Sophie~M. Heringa,
  Huiberdina~L. Koek, Geert~Jan Biessels, and On~behalf of~the Utrecht Vascular
  Cognitive Impairment Study~Group.
\newblock Disruption of cerebral networks and cognitive impairment in alzheimer
  disease.
\newblock {\em Neurology}, 80(15):1370--1377, 2013.

\bibitem{SeoPLOSOne2013}
Eun~Hyun Seo, Dong~Young Lee, Jong-Min Lee, Jun-Sung Park, Bo~Kyung Sohn,
  Dong~Soo Lee, Young~Min Choe, and Jong~Inn Woo.
\newblock Whole-brain functional networks in cognitively normal, mild cognitive
  impairment, and alzheimer’s disease.
\newblock {\em PLOS ONE}, 8(1):1--11, 01 2013.

\bibitem{StamNatRevNeuro2014}
Cornelis~J. Stam.
\newblock Modern network science of neurological disorders.
\newblock {\em Nat Rev Neurosci}, 15(10):683--695, 10 2014.

\bibitem{WarrenPNAS2014}
David~E. Warren, Jonathan~D. Power, Joel Bruss, Natalie~L. Denburg, Eric~J.
  Waldron, Haoxin Sun, Steven~E. Petersen, and Daniel Tranel.
\newblock Network measures predict neuropsychological outcome after brain
  injury.
\newblock {\em Proceedings of the National Academy of Sciences},
  111(39):14247--14252, 2014.

\bibitem{OldeDubbelinkBrain2014}
Kim T.~E. Olde~Dubbelink, Arjan Hillebrand, Diederick Stoffers, Jan~Berend
  Deijen, Jos W.~R. Twisk, Cornelis~J. Stam, and Henk~W. Berendse.
\newblock Disrupted brain network topology in parkinson{\textquoteright}s
  disease: a longitudinal magnetoencephalography study.
\newblock {\em Brain}, 137(1):197--207, 2014.

\bibitem{BernhardEpBeh2015}
Boris~C. Bernhardt, Leonardo Bonilha, and Donald~W. Gross.
\newblock Network analysis for a network disorder: The emerging role of graph
  theory in the study of epilepsy.
\newblock {\em Epilepsy \& Behavior}, 50:162 -- 170, 2015.

\bibitem{MedagliaArxiv2017}
J.~D. {Medaglia} and D.~S. {Bassett}.
\newblock {Network Analyses and Nervous System Disorders}.
\newblock {\em ArXiv e-prints}, January 2017.

\bibitem{HoneyNeuroImage2010}
Christopher~J. Honey, Jean-Philippe Thivierge, and Olaf Sporns.
\newblock Can structure predict function in the human brain?
\newblock {\em NeuroImage}, 52(3):766 -- 776, 2010.
\newblock Computational Models of the Brain.

\bibitem{Simoncelli2004}
EP~Simoncelli, L~Paninski, JW~Pillow, and O~Schwartz.
\newblock {Characterization of Neural Responses with Stochastic Stimuli}.
\newblock In M~Gazzaniga, editor, {\em The Cognitive Neurosciences}, pages
  327--338. MIT Press, 3rd edition, 2004.

\bibitem{Paninski2004}
L~Paninski.
\newblock {Maximum likelihood estimation of cascade point-process neural
  encoding models}.
\newblock {\em Network: Computation in Neural Systems}, 15(4):243--262,
  November 2004.

\bibitem{Pillow2005}
JW~Pillow, L~Paninski, Valerie~J Uzzell, EP~Simoncelli, and E~J Chichilnisky.
\newblock {Prediction and decoding of retinal ganglion cell responses with a
  probabilistic spiking model.}
\newblock {\em The Journal of neuroscience : the official journal of the
  Society for Neuroscience}, 25(47):11003--13, November 2005.

\bibitem{KulkarniNetworkCompNeuralSys2007}
Jayant~E. Kulkarni and Liam Paninski.
\newblock Common-input models for multiple neural spike-train data.
\newblock {\em Network: Computation in Neural Systems}, 18(4):375--407, 2007.
\newblock PMID: 17943613.

\bibitem{Pillow2008}
JW~Pillow, J~Shlens, L~Paninski, A~Sher, AM~Litke, EJ~Chichilnisky, and
  EP~Simoncelli.
\newblock {Spatio-temporal correlations and visual signalling in a complete
  neuronal population.}
\newblock {\em Nature}, 454(7207):995--9, 2008.

\bibitem{Field2010}
GD~Field, J.~L. Gauthier, A~Sher, Martin Greschner, Timothy~a Machado, Lauren~H
  Jepson, J~Shlens, Deborah~E Gunning, Keith Mathieson, Wladyslaw Dabrowski,
  L~Paninski, AM~Litke, and E~J Chichilnisky.
\newblock {Functional connectivity in the retina at the resolution of
  photoreceptors.}
\newblock {\em Nature}, 467(7316):673--7, October 2010.

\bibitem{Vidne2012}
M~Vidne, Y~Ahmadian, J~Shlens, JW~Pillow, JE~Kulkarni, AM~Litke,
  EJ~Chichilnisky, EP~Simoncelli, and L~Paninski.
\newblock {Modeling the impact of common noise inputs on the network activity
  of retinal ganglion cells}.
\newblock {\em J Comput Neurosci}, 33(1):97--121, 2012.

\bibitem{Paninski2015}
Liam Paninski.
\newblock Maximum likelihood estimation of cascade point- process neural
  encoding models.
\newblock {\em Network : Computation in Neural Systems}, 6536(October), 2015.

\bibitem{HuangJPhysA2015}
Haiping Huang.
\newblock Effects of hidden nodes on network structure inference.
\newblock {\em Journal of Physics A: Mathematical and Theoretical},
  48(35):355002, 2015.

\bibitem{DahlhausMetrika2000}
Rainer Dahlhaus.
\newblock Graphical interaction models for multivariate time series1.
\newblock {\em Metrika}, 51(2):157--172, 2000.

\bibitem{EichlerBiolCyber2003}
Michael Eichler, Rainer Dahlhaus, and J{\"u}rgen Sandk{\"u}hler.
\newblock Partial correlation analysis for the identification of synaptic
  connections.
\newblock {\em Biological Cybernetics}, 89(4):289--302, 2003.

\bibitem{PillowNIPS2007}
Jonathan~W. Pillow and Peter~E. Latham.
\newblock Neural characterization in partially observed populations of spiking
  neurons.
\newblock In J.c. Platt, D.~Koller, Y.~Singer, and S.~Roweis, editors, {\em
  Advances in Neural Information Processing Systems 20}, pages 1161--1168. MIT
  Press, Cambridge, MA, 2007.

\bibitem{StevensonCurrOpBio2008}
Ian~H Stevenson, James~M Rebesco, Lee~E Miller, and Konrad~P Körding.
\newblock Inferring functional connections between neurons.
\newblock {\em Current Opinion in Neurobiology}, 18(6):582 -- 588, 2008.

\bibitem{StevensonPLOSCB2012}
Ian~H. Stevenson, Brian~M. London, Emily~R. Oby, Nicholas~A. Sachs, Jacob
  Reimer, Bernhard Englitz, Stephen~V. David, Shihab~A. Shamma, Timothy~J.
  Blanche, Kenji Mizuseki, Amin Zandvakili, Nicholas~G. Hatsopoulos, Lee~E.
  Miller, and Konrad~P. Kording.
\newblock Functional connectivity and tuning curves in populations of
  simultaneously recorded neurons.
\newblock {\em PLOS Computational Biology}, 8(11):1--14, 11 2012.

\bibitem{LiegeoisIEEE2015}
R.~Li\'{e}geois, B.~Mishra, M.~Zorzi, and R.~Sepulchre.
\newblock Sparse plus low-rank autoregressive identification in neuroimaging
  time series.
\newblock In {\em 2015 54th IEEE Conference on Decision and Control (CDC)},
  pages 3965--3970, Dec 2015.

\bibitem{PetersRSSB2016}
Jonas Peters, Peter Bühlmann, and Nicolai Meinshausen.
\newblock Causal inference by using invariant prediction: identification and
  confidence intervals.
\newblock {\em Journal of the Royal Statistical Society: Series B (Statistical
  Methodology)}, 78(5):947--1012, 2016.

\bibitem{FotiMILETS2016}
Nicholas~J. Foti, Rahul Nadkarni, Adrian~KC Lee, and Emily~B. Fox.
\newblock Sparse plus low-rank graphical models of time series for functional
  connectivity in meg.
\newblock In {\em 2nd KDD Workshop on Mining and Learning from Time Series}.
  2016.

\bibitem{Prinz2004}
Astrid~a Prinz, Dirk Bucher, and Eve Marder.
\newblock {Similar network activity from disparate circuit parameters.}
\newblock {\em Nature neuroscience}, 7(12):1345--52, December 2004.

\bibitem{GutenkunstPLOSCB2007}
Ryan~N Gutenkunst, Joshua~J Waterfall, Fergal~P Casey, Kevin~S Brown,
  Christopher~R Myers, and James~P Sethna.
\newblock Universally sloppy parameter sensitivities in systems biology models.
\newblock {\em PLOS Computational Biology}, 3(10):1--8, 10 2007.

\bibitem{ApgarMolBioSyst2010}
Joshua~F. Apgar, David~K. Witmer, Forest~M. White, and Bruce Tidor.
\newblock Sloppy models{,} parameter uncertainty{,} and the role of
  experimental design.
\newblock {\em Mol. BioSyst.}, 6:1890--1900, 2010.

\bibitem{GutierrezNeuron2013}
Gabrielle~J. Gutierrez, Timothy O'Leary, and Eve Marder.
\newblock Multiple mechanisms switch an electrically coupled, synaptically
  inhibited neuron between competing rhythmic oscillators.
\newblock {\em Neuron}, 77(5):845--858, 2013.

\bibitem{FisherNeuron2013}
Dimitry Fisher, Itsaso Olasagasti, David~W. Tank, Emre R.~F. Aksay, and Mark~S.
  Goldman.
\newblock A modeling framework for deriving the structural and functional
  architecture of a short-term memory microcircuit.
\newblock {\em Neuron}, 79(5):987--1000, 2013.

\bibitem{MarderDevNeurobio2017}
Eve Marder, Gabrielle~J. Gutierrez, and Michael~P. Nusbaum.
\newblock Complicating connectomes: Electrical coupling creates parallel
  pathways and degenerate circuit mechanisms.
\newblock {\em Developmental Neurobiology}, 77(5):597--609, 2017.

\bibitem{DunnPRE2013}
Benjamin Dunn and Yasser Roudi.
\newblock Learning and inference in a nonequilibrium ising model with hidden
  nodes.
\newblock {\em Phys. Rev. E}, 87:022127, Feb 2013.

\bibitem{TyrchaMathBioEng2014}
Joanna Tyrcha and John Hertz.
\newblock Network inference with hidden units.
\newblock {\em Mathematical Biosciences and Engineering}, 11(1):149--165, 2014.

\bibitem{DunnArxiv2016}
B.~{Dunn} and C.~{Battistin}.
\newblock {The appropriateness of ignorance in the inverse kinetic Ising
  model}.
\newblock {\em ArXiv e-prints}, December 2016.

\bibitem{PouilleScience2001}
Fr{\'e}d{\'e}ric Pouille and Massimo Scanziani.
\newblock Enforcement of temporal fidelity in pyramidal cells by somatic
  feed-forward inhibition.
\newblock {\em Science}, 293(5532):1159--1163, 2001.

\bibitem{vanVreeswijkScience1996}
C.~van Vreeswijk and H.~Sompolinsky.
\newblock Chaos in neuronal networks with balanced excitatory and inhibitory
  activity.
\newblock {\em Science}, 274(5293):1724--1726, 1996.

\bibitem{LitwinKumarNatNeuro2012}
Ashok Litwin-Kumar and Brent Doiron.
\newblock Slow dynamics and high variability in balanced cortical networks with
  clustered connections.
\newblock {\em Nat Neurosci}, 15(11):1498--1505, 11 2012.

\bibitem{RosenbaumPRX2014}
Robert Rosenbaum and Brent Doiron.
\newblock Balanced networks of spiking neurons with spatially dependent
  recurrent connections.
\newblock {\em Phys. Rev. X}, 4:021039, May 2014.

\bibitem{DeneveNatNeuro2016}
Sophie Deneve and Christian~K Machens.
\newblock Efficient codes and balanced networks.
\newblock {\em Nat Neurosci}, 19(3):375--382, 03 2016.

\bibitem{BarralNatNeuro2016}
Jeremie Barral and Alex D~Reyes.
\newblock Synaptic scaling rule preserves excitatory-inhibitory balance and
  salient neuronal network dynamics.
\newblock {\em Nat Neurosci}, 19(12):1690--1696, 12 2016.

\bibitem{NykampSIAMJAM2005}
Duane~Q. Nykamp.
\newblock Revealing pairwise coupling in linear-nonlinear networks.
\newblock {\em SIAM Journal on Applied Mathematics}, 65(6):2005--2032, 2005.

\bibitem{NykampMathBiosci2007}
Duane~Q. Nykamp.
\newblock A mathematical framework for inferring connectivity in probabilistic
  neuronal networks.
\newblock {\em Mathematical Biosciences}, 205(2):204 -- 251, 2007.

\bibitem{NykampSIAMJAM2007}
Duane~Q. Nykamp.
\newblock Exploiting history-dependent effects to infer network connectivity.
\newblock {\em SIAM Journal on Applied Mathematics}, 68(2):354--391, 2007.

\bibitem{NykampPRE2008}
Duane~Q. Nykamp.
\newblock Pinpointing connectivity despite hidden nodes within stimulus-driven
  networks.
\newblock {\em Phys. Rev. E}, 78:021902, Aug 2008.

\bibitem{OckerPLOSCB2017}
Gabriel~Koch Ocker, Krešimir Josić, Eric Shea-Brown, and Michael~A. Buice.
\newblock Linking structure and activity in nonlinear spiking networks.
\newblock {\em PLOS Computational Biology}, 13(6):1--46, 06 2017.

\bibitem{ChornoboyBiolCyber1988}
E.~S. Chornoboy, L.~P. Schramm, and A.~F. Karr.
\newblock Maximum likelihood identification of neural point process systems.
\newblock {\em Biological Cybernetics}, 59(4):265--275, 1988.

\bibitem{GerstnerBook}
Wulfram Gerstner, Werner~M. Kistler, Richard Naud, and Liam Paninski.
\newblock {\em Neuronal Dynamics: From single neurons to networks and models of
  cognition}.
\newblock Cambridge University Press, Cambridge, U.K., 2014.

\bibitem{OstojicPLOSCB2011}
Srdjan Ostojic and Nicolas Brunel.
\newblock From spiking neuron models to linear-nonlinear models.
\newblock {\em PLOS Computational Biology}, 7(1):1--16, 01 2011.

\bibitem{BraviArxiv2016}
B.~{Bravi} and P.~{Sollich}.
\newblock {Statistical physics approaches to subnetwork dynamics in biochemical
  systems}.
\newblock {\em ArXiv e-prints}, November 2016.

\bibitem{BraviArxiv2016b}
B.~{Bravi} and P.~{Sollich}.
\newblock {Critical scaling in hidden state inference for linear Langevin
  dynamics}.
\newblock {\em ArXiv e-prints}, December 2016.

\bibitem{BraviPRE2017}
B.~Bravi, M.~Opper, and P.~Sollich.
\newblock Inferring hidden states in langevin dynamics on large networks:
  Average case performance.
\newblock {\em Phys. Rev. E}, 95:012122, Jan 2017.

\bibitem{PernicePLOSCB2011}
Volker Pernice, Benjamin Staude, Stefano Cardanobile, and Stefan Rotter.
\newblock How structure determines correlations in neuronal networks.
\newblock {\em PLOS Computational Biology}, 7(5):1--14, 05 2011.

\bibitem{HuJStatMechThExp2013}
Yu~Hu, James Trousdale, Krešimir Josić, and Eric Shea-Brown.
\newblock Motif statistics and spike correlations in neuronal networks.
\newblock {\em Journal of Statistical Mechanics: Theory and Experiment},
  2013(03):P03012, 2013.

\bibitem{HuPRE2014}
Yu~Hu, James Trousdale, Kre{\v{s}}imir Josi{\'{c}}, and Eric Shea-Brown.
\newblock Local paths to global coherence: Cutting networks down to size.
\newblock {\em Phys. Rev. E}, 89:032802, Mar 2014.

\bibitem{GoldenfeldBook1992}
Nigel Goldenfeld.
\newblock {\em {L}ectures on {P}hase {T}ransitions and the {R}enormalization
  {G}roup}.
\newblock Westview Press, 1992.

\bibitem{MachtaScience2013}
Benjamin~B. Machta, Ricky Chachra, Mark~K. Transtrum, and James~P. Sethna.
\newblock Parameter space compression underlies emergent theories and
  predictive models.
\newblock {\em Science}, 342(6158):604--607, 2013.

\bibitem{CaycoGajicFrontiersCompNeuro2015}
Natasha~A. Cayco-Gajic, Joel Zylberberg, and Eric Shea-Brown.
\newblock Triplet correlations among similarly tuned cells impact population
  coding.
\newblock {\em Frontiers in Computational Neuroscience}, 9:57, 2015.

\bibitem{SuppInfo}
{\em See Supplementary Information.}

\bibitem{ErdosRussianMathSurv2011}
Laszlo Erdős.
\newblock Universality of wigner random matrices: a survey of recent results.
\newblock {\em Russian Mathematical Surveys}, 66(3):507, 2011.

\bibitem{AhmadianPhysRevE2015}
Yashar Ahmadian, Francesco Fumarola, and Kenneth~D. Miller.
\newblock Properties of networks with partially structured and partially random
  connectivity.
\newblock {\em Phys. Rev. E}, 91:012820, Jan 2015.

\bibitem{TaoCommContMath2008}
TERENCE TAO and VAN VU.
\newblock Random matrices: The circular law.
\newblock {\em Communications in Contemporary Mathematics}, 10(02):261--307,
  2008.

\bibitem{LatimerNIPS2014}
{Kenneth W.} Latimer, {E. J.} Chichilnisky, Fred Rieke, and {Jonathan W.}
  Pillow.
\newblock {\em Inferring synaptic conductances from spike trains under a
  biophysically inspired point process model}, volume~2, pages 954--962.
\newblock Neural information processing systems foundation, january edition,
  2014.

\bibitem{SompolinskyPRL1988}
H.~Sompolinsky, A.~Crisanti, and H.~J. Sommers.
\newblock Chaos in random neural networks.
\newblock {\em Phys. Rev. Lett.}, 61:259--262, Jul 1988.

\bibitem{MatrixAnalysisHornJohnson}
Roger~A. Horn and Charles~R. Johnson, editors.
\newblock {\em Matrix Analysis}.
\newblock Cambridge University Press, New York, NY, USA, 1986.

\bibitem{SongPRE2014}
H.~Francis Song and Xiao-Jing Wang.
\newblock Simple, distance-dependent formulation of the watts-strogatz model
  for directed and undirected small-world networks.
\newblock {\em Phys. Rev. E}, 90:062801, Dec 2014.

\bibitem{HawkesBiometrika1971}
ALAN~G. HAWKES.
\newblock Spectra of some self-exciting and mutually exciting point processes.
\newblock {\em Biometrika}, 58(1):83, 1971.

\bibitem{bremaudAnnProb1996}
Pierre Br\'{e}maud and Laurent Massouli\'{e}.
\newblock Stability of nonlinear hawkes processes.
\newblock {\em Ann. Probab.}, 24(3):1563--1588, 07 1996.

\end{thebibliography}

\end{document}